\documentclass[manuscript,nonacm,table,acsmall]{acmart}
\usepackage{array}
\usepackage[utf8]{inputenc}
\usepackage{graphicx}
\usepackage{hyperref}
\usepackage{pdfpages}
\usepackage{multirow}
\usepackage{longtable}

\usepackage{caption,subcaption}
\usepackage{cleveref}
\usepackage{vcell}

\PassOptionsToPackage{numbers, compress}{natbib}

\AtBeginDocument{%
  }

\setcopyright{none}
\begin{document}

\title[Limits to AI Growth]{Limits to AI Growth: The Ecological and Social Consequences of Scaling}

\author{Eshta Bhardwaj}
\email{eshta.bhardwaj@mail.utoronto.ca}
\orcid{}
\affiliation{%
  \institution{University of Toronto}
  \streetaddress{}
  \city{Toronto}
  \state{Ontario}
  \country{Canada}
  \postcode{}
}
\author{Rohan Alexander}
\email{rohan.alexander@utoronto.ca}
\orcid{}
\affiliation{%
  \institution{University of Toronto}
  \streetaddress{}
  \city{Toronto}
  \state{Ontario}
  \country{Canada}
  \postcode{}
}

\author{Christoph Becker}
\email{christoph.becker@utoronto.ca}
\orcid{}
\affiliation{%
  \institution{University of Toronto}
  \streetaddress{}
  \city{Toronto}
  \state{Ontario}
  \country{Canada}
  \postcode{}
}

\renewcommand{\shortauthors}{Bhardwaj et al.}

\begin{abstract}

The accelerating development and deployment of AI technologies depend on the continued ability to scale their infrastructure. This has implied increasing amounts of monetary investment and natural resources. Frontier AI applications have thus resulted in rising financial, environmental, and social costs. While the factors that AI scaling depends on reach its limits, the push for its accelerated advancement and entrenchment continues. In this paper, we provide a holistic review of AI scaling using four lenses (technical, economic, ecological, and social) and review the relationships between these lenses to explore the dynamics of AI growth. We do so by drawing on system dynamics concepts including archetypes such as ``limits to growth'' to model the dynamic complexity of AI scaling and synthesize several perspectives. Our work maps out the entangled relationships between the technical, economic, ecological and social perspectives and the apparent limits to growth. The analysis explains how industry’s responses to external limits enables continued (but temporary) scaling and how this benefits Big Tech while externalizing social and environmental damages. To avoid an “overshoot and collapse” trajectory, we advocate for realigning priorities and norms around scaling to prioritize sustainable and mindful advancements.

\end{abstract}





\maketitle

\section{Introduction} \label{introduction}

In the past decade, the scaling of AI models has accelerated rapidly due to a combination of foundational research advances, improvements in technical infrastructure, access to huge amounts of Internet-based data, and significant capital investment. Data storage alone requires considerable amounts of energy in order to maintain data centers, equivalent to the energy usage of small towns \cite{dayarathna_data_2015,Eric_Masanet_Nuoa_Lei_2020} and predicted to increase to the usage of entire countries \cite{bourzac_fixing_2024,calvert_ai_2024}. The amount of capital invested into driving further advancements in the AI industry yields numerous social implications. Large Language Models (LLMs) in particular have become pervasive and their environmental impact has become a significant concern. On the other hand, recent debates have raised doubts over the feasibility of continued scaling including concerns over the end of training data \cite{Sutskever_2024}, industry profitability \cite{Al-Sibai_2025}, and the industry's mounting climate impact \cite{Temple_2024}. Critics have variously examined the social harms \cite{acemoglu_harms_2021,hagerty_global_2019,lacroix_deep_2023,rogers_position_2024,smuha_beyond_2021}, emphasized the environmental damages wrought by generative AI \cite{bashir_climate_2024,luccioni_counting_2023}, and questioned industry commitments to ethical principles \cite{abdalla_grey_2021,bietti_ethics_2020,unruh_human_2022,van_maanen_ai_2022}. 

It is thus an important moment to ask whether this scaling can and should continue, at what cost, and to whom \cite{varoquaux_hype_2024,widder_watching_2024}. This paper asks: ``\textbf{How can we characterize the dynamics of AI growth to identify and analyze its limits?}''. We use `AI' as an encompassing term including LLMs, generative AI, applications of machine learning models including deep learning, and we use `AI industry' to refer to large or frontier companies in the development of AI technologies and hardware. 

% We focus here on how these perspectives interrelate to explore the \textit{limits to growth}  for AI. 

% Our goal is to map the current landscape of AI expansion and systematically examine the relationships between several key perspectives. Accordingly, our paper investigates, “How should we characterize the dynamics of AI growth to identify its limits to growth and navigable paths forward?”. 
% Our work thus offers an approach for reasoning and understanding the relationships between these lenses.

We first establish a baseline by reviewing recent discussions about AI scaling and introduce the basic modeling constructs of system dynamics (Section \ref{sec:background}). We then examine AI scaling from four perspectives – technical, economical, ecological, and social (Section \ref{sec:map}). For each, we highlight \textit{barriers} that are surfacing. This allows us to model how these factors interact by using system dynamics to represent causal relationships across the four perspectives and demonstrate \textbf{limits to growth} \cite{meadows_limits_1972,meadows_limits_2004} in AI scaling. We model systemic relationships between the technical aspects of scaling models and data volumes, the economic implications, the ecological consequences, and the ethical quandaries this raises in Section \ref{sec:dynamics}. In Section \ref{sec:limits}, we use \textit{archetypes} (recurring patterns) to show how common attempts to overcome limits to growth are prone to fail or shift the burden and are thus likely to increase damages. In Section \ref{sec:discussion}, we discuss social and ecological harms that have resulted from AI scaling, explore the risks of following the trajectory of \textbf{overshoot and collapse} (archetype), and identify alternative paths for development. 

\section{Background} \label{sec:background}

\noindent\textbf{Related Work.} In 2021, Bender et al. asked “can language models be too big?”. They highlighted that “... increasing the environmental and financial costs of these models doubly punishes marginalized communities that are least likely to benefit from the progress achieved by large LMs and most likely to be harmed by negative environmental consequences of its resource consumption.” \citep[p.~610]{bender_dangers_2021}. The uneven distribution of benefits and harms of large models has been acknowledged in discussions of intrinsic and extrinsic harms \cite{Liang_Wu_Morency_Salakhutdinov_2021}, labour and wage inequality \cite{bommasani_opportunities_2022}, surveillance through data collection \cite{zuboff_age_2023}, centralization of the development of large models \cite{whittaker_steep_2021}, and increased usage of decision-making systems that may cause a “monoculture” \cite{oneil_weapons_2017,Kleinberg_Raghavan_2021}. It has also been observed that these models exacerbate societal and social concerns like the deterioration of content quality through misinformation \cite{monteith_artificial_2024,xu_combating_2023,zhou_synthetic_2023}, harmful content \cite{varantsou_image_2024}, and security and privacy \cite{gupta_chatgpt_2023}. When a technology is introduced, it does not exist in a vacuum that only advances that technology or frontier, but rather has multi-dimensional impacts to both society and the environment. 

Several taxonomies of harms or risks within the AI landscape exist, (e.g., \cite{bommasani_opportunities_2022,Diamond_Banerjee_2024,Rillig_Ågerstrand_Bi_Gould_Sauerland_2023, Weidinger_2022}). Each of these outline repercussions that occur beyond the confines of the surroundings that the AI technology or model is produced in. Particularly, a social-ecological-technical (SET) framework to analyze the outcomes of algorithmic technologies has been proposed “...to [study] complex dynamical systems and their constitutive relations...” \citep[p.~3]{rakova_algorithms_2023}. Thus by examining technological systems in larger social, economical, and ecological contexts, we can better understand how they interact with and depend on non-technical and mathematical aspects and permeate into our broader ecosystem. 

Other works have similarly highlighted that a wider lens is required to address “algorithmic injustice” such as using relational ethics as a framework to better understand and act on “how we can re-examine our underlying working assumptions... interrogate hierarchical power asymmetries, and... consider the broader, contingent, and interconnected background that algorithmic systems emerge from
(and are deployed to) in the process of protecting the welfare of the most vulnerable” \citep[p.~2]{birhane_algorithmic_2021}.

%It similarly highlights that relational ethics is about considering social, political, and historical contexts \cite{birhane_algorithmic_2021}.

A recent paper \cite{varoquaux_hype_2024} examines the scaling hype in LLMs and outlines three negative consequences: environmental unsustainability, data quality issues at scale, and a centralization of power with harmful consequences on innovation and society. Highlighting that “small can also be beautiful”, the authors advocate for increased transparency about size and cost and a focus on progress in terms of resource efficiency. This is a timely call considering that the global scale of LLM R\&D is such that it now seriously impacts planetary ecosystems on a global scale.

\noindent\textbf{System Dynamics.} In 1972, the well-known \textit{Limits to Growth} report developed a system dynamics model of key global parameters and relationships connecting population dynamics, industrialization, agriculture, CO\textsubscript{2} emissions, and temperature increases. The conclusions are well known: the planet would reach its limit for unlimited growth within the next 100 years after which there would be a drastic decline, but it was possible to achieve global equilibrium if ecological and economic stability was attained, and there was greater likelihood of success the sooner we strived for this stability \cite{meadows_limits_1972}. While it was originally criticized and doubted for its technical feasibility, the 30-year update demonstrated its general robustness \cite{meadows_limits_2004}. The now influential report laid a crucial foundation for climate frameworks such as the planetary boundaries model \cite{rockstrom_planetary_2009,Steffen__2015} and economic theories like doughnut economics \cite{raworth_doughnut_2017}. The \textit{Computing within Limits} research community similarly recognizes that a growth trajectory poses risks to the environment and society and instead “...[explores] ways that new forms of computing [support] well-being while enabling human civilizations to live within global ecological and material limits” \cite[p.~86]{nardi_computing_2018}. 

\begin{figure*} 
    \centering
    \includegraphics[width=0.75\linewidth]{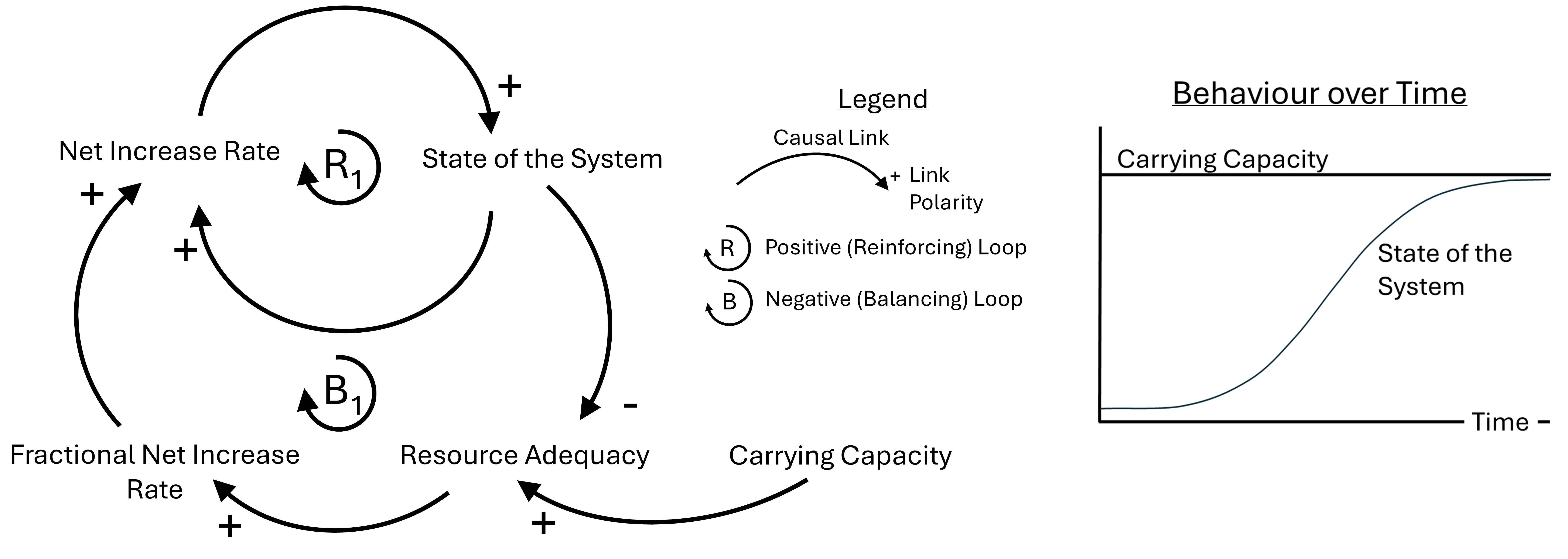}
    \caption{Limits to growth \cite{meadows_thinking_2011,meadows_limits_1972} archetype \cite{senge_fifth_1997} template, adapted from \cite{sterman_business_2000}. The archetype demonstrates that growth cannot continue forever because while the state of the system will grow at first, it will then slow until it reaches a state of equilibrium due to the limits dictated by the carrying capacity.}
    \label{fig:fig1}
\end{figure*}

In our paper, we draw on the concept of ``limits to growth'' and use it as a lens with which to view AI scaling, motivated by the recognition that the environmental impact of AI is large enough to play a significant role in accelerating climate change \cite{dominguez_hernandez_mapping_2024, creutzig_digitalization_2022}. While the underpinning models are of considerable complexity, the modeling constructs of system dynamics are minimal. The patterns of structural dynamics, or \textit{archetypes}, are a common vehicle of analyzing the trajectories of complex systems and their underpinning dynamics.  Fig. \ref{fig:fig1} provides an illustration of the basic modeling constructs, in the form of the core archetype \textit{limits to growth}, which represents the structural relations between key factors in a resource-constrained system through a causal loop diagram (CLD). A reinforcing feedback loop powers a growth dynamics that eventually, as it approaches the carrying capacity of the environment, meets barriers and slows down. It can be used to describe scenarios ranging from ecosystem populations to market saturation. The CLD in Fig. \ref{fig:fig1} models the structural factors that produce the dynamic behaviours observed (and charted on the right). This link between observable dynamics and structural factors is the crucial contribution of system dynamics modelling. Further below, we will draw on these models to map out potential growth dynamics of AI and consequences of eroding the carrying capacity. We provide additional information on the system dynamics archetypes used in this paper in Appendix \ref{appendixa}.

\section{Mapping the Perspectives of AI Scaling} \label{sec:map}
AI scaling  can be examined from several perspectives \cite{dominguez_hernandez_mapping_2024}. The trajectory of each and how they intersect evolves rapidly. In compiling an evidence base for modeling dynamic relationships, we triangulate between academic literature that provides robust arguments on a comparably slow timescale; company reports and industry conversations that illustrate important facets rapidly but lack the rigor and representativeness of peer-reviewed research; and investigative journalism that, for example, demonstrates the intentional opacity of omissions in reports \cite{gabbott_why_2024}. Here we provide a distilled synthesis as basis for the dynamic models below. 

% An extensive review is found in Appendix \ref{appendixb}. 

\begin{figure*}
    \centering
    \begin{subfigure}{0.9\textwidth}
        \includegraphics[width=\linewidth]{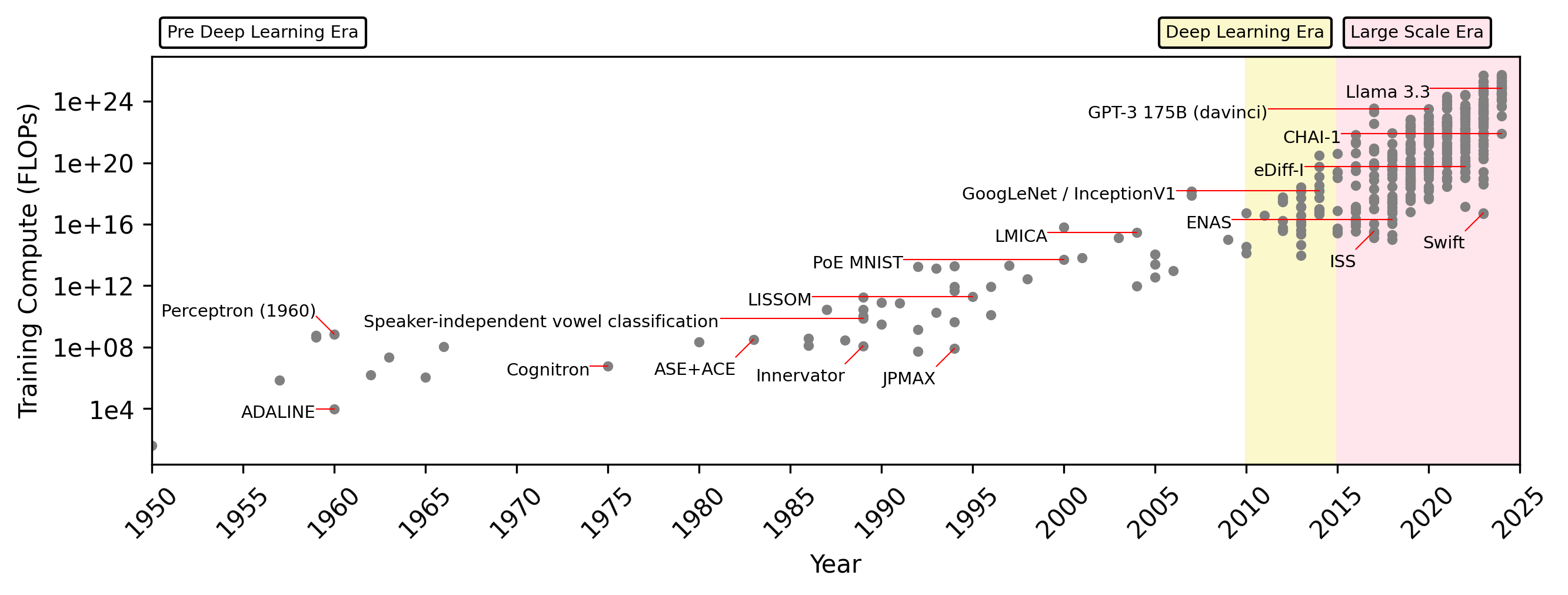}
        %\caption{Enter Caption}
    \end{subfigure}
    \vspace{-7mm}
     \begin{subfigure}{0.9\textwidth}
        \includegraphics[width=\linewidth]{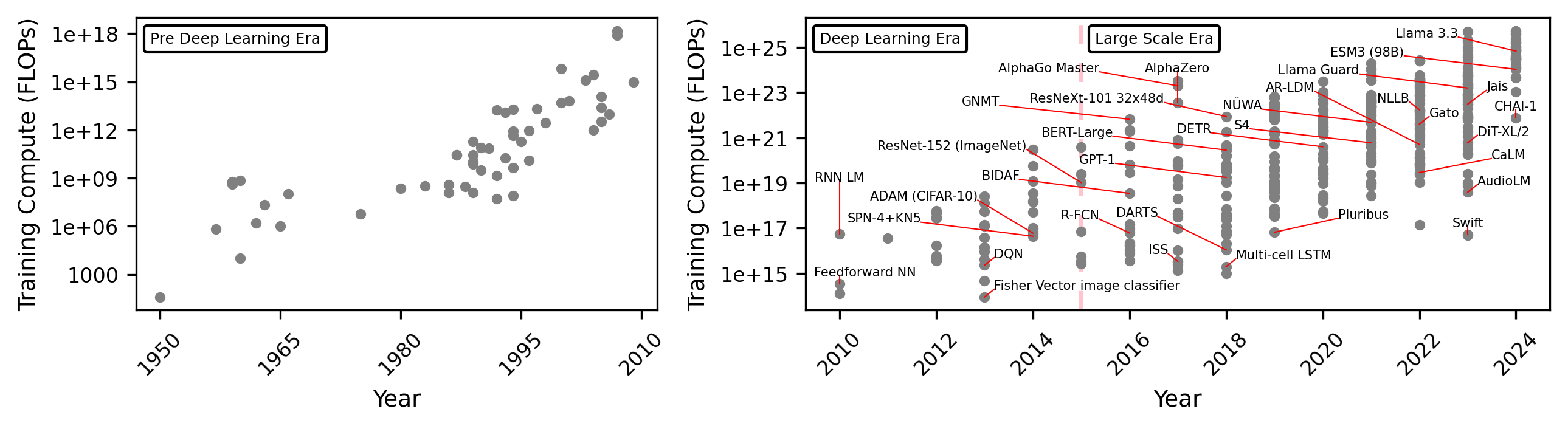}
        %\caption{Enter Caption}
    \end{subfigure}
    \vspace{3mm}
    \caption{Compute of ML models over time. 
    \textbf{Top}: Select models from 1950-2024; \textbf{Bottom left:} Filtered for models in the pre deep learning era (1950-2010); \textbf{Bottom right:} Filtered for models in the the deep learning and large scale era (2010-2024). Data: \cite{epoch_ai_data_2024,sevilla_compute_2022}.
    }
    \label{fig:fig2}
\end{figure*}

\subsection{Technical} \label{technical}
%\textbf{Scaling Laws.} 
Language models and transformer architectures continue computing's history of exponential growth curves \cite{nardi_computing_2018} described in a succession of ``laws'' for integrated circuits, disk storage, and network bandwidth \cite{vaswani_attention_2017}. Scaling laws for Transformer architecture follow a power law relationship \cite{dan_hendrycks_ai_2024,kaplan_scaling_2020}, i.e., a model's performance depends on the scale of its parameters, volume of data, and amount of compute \cite{kaplan_scaling_2020}. Scaling laws dictate that either the size of model, training data, or compute budget must continually be scaled to improve \cite{bahri_explaining_2024,dohmatob_tale_2024,tunstall_natural_2022}. With current architectures, larger models outperform bespoke models, so technical advances have focused on increasing size over all else \cite{mollick_scaling_2024,varoquaux_hype_2024}. Fig. \ref{fig:fig2} shows how training compute has doubled every 18 months between 1952-2010 (pre deep learning era), 6 months between 2010-2022 (deep learning era) and 10 months between 2015-2022 (large scale era) \cite{sevilla_compute_2022}. However, an increase in model size and complexity does not yield an equivalent increase in model performance \cite{serokell_what_2023}. The larger a model is, robustness and generalizability become difficult to attain while overfitting becomes more likely \cite{serokell_what_2023}.

\noindent\textbf{Barrier 1: Diminishing returns.} While efficiency improvements and data center size enable further scaling, these imply economic and environmental impacts, and marginal returns diminish: “increasing the computing budget from \$10 million to \$100 million increased the pass rate for AI-generated computer programs from about 65\% to about 75\%” and “a trillion-dollar model would increase those odds to just over 91\%” \cite{lohn_scaling_2023}. Even with a trillion-dollar model, “record performance” is not guaranteed. In other words, scaling alone may not lead to gains in model performance \cite{lohn_scaling_2023}.  

\noindent\textbf{Barrier 2: No more data.}
The exploding volumes of extracted data available for training were a driving force for model improvements, growing at 0.22 orders of magnitude per year for language datasets (see Fig. \ref{fig:fig3a}) but seem to be reaching a peak as the amount of public-generated and generally-available data is no longer sufficient to drive further progress of AI models \cite{villalobos_will_2024,Sutskever_2024}.

\subsection{Economic} \label{economic}

\begin{figure*}[h!]
    \centering
    \begin{subfigure}{0.33\textwidth}
        \includegraphics[width=5cm,height=4cm]{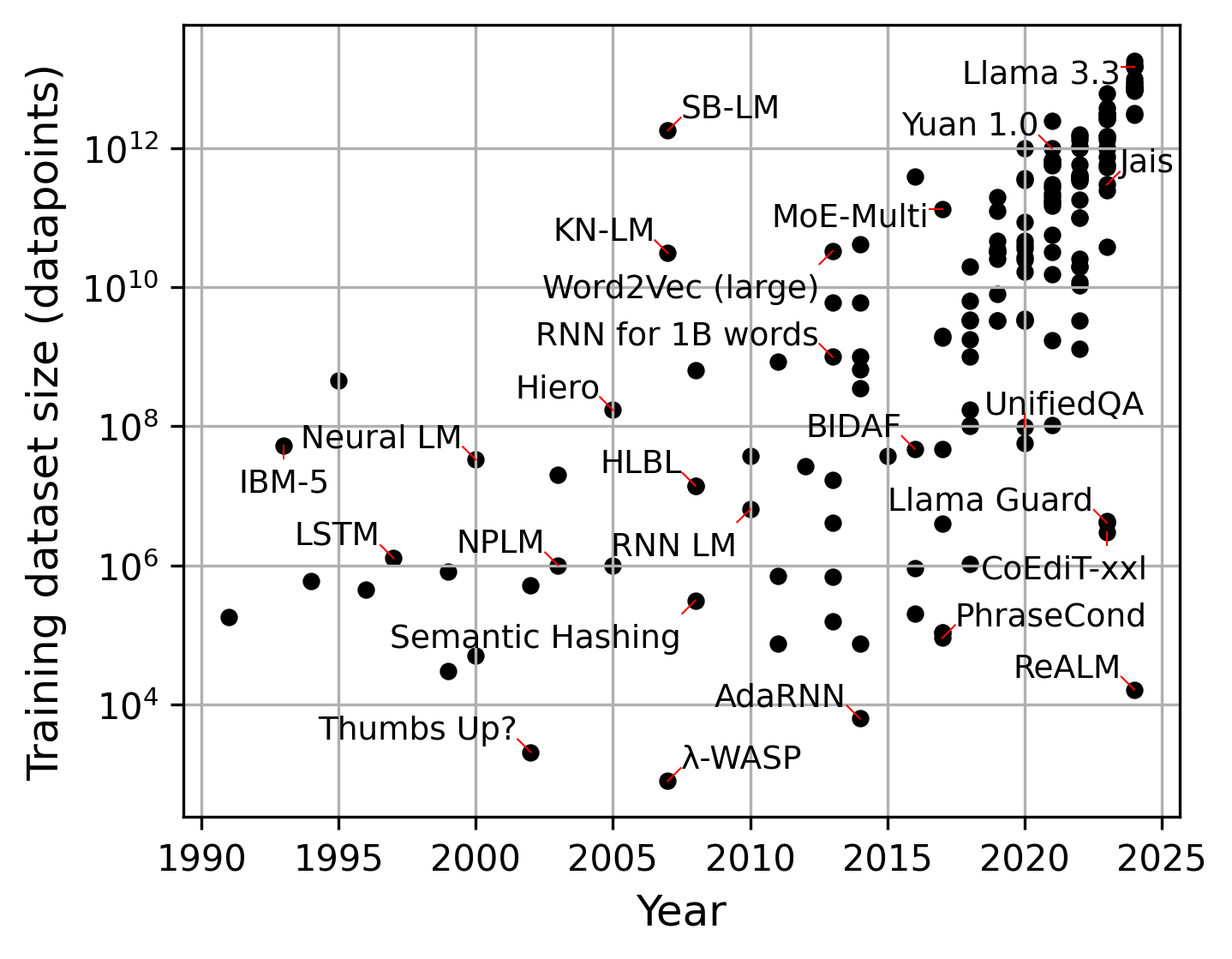}
        \vspace{-7mm}
        \caption{}
        \label{fig:fig3a}
    \end{subfigure}
    %\vspace{-7mm}
     \begin{subfigure}{0.33\textwidth}
        \includegraphics[width=5cm,height=4.1cm]{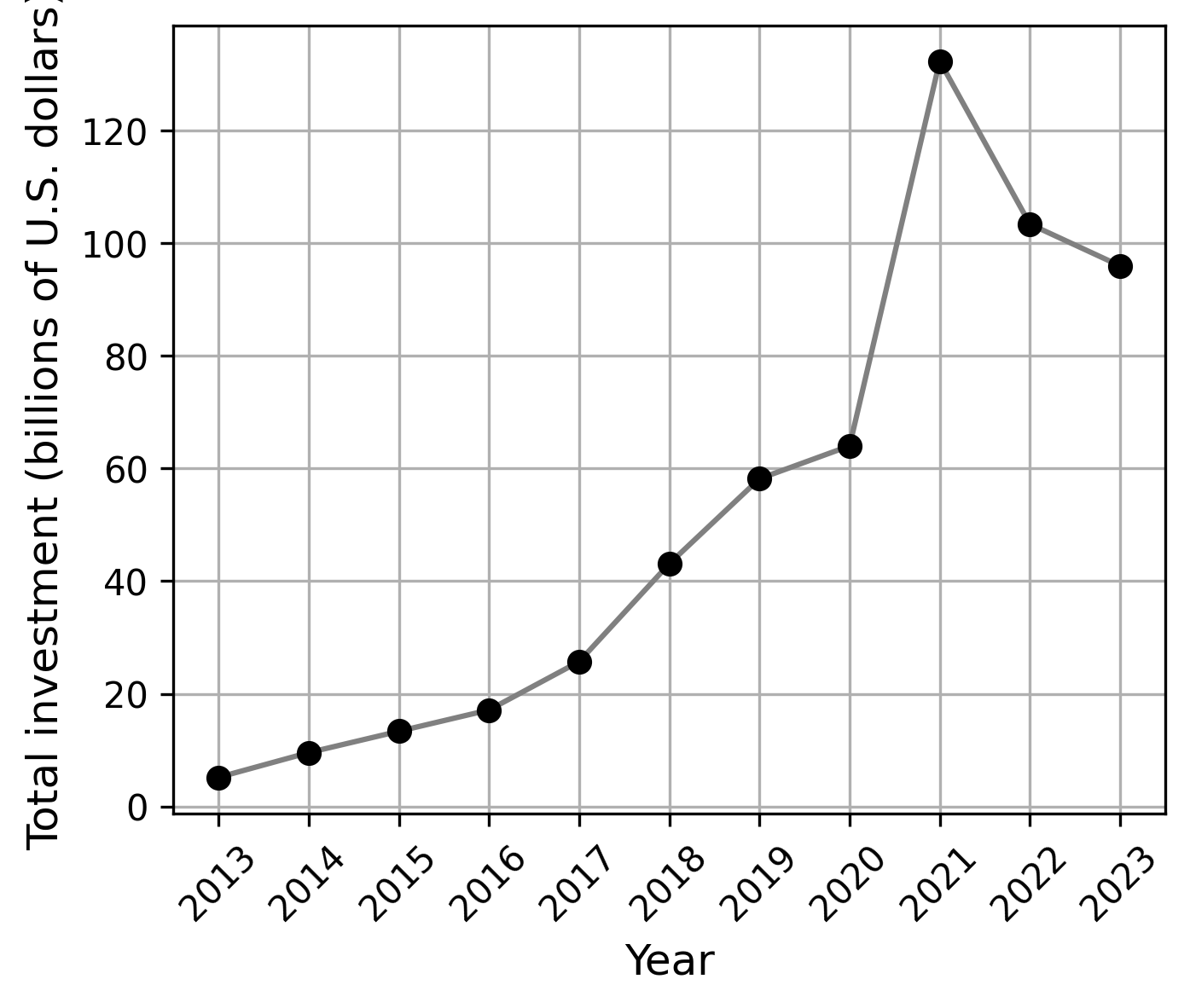}
        \vspace{-7mm}
        \caption{}
        \label{fig:fig3b}
    \end{subfigure}
    %\vspace{-7mm}
     \begin{subfigure}{0.33\textwidth}
        \includegraphics[width=5cm,height=4cm]{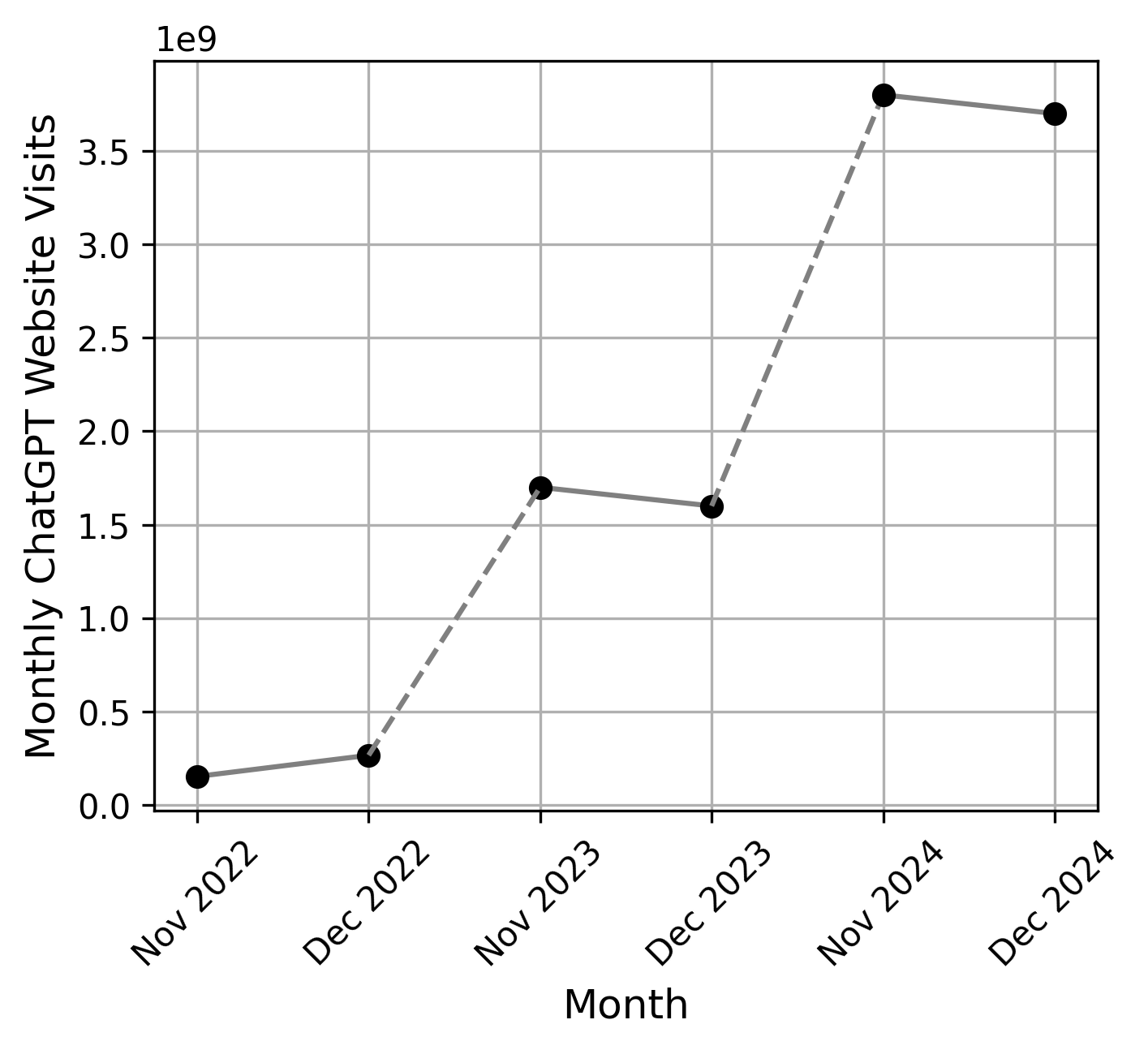}
        \vspace{-7mm}
        \caption{}
        \label{fig:fig3c}
    \end{subfigure}
    %\vspace{-7mm}
     \begin{subfigure}{0.33\textwidth}
        \includegraphics[width=5cm,height=4cm]{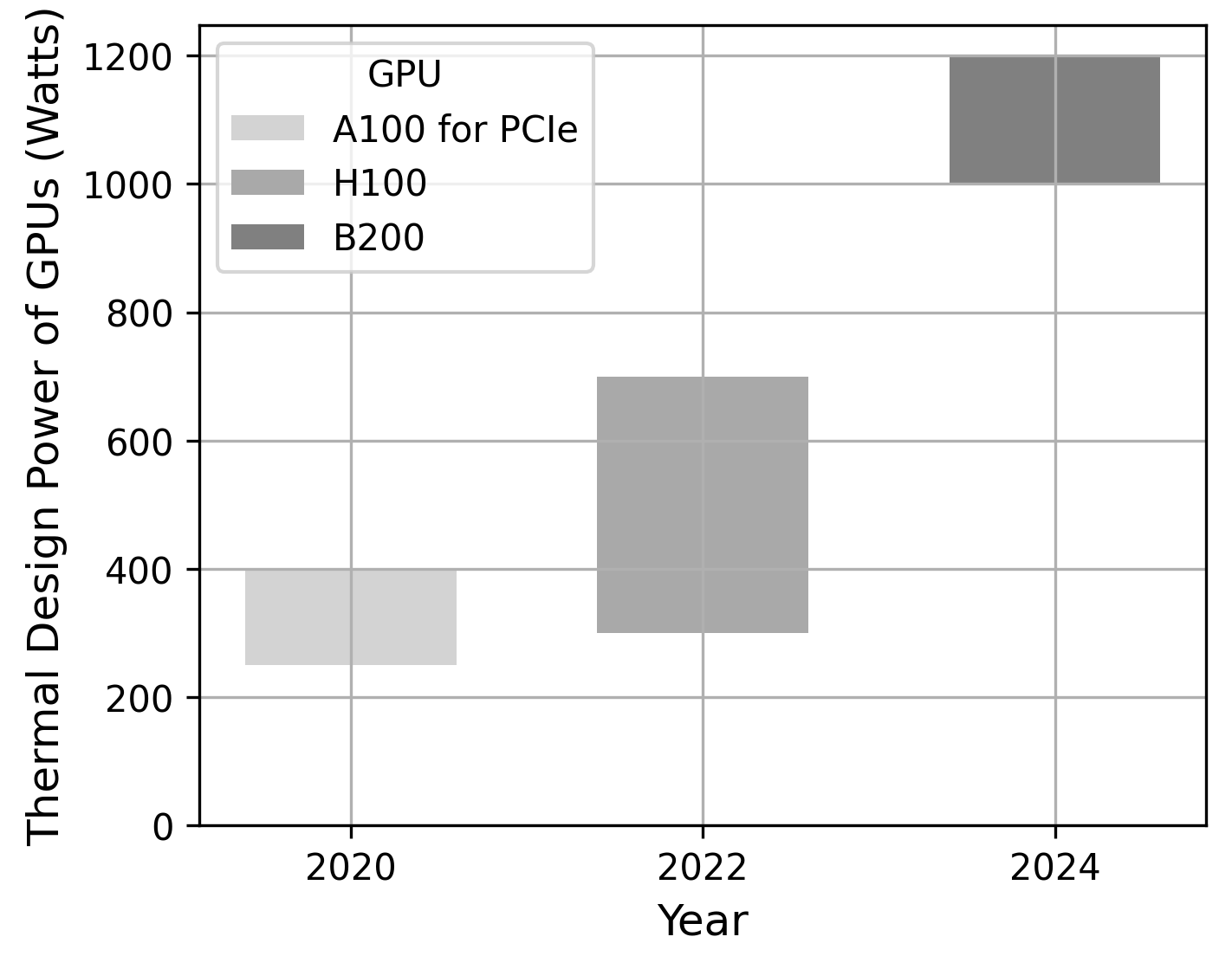}
        \vspace{-7mm}
        \caption{}
        \label{fig:fig3d}
    \end{subfigure}
    %\vspace{-7mm}
    ~
     \begin{subfigure}{0.33\textwidth}
        \includegraphics[width=5cm,height=4cm]{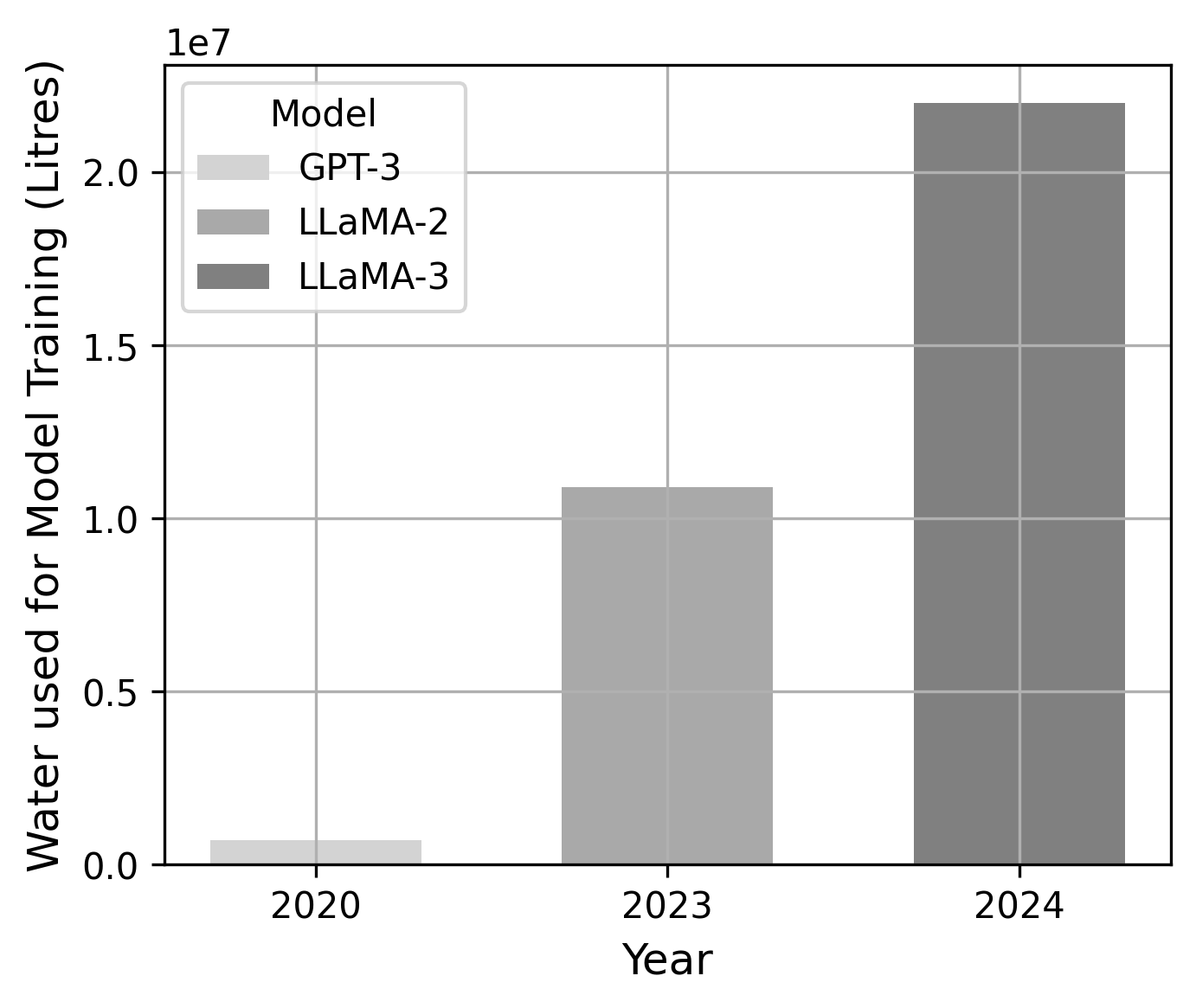}
        \vspace{-7mm}
        \caption{}
        \label{fig:fig3e}
    \end{subfigure}
    %\vspace{-7mm}
     \begin{subfigure}{0.33\textwidth}
        \includegraphics[width=5cm,height=4cm]{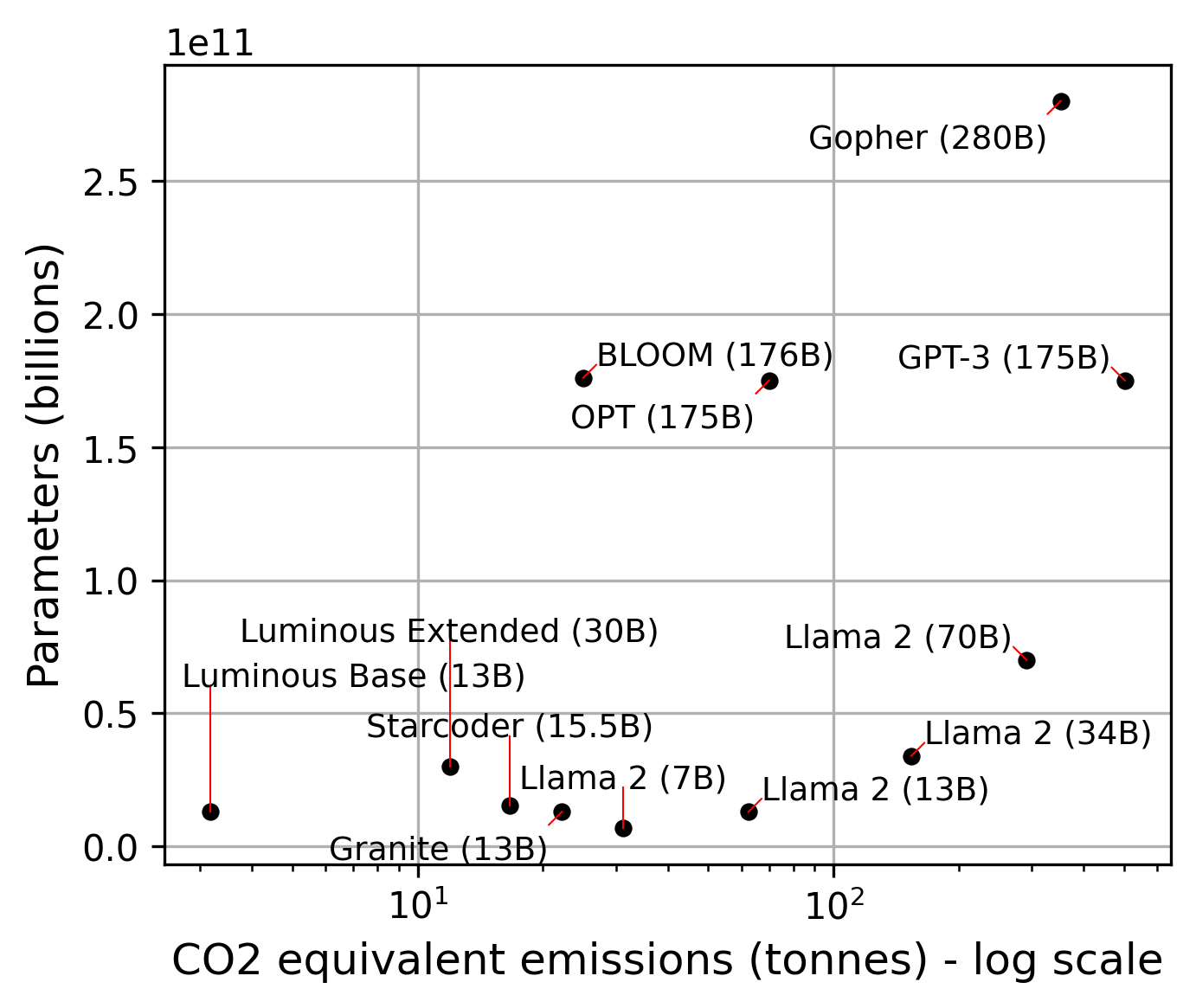}
        \vspace{-7mm}
        \caption{}
        \label{fig:fig3f}
    \end{subfigure}
    %\vspace{3mm}
    \caption{Scaling of training data, investment, usage, energy, water, and carbon emissions of AI models. \textbf{A:} Language training dataset size over time, data: \cite{epoch_ai_data_2024,sevilla_compute_2022}. \textbf{B:} Private investment in AI, data: \cite{nestor_maslej_ai_2024}. \textbf{C:}  ChatGPT website visits (2022-24), data: Similarweb \cite{Similarweb}. \textbf{D:} Maximum power consumption enabled by GPUs, data: \cite{kindig_ai_2024}. \textbf{E:} Water usage for training models, data: \cite{li_making_2023,shaolei_ren_how_2023,pranshu_verma_bottle_2024,barr_llamas_2023}. \textbf{F:} Emissions and parameters for AI models, data: \cite{nestor_maslej_ai_2024} . 
    }
    \label{fig:fig3}
\end{figure*}

Technical advances are tied to capital investments. Fig. \ref{fig:fig3b} shows how billions of dollars in investments have poured into the AI industry since 2013, with exponential growth in 2020. Similarly, the global semiconductor market is forecast to double from \$590 billion USD by 2030 \cite{Ondrej_2022}. The undisputed winners in that market so far are hardware companies like NVidia \cite{Mickle_2024}. In total, Alphabet, Amazon, Apple, Meta and Microsoft have budgeted to spend approximately \$400 billion on AI capital expenditures in 2024 \cite{noauthor_what_2024}. But, OpenAI expected to lose \$5 billion in 2024 to rapid expansion efforts \cite{metz_how_2024,isaac_openai_2024}. It has been predicted that developing a model in 2027 could cost \$100 billion \cite{wong_silicon_2024}. There are also operational costs which are dependent on query length, number of queries, and response speeds \cite{lohn_scaling_2023}. For example, ChatGPT's increasing website visits from 2022 to 2024 shown in Fig. \ref{fig:fig3c}, accordingly scaled OpenAI's operational costs. 

\noindent\textbf{Barrier 3: Returns on investments.}  Investors increasingly notice a gap between the revenue prospects generated by AI models and the capital required to power it \cite{herrman_ai_2024}. Genuine business adoption rates for AI of ~5\% in the U.S. are much lower than the widespread rate of experimentation often cited \cite{noauthor_what_2024}, and comparisons are made to the dot-com boom when over-production without genuine demand led to short-term collapse \cite{herrman_ai_2024}. While investments upwards of trillions of dollars are being sought to further scale the semiconductor industry \cite{keach_hagey_sam_2024}, “some of the largest tech companies’ current spending on AI data centers will require roughly \$600 billion of annual revenue to break even, of which they are currently about \$500 billion short.” \cite{wong_silicon_2024}. While many see disruptive potential in generative AI, ``no one knows what its main uses will be, or how it will make money.'' \cite{noauthor_history-lovers_2024}.

\subsection{Ecological} \label{ecological}

While the list of concerns is broad and includes waste, air pollution, minerals, and the impact of mining, below we focus on energy, water, and CO\textsubscript{2} emissions to illustrate the ecological perspective. 

%\noindent\textbf{Energy:} 
AI requires energy and materials to manufacture, power, run, and cool hardware. Dedicated AI data centers are increasingly specialized for training (data processing and ML model training) and inference (deployed models used in applications) \cite{sayegh_billion-dollar_2024}. While inference is tiny when compared to training, it occurs much more frequently \cite{kaack_aligning_2022,samsi_words_2023} and its energy demand varies based on task complexity (e.g., image generation versus text classification \cite{luccioni_power_2024}). In the case of the BLOOMz-7B model, approximately 590 million inferences would be needed for the energy cost of inference to match that of training \cite{luccioni_power_2024}. But for popular models like GPT with 100 million monthly users \cite{malik_openais_2023}, inference is so frequent that it heavily outweighs the energy impact of training \cite{luccioni_power_2024}. About 50\% of energy consumption of data centers is used to power the hardware and another 40\% to cool it \cite{luccioni_environmental_2024}. For example, the energy demand of the LLaMa 3.1 405B model (released in July 2024) for powering and cooling peaks at 1.7kW per GPU \cite{noauthor_introducing_2024}. At 16,000 GPUs, this amounted to roughly 27.2MW \cite{sevilla_can_2024}. Newer chip generations have also steadily increased power consumption (Fig. \ref{fig:fig3d} shows the tripling of maximum power consumption enabled by NVidia chips within four years). 

% GPU manufacturing  itself is  material-intensive \cite{wang_environmental_2023}. 

%\noindent\textbf{Water:} 
While clean water is needed to cool data centers and to generate electricity (through water-intensive thermoelectric plants) \cite{luccioni_environmental_2024,shaolei_ren_how_2023}, chip manufacturing requires ``ultrapure'' water \cite{heilweil_want_2023, li_making_2023} and one facility ``can use 37 million litres of ultrapure water per day'' \cite{kirsten_james_semiconductor_2024}. Google's reported data center water consumption rose from 16 billion litres in 2021 to 28 billion in 2023 \cite{urs_holzle_our_2022,google_2024_2024} and it's been found that larger models need more water (see Fig. \ref{fig:fig3e}).

%and ChatGPT `drinks' 500ml of water for every 10-50 queries \cite{li_making_2023,shaolei_ren_how_2023}. 

%\noindent\textbf{Carbon:} 
CO\textsubscript{2} is emitted at all stages of the AI lifecycle. For the BLOOM language model, “of the total 50 tonnes of CO\textsubscript{2eq} emissions emitted during model training, only half was due to the energy consumption of the GPUs used for training BLOOM (‘dynamic consumption’), with a further 29\% stemming from the idle consumption of the data center (i.e., the energy used for heating/cooling, networking, storage, etc.), and a final 22\% ...from the GPU manufacturing process.” \cite[p.~6]{luccioni_environmental_2024}. Aggregate emissions are increasing over time as models become larger (see Fig. \ref{fig:fig3f}), require more data centers, and need chips with greater processing power. Google’s emissions increased nearly 50\% from 2019 to 2023, Microsoft’s emissions increased 29\% from 2020 to 2023, and Meta’s increased 66\% from 2021 to 2023 \cite{gelles_is_2024}. Efforts towards compute efficiency also do not translate to energy efficiency and hence do not result in savings in carbon emissions \cite{wright_efficiency_2023}. 

% A 2019 paper showed how a transformer model with 213M parameters produced over 272 tonnes of CO\textsubscript{2} emissions \cite{strubell_energy_2019}. In comparison, LLaMA 3 (released in 2024) is estimated to produce nearly 2290 tonnes of CO\textsubscript{2eq} emissions \cite{meta_meta-llamameta-llama-3-70b_2024}.

\noindent\textbf{Barrier 4: Shortage in energy supply.} Building more data centers and developing new, advanced GPUs is energy-, material- and water-intensive and generates carbon emissions and other pollutants. Data center capacity (in megawatts) has seen nearly a 2000\% increase in the U.S. alone from 2019 to 2024 \cite{eleni_kemene_ai_2024}. Of particular concern to the AI industry is the shortage in the power supply that places constraints on further expansion due to pre-existing contractual commitments for energy and time requirements for expanding or acquiring new sources of it \cite{sevilla_can_2024}. 
% Similarly, it’s estimated that water consumption of AI will reach 4.2-6.6 billion cubic meters by 2027 \cite{li_making_2023} and emissions are predicted to triple by 2030 owing to generative AI totalling 2.5 billion tonnes of (CO\textsubscript{2eq}) \cite{jenkins_will_2024}.

\noindent\textbf{Barrier 5: Accountability for environmental impact.} Companies have claimed sustainability goals such as to be carbon neutral, but AI advancements require rising electricity and water consumption. As for growing emissions \cite{akshat_rathi_google_2024,milmo_googles_2024,widder_watching_2024,corps_abandon_esg_2024}, companies have made future promises only to renege, postpone, or walk back on them, e.g., \cite{Akshat_Rathi_Dina_Bass_2024,Hao_2024}. While promises of sustainable practices continue, so does the development of increasingly complex models that require more energy, more water, and cause more emissions. These operations reveal business priorities. At the moment, real transparency and accountability are not in place, such as in the case of data centers that are not legally required to report their water usage figures while companies continue to lobby to keep this information a trade secret \cite{gabbott_why_2024}.  

\subsection{Social} \label{social}

\noindent\textbf{AI’s Scale of Impact.} The widespread adoption of AI poses old questions in new forms as well as new ethical challenges \cite{gebru_tescreal_2024,Gebru_2020,Coeckelbergh_2021}. Its scale of impact is larger than ever before in the history of computing ethics due to the opacity of the models, the number of people it can impact, its potential for damage, and the feedback loops that increase their scale of harm \cite{oneil_weapons_2017}. A major consequence of the prevalence of AI in our society is the standardization and lack of autonomy in deciding whether you want to be part of the data economy \cite{mejias_data_2024} as power centralizes in the hands of monopolies driving the “social quantification sector” \cite{couldry_data_2019,mejias_data_2024,melissa_heikkila_this_2024}. Datafication enables surveillance capitalism in a “transformation of human life so that its elements can be a continual source of data” \cite[p.~2]{mejias_data_2024},\cite{zuboff_age_2023}. The ability to datafy remains largely with the big AI companies. “Inequality resides... in having or not having the power to decide what kind of data is being generated... by whom, for what purpose, and for whose benefit.” \cite[p.~831]{fisher_confronting_2022}. This also enables the exploitation of labour. Millions of gig workers, often employed from India, Kenya, Philippines, or Mexico, are paid \$1.46/hour to enable the development of AI models \cite{ilr_school_and_the_aspen_institute_how_nodate}. The poor labour conditions of the workers include monitoring through automated tools \cite{williams_exploited_2022}. The lack of equity in the actors that can afford and access data, and the monopoly they exercise signals that “...capitalism is premised upon the preservation of unequal power: of the enforcement of the racial, gendered, and other social hierarchies which enable the extraction of labor, and therefore value, from the many for the profit of the few.” \cite[p.~102]{klein_data_2024},\cite{couldry_data_2019}. Thus the computing industry offloads its ethical debts to distant stakeholders \cite{becker_insolvent_2023}.

\noindent\textbf{Environmental Justice.} It has long been recognized that ecological and social impacts are closely linked, and environmental sustainability is always an issue of equity and social justice \cite{agyeman_just_2003}. A few examples follow. The globally dominant chip manufacturer TSMC is estimated to consume as much electricity as New Zealand by 2030 \cite{hilton_taiwan_2024}. The consequences were experienced by over 5 million residents impacted by  blackouts caused by a grid operating much closer to capacity than it should \cite{hilton_taiwan_2024}. Data centers impact communities globally \cite{crawford_generative_2024,jeff_clabaugh_northern_2024}, including by distorting the electric grid and causing billions of dollars in damages to appliances \cite{leonardo_nicoletti_ai_2024} and by air pollution, which alone is estimated to cause 1300 deaths annually in the U.S. by 2030 \cite{han_unpaid_2024}. 
In Uruguay, Google plans to construct a data center predicted to produce 25,000 tonnes of CO\textsubscript{2} while releasing 86 tonnes of hazardous waste with no plans in place of how the waste matter will be disposed \cite{livingstone_anger_2024}. This construction is estimated to use 7.6 million litres of water at a time when Uruguay is experiencing a drought, water that could be used by 55,000 people \cite{livingstone_its_2023}.

\noindent\textbf{Barrier 6: Organized resistance.} In response to these environmental injustices, the erosion of worker autonomy, and injustices resulting from data colonialism \cite{couldry_data_2019}, organized public resistance has increased \cite{mcquillan_resisting_2022}, including residents mobilizing against data centers in many locations \cite{odonovan_fighting_2024,dwoskin_google_2019}, calls to defund Big Tech \cite{barendregt_defund_2021}, environmental activism by Amazon workers \cite{glaser_how_2019}, campaigns against military projects by Google employees \cite{tarnoff_tech_2020}, legal action by content moderators \cite{booth_more_2024}, and calls to democratize decision-making about how AI is developed and deployed \cite{jeni_tennison_lets_2024}. 

\begin{figure*}
    \centering
    \includegraphics[width=1\linewidth]{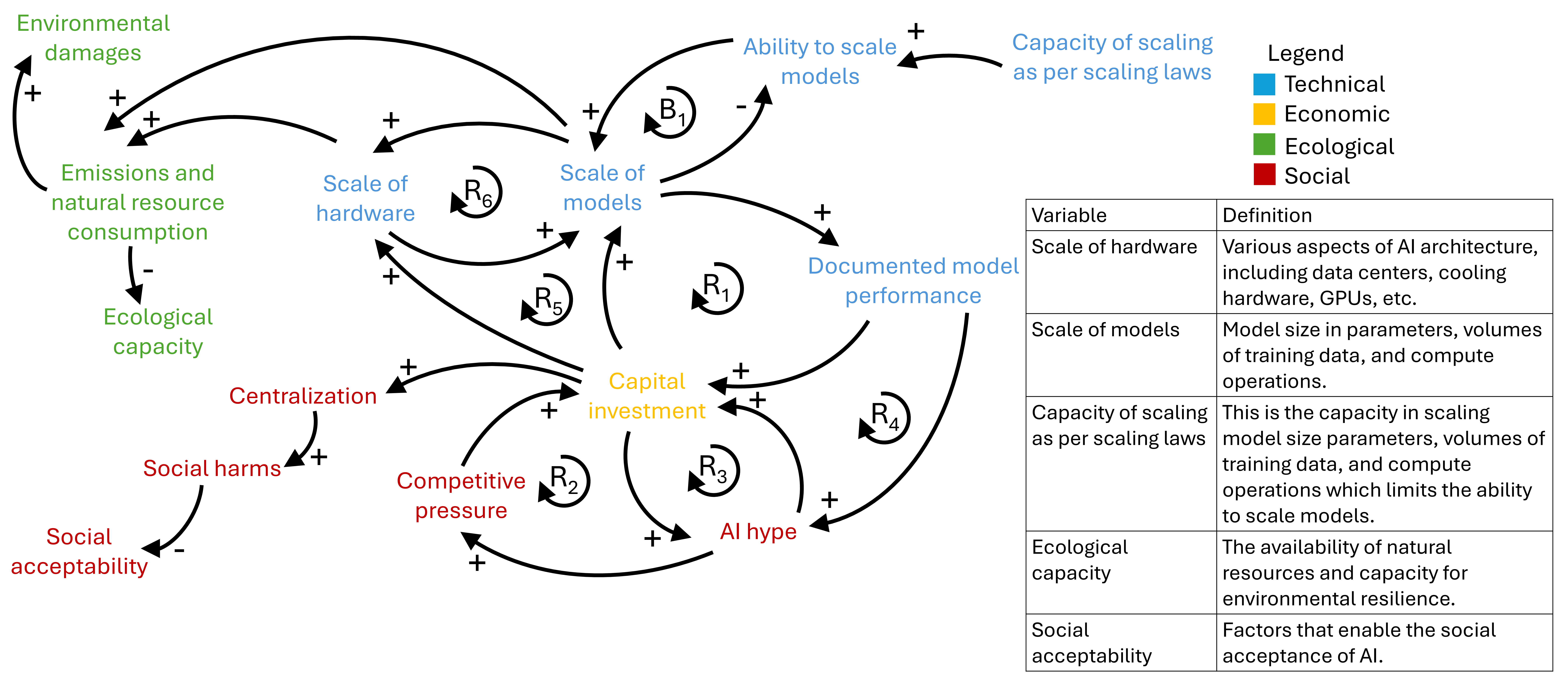}
    \caption{The dynamics of AI scaling across perspectives (colour-coded) with social and environmental damages, showing that multiple reinforcing loops drive scaling while only one feedback loop limits scaling (i.e., the capacity of model scaling). The loops are R\textsubscript{1}: Economies of scale, R\textsubscript{2}:  Competitive pressure loop, R\textsubscript{3}: AI hype and speculation, R\textsubscript{4}: Scaling driven by hype,   R\textsubscript{5}: Scaling driven by capital investment, R\textsubscript{6}: Infrastructure development, B\textsubscript{1}: Capacity of model scaling. }
    \label{fig:fig4}
\end{figure*}

\subsection{Dynamics of AI Scaling} \label{sec:dynamics}

The four perspectives illustrated in the previous sections intersect with each other and result in the complex dynamics of AI scaling. We present a simplified version of this in Fig. \ref{fig:fig4}. We define and justify each causal relationship of Fig. \ref{fig:fig4} and subsequent CLDs in Appendix \ref{appendixb}. 

The reinforcing loops in the CLD demonstrate how technical scaling is reinforced by socially recognized and documented model performance, i.e., that scaling leads to improved performance, which spurs more capital investment  (R\textsubscript{1}). Competitive pressure (R\textsubscript{2}), AI hype  (R\textsubscript{3}), and model performance (R\textsubscript{4}) drive investment cycles which continue to power scaling of AI models and hardware (R\textsubscript{5}) which are also mutually reinforcing (R\textsubscript{6}). The development of GPUs and data centers (among other infrastructure) consumes natural resources while continuous capital investment leads to centralization of power. The reinforcing loops together match the reinforcing growth loop of Fig. \ref{fig:fig1}.

\noindent\textbf{Limits to growth.} The capacity of scaling, dictated by AI scaling laws, limits the \textit{ability} to scale models and consequently the scale of the models themselves (B\textsubscript{1}). Note that environmental damages and social harms do not currently feed back into the growth loop, since there are no genuine accountability mechanisms that would justify such links. That means that three of the barriers represent actively recognized limits that are in the focus of AI industry and research: (1) limits to performance, (2) limits to data, (3) limits to energy. We discuss these responses in Section \ref{sec:limits}.
On the other hand, capital is not yet scarce, environmental transparency and regulations are not in place, and social accountability is virtually nonexistent. We discuss these \textit{externalities} in Section \ref{sec:discussion}.

\section{Limits to AI Scaling} \label{sec:limits}

In this section, we revisit three of the barriers surfaced previously and map a (non-comprehensive) set of responses by the AI industry to each of them using system dynamics archetypes. 

\subsection{Limits to Performance: Rebound Effects} \label{limits_performance}

Continuous growth has yielded diminishing returns to performance. The industry response has been to force further performance scaling by investments into data centers and GPU technology. While this addresses the technical barrier (that new scale must be reached in order to improve performance), it has significant ecological and social implications and relies on economic scaling.

It is estimated that there will be increased power-efficiency of hardware, more efficient hardware usage, and longer training periods which distribute the energy consumption over time \cite{sevilla_compute_2022}. “Given all of the above, we expect training runs in 2030 will be 4x (hardware efficiency) * 2x (FP8) * 3x (increased duration) = 24x more power-efficient than the LLaMa 3.1 405B training run.” \cite{sevilla_compute_2022}. However, “2e29 FLOP training runs in 2030 will require 5,000x (increased scale) / 24x [efficiency] $\approx$ 200x more power than was used for training of LLaMa 3.1 405B” \cite{sevilla_compute_2022}.

\begin{figure*}
    \centering
    \includegraphics[width=1\linewidth]{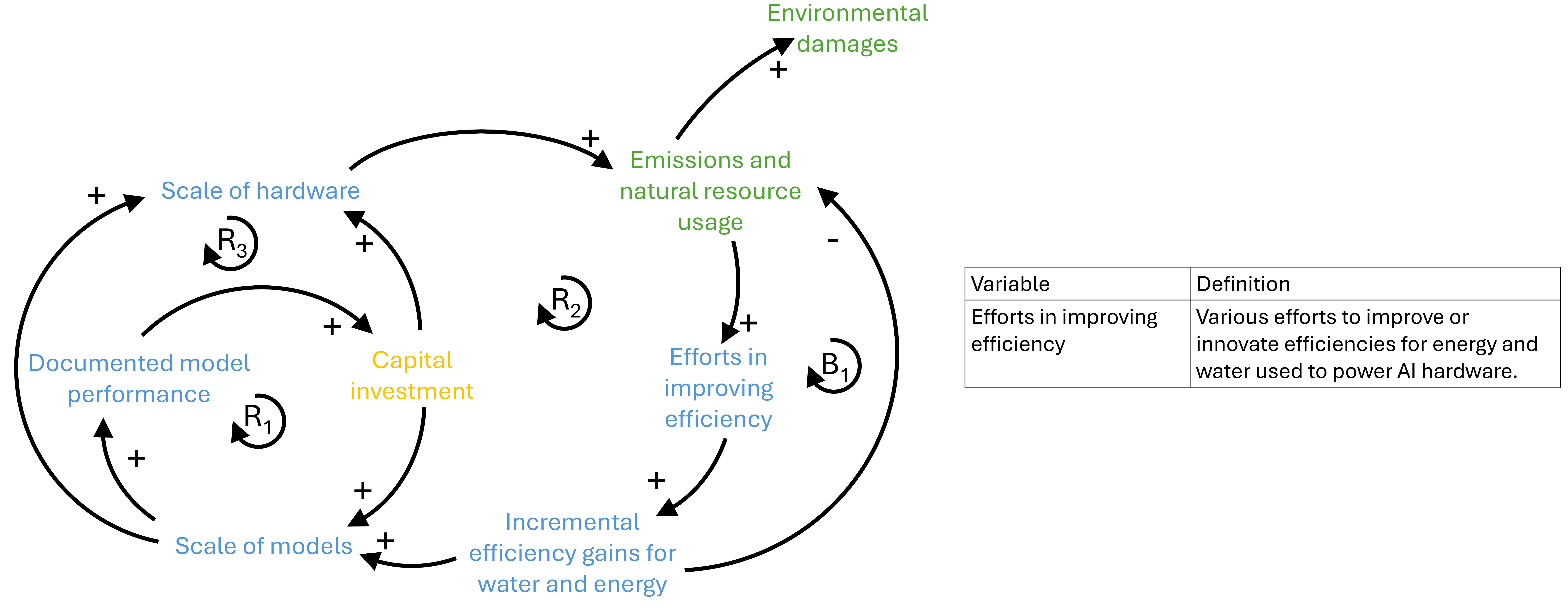}
    \caption{As models scale, the rise in hardware development implies increased energy and water usage which leads to efforts to make resource consumption more efficient. While this does lead to some efficiency gains for energy and water, there is also a rebound effect that contributes to further scaling.  The loops are R\textsubscript{1}: AI model scaling, R\textsubscript{2}: Rebound effect of water and energy efficiency, R\textsubscript{3}: AI hardware scaling, B\textsubscript{1}: Water and energy efficiency gains, adapted from \cite{laurenti_unintended_2016}.}
    \label{fig:fig5}
\end{figure*}

Efforts focused on efficiencies always have to be conscious of resulting \textit{rebound effects}. Rebound effects are “behavioural or other systemic responses to the introduction of new technologies that increase the efficiency of resource use. These responses tend to offset the beneficial effects of the new technologies/measures.” \cite[p.~107]{borjesson_rivera_including_2014}. Rebound effects have been studied widely for computing as well as for sustainability-related interventions. For example, it has been found that efficient energy developments are often met with increasing usage \cite{ozsoy_energy_2024} and the same is seen for renewable sources like solar energy \cite{nguyen_solar_2024}. The consequence of rebound effects is that increased efficiency often does not result in an aggregate decrease of resources but instead an \textit{increase} in aggregate use or production \cite{tsao_solid-state_2010,saunders_rebound_2012}. Accordingly, similar effects will be true of haphazard interventions to modernize and “green” the energy requirement of AI. The strength of the rebound effect varies and needs to be established. 

% Similarly greater efficiency in chip manufacturing capacity or overall hardware efficiency \cite{sevilla_can_2024} can further lead to a rise in uptake. 

These dynamics are presented in Fig. \ref{fig:fig5}. Incremental efficiency gains reduce natural resource usage (B\textsubscript{1}), but improvements in performance spur further data center and GPU development (i.e., hardware) (R\textsubscript{3}) and result in rebound effects (R\textsubscript{2}) that cause further scaling (R\textsubscript{1}). This is linked to increased emissions and usage of energy and water, which ultimately has environmental impacts.

\subsection{Limits to Data: Fixes that Fail} \label{limits_data}

The “data cliff” is considered a soon-impending barrier to further scaling of AI models, particularly LLMs, as publicly available, human-generated data on the Internet runs out \cite{villalobos_will_2024}. To avoid this stagnation, many are exploring the creation of \textit{synthetic data} as a way of ensuring continued performance scaling of LLMs \cite{lee_synthetic_2024,Long_Wang_Xiao_Zhao_Ding_Chen_Wang_2024}. One perspective on synthetic data is that it enables the creation of endless amounts of high quality, error-free, complete data while simultaneously democratizing the landscape of proprietary data. It has fewer privacy concerns and can be less biased because of the ability to ensure greater diversity, and prevents copyright issues \cite{lee_synthetic_2024}. Others have argued that continued progress is possible through improvements to the quality of data by having diverse, unique, and more smartly selected data instead \cite{edward_beeching_scaling_2024}. However, far fewer contributions apply small datasets for deep learning than the creation and uptake of large synthetic data \cite{luccioni_power_2024,rather_breaking_2024}. There are multiple difficulties with this approach, however, and we discuss them under misinformation, authenticity, and data quality, before interpreting the pattern that emerges.
%, in addition to alternative approaches such as scaling test-time compute mentioned earlier. 

% AI companies are trying to find more avenues to capture human-generated text on the Internet, however these approaches are ethically questionable. Commonly used domains to train datasets are taking measures to ensure that their data is not scraped \cite{longpre_consent_2024,roose_data_2024,roose_is_nodate,verhulst_are_2024}. Publishers like The New York Times have sued OpenAI and Microsoft for copyright infringement while Reddit, Stack Overflow, and The Wall Street Journal are now charging AI companies for access to their content \cite{roose_data_2024}. Simultaneously companies are trying to increase data volume through legally gray avenues. For instance, OpenAI developed Whisper, a tool that transcribes YouTube videos’ audio into text to be used as training data for GPT-4 which may be in violation of the copyrights of the creators of those videos \cite{metz_how_2024}. Google’s updated terms of services now also allows Google to use public Google Docs and reviews on Google Maps to get access to more data \cite{creamer_meta_2024,metz_how_2024}. Meta’s scraping of articles, books, essays from the Internet takes intellectual property from their creators \cite{metz_how_2024}. Many of these legal concerns and lawsuits are met with the AI industry’s ability to absorb monetary losses given its market size, e.g., \cite{jacob_apple_2025}. 

\noindent\textbf{Misinformation.} Studies have highlighted that synthetic data can increase the amount of misinformation \cite{zhou_synthetic_2023, Jeng_Chang_Duffy_Lam_Maercklein_2024} if it creates a false belief of increased data diversity and representativeness \cite{whitney_real_2024}. For instance, when examining the diversity and representativeness of a dataset of human faces, it was found that a synthetic dataset generated from a set of base scans lead to “statistical diversity without representational diversity”, i.e., that the underlying distribution of diversity of the base scans meant that the generated images were defined based on manipulations of facial features of a small set of images \cite{whitney_real_2024}. 

\noindent\textbf{Authenticity.} Current methods to establish authenticity include providing provenance information that indicates that the data was AI-generated and/or “watermarking” the data as AI-generated \cite{Jeng_Chang_Duffy_Lam_Maercklein_2024,kirchenbauer_watermark_2024,xing_when_2024}. Both of these have been found to be inadequate because “any pixel-based invisible watermark can be removed and new images that look similar to the original image can be regenerated without the watermark” and metadata labelling puts the onus on the creator to provide provenance \cite[p.~3]{Jeng_Chang_Duffy_Lam_Maercklein_2024}. Particularly, Zhao et al. find that their “regeneration attack” “...guarantees the removal of any invisible watermark such that no detection algorithm could work” \cite[p.~3]{zhao_invisible_2023}. 

% \textbf{Consent:} It was found that “deceptive data practices” by ML companies (i.e., where consent was not obtained) could go unchecked because the regulations do not adequately cover cases in which synthetic data was used \cite{whitney_real_2024}. This is because data practices used to generate synthetic data have unclear origins and therefore information on consent can be concealed. Thus enforcement (like deletion of the model and other work products) on dataset derivatives is dependent on the documentation provided by the companies themselves \cite{whitney_real_2024}. Furthermore, the use of public image data to create synthetic dataset derivatives raises concerns of ambiguous (at best) consent. It has even been found that “...having your face included in such a dataset increases the accuracy of facial recognition models on your specific face” \cite[p.~1740]{whitney_real_2024},\cite{dulhanty_investigating_2020}. 

\noindent\textbf{Data quality.} When considering the Internet as an information ecosystem, the use of LLMs to replace information search tasks poses a risk to transparency and provenance while also changing the nature of the activity \cite{shah_situating_2022,shah_envisioning_2024}. Proposals for future uses of LLMs include for them to act as ``experts'' that provide responses to search queries as compared to the information-seeking interactions that we currently engage in. Not only does this limit searching behavior by removing browsing and sense-making processes but it can also synthesize information that is not correct \cite{shah_situating_2022}. Synthetic media can further “endanger the information ecosystem” when it generates biased text and presents it as a response from a “neutral” machine \cite{shah_envisioning_2024}. That this is already happening online is illustrated by the shutdown of language observatory \textit{Wordfreq} in Fall 2024, citing as reason that ``generative AI has polluted the data.'' \cite{koebler_project_2024}.

\begin{figure*}
    \centering
    \includegraphics[width=0.5\linewidth]{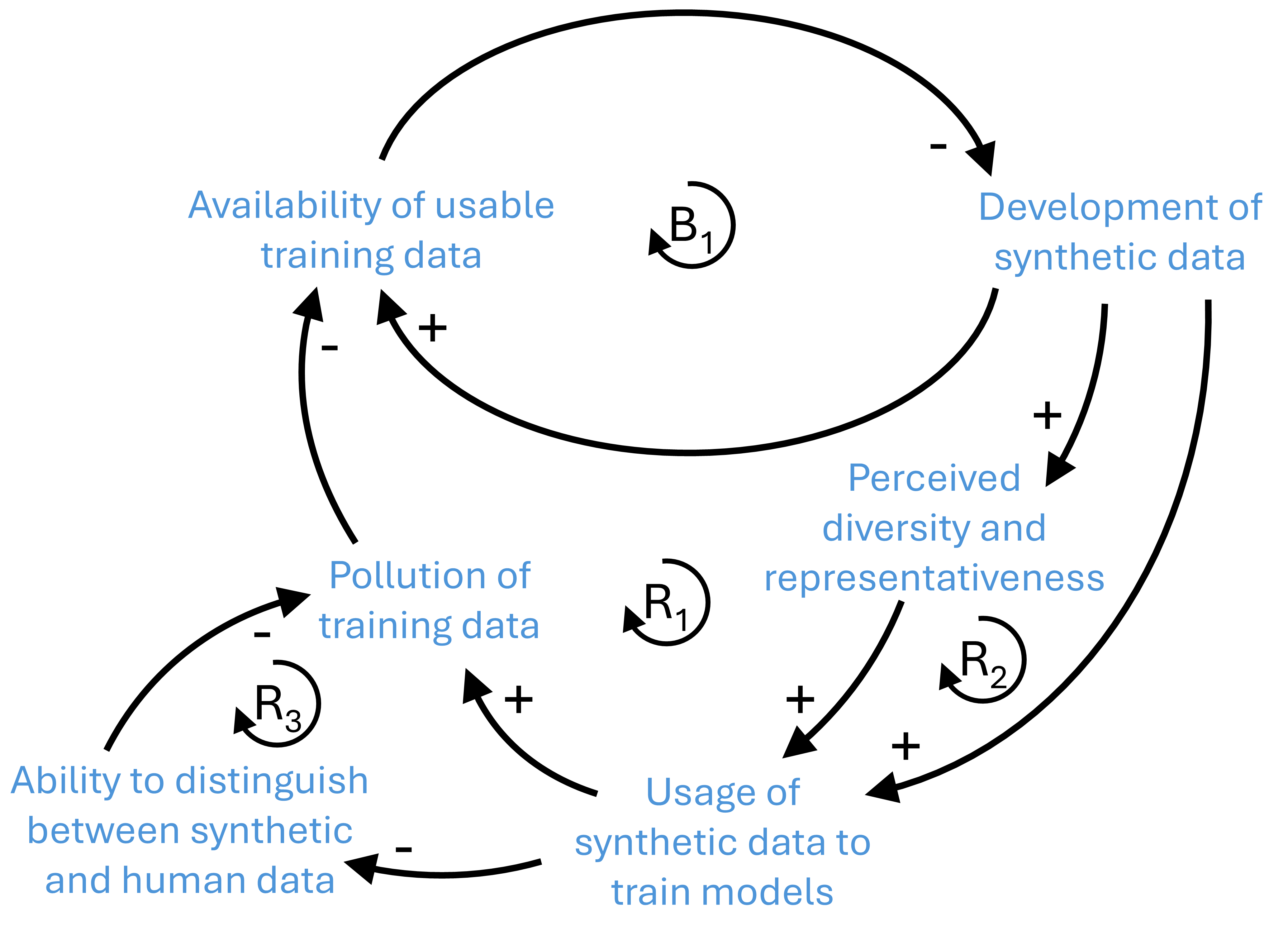}
    \caption{To improve the availability of usable training data, development of synthetic data is a response that is currently pursued. While it addresses the availability, it has side effects including increased misinformation, model collapse, and decline in data authenticity. These reinforcing loops will cause the fix to fail. The loops are: B\textsubscript{1}: Data scaling, R\textsubscript{1}: Misinformation, R\textsubscript{2}: Model collapse, R\textsubscript{3}: Impact on data authenticity. }
    \label{fig:fig6}
\end{figure*}

When such polluted data continues to be used for training, a feedback loop sets in. The use of synthetic data to train models is known as model collapse for generative models and unfairness feedback loops for supervised models \cite{wyllie_fairness_2024}. “Model collapse is a degenerative process affecting generations of learned generative models, in which the data they generate end up polluting the training set of the next generation. Being trained on polluted data, they then misperceive reality.” \cite[p.~755]{shumailov_ai_2024}. This is demonstrated for text outputs \cite{shumailov_ai_2024} and for synthetically generated images \cite{hataya_will_2023}. It therefore becomes vital to preserve human-generated data and be able to distinguish synthetic from real data.

The trajectory of scaling data follows the \textit{fixes that fail} archetype \cite{senge_fifth_1997}: the ``fix'' of using synthetic data to address the shortage of training data only temporarily enables further scaling (B\textsubscript{1}) but does not change the underlying issue that human-generated, high quality data is limited. Instead, over time this intervention exacerbates several unintended consequences, such as misinformation (R\textsubscript{1}), model collapse (R\textsubscript{2}), and loss of authenticity (R\textsubscript{3}). In Fig. \ref{fig:fig6}, we show how scaling the amount of usable data (the problem) is met with the development of synthetic data (the perceived fix) but ultimately results in worsening the quality of training data (unintended consequences). 

\subsection{Limits to Energy Supply: Shifting the Burden} \label{limits_energy}

Given the push towards larger and more complex models, due to the need to continue scaling in order to meet revenue demands and the ingrained industry norm of exponential growth, resource consumption is further increasing. This is a limiting factor because energy and water requirements are becoming staggering and companies are facing backlash for their GHG emissions. In response, there is “energy hunger” \cite{halper_ais_2024,halper_energy-hungry_2024}: AI companies continue to invest in gas, coal, and nuclear \cite{Lawson_2024} sources, invest in renewable energies because it is cheaper to scale, upgrade or create new electricity grids, and try to make AI development more efficient. 

\begin{figure*}[h!]
    \centering
    \includegraphics[width=0.65\linewidth]{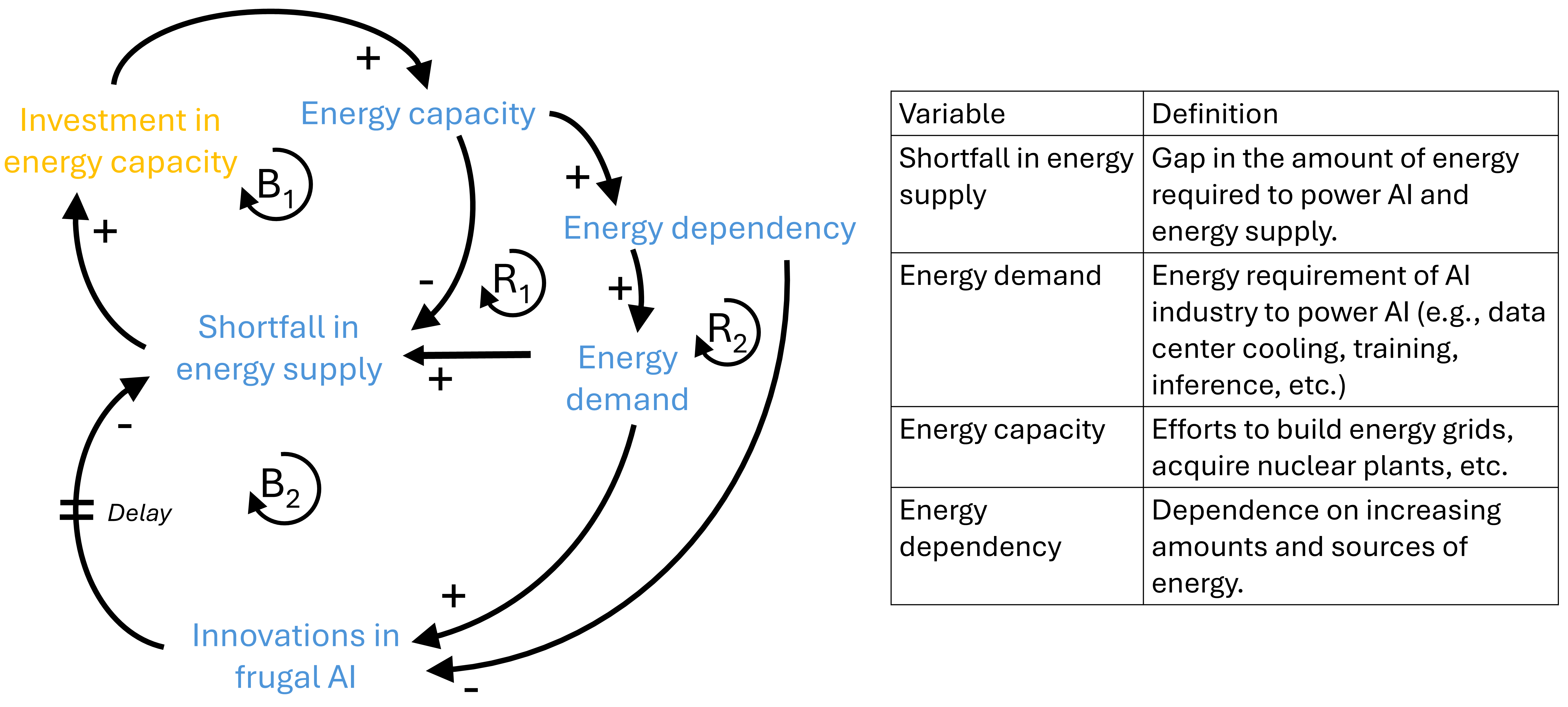}
    \caption{In order to address the shortfall in energy supply, the AI industry creates more energy capacity. This results in an energy dependency which shifts the burden to energy capacity development rather than the foundational solution of frugal AI. The loops are B\textsubscript{1}: Energy scaling, B\textsubscript{2}: Sustainable solution, R\textsubscript{1}: Reinforced energy hunger from capacity development, R\textsubscript{2}: Energy dependency that diverts from frugal AI.}
    \label{fig:fig7}
\end{figure*}

This trajectory follows the \textit{shifting the burden} archetype (see Fig. \ref{fig:fig7}) \cite{senge_fifth_1997}. The temporary solution of developing more energy capacity (e.g., nuclear) to meet energy demands (B\textsubscript{1}) creates a dependence on scaling energy sources (R\textsubscript{1}) to fuel more powerful AI models and diverts attention (R\textsubscript{2}) from the better solution of more responsible energy consumption through frugal AI (B\textsubscript{2}). 

An example of shifting the burden is seen with AI companies' investments in renewable energies to help power their data centers as purchasing further electricity becomes economically unfeasible as the price to meet power demand starts to surge at 130GW \cite{reed_cassady_challenge_2024}. Microsoft has signed a \$10 billion deal for 10.5GW of renewable energy to be delivered by 2030 \cite{kimball_microsoft_2024} and Meta has previously invested in 12GW of renewable energy projects globally over the last 10 years \cite{meta_sustainability_our_2024}. Renewable energies are a long-term investment and in the short-term, companies are turning towards upgrading or building electricity grids and continuing the use of fossil-fueled back-up generators and power sources for data centers, including coal plants \cite{Olivo_2024,Fanger_2024}. Amazon claims to be supporting U.S. power grid development by creating grid-enhancing technologies \cite{grene_microsoft_2023}. Despite these efforts, power generated from renewable energies is not consistent enough to power AI models and comes with a significant material footprint \cite{international_energy_agency_mineral_2021}.

% Thus moving towards renewable energies enables continued scaling and decreases expenditure on energy. 

% In recent times, companies have moved towards making computing more efficient such as by recycling water used by chip manufacturing facilities \cite{alex_irwin-hunt_thirsty_2023} or being more compute efficient in order to reduce emissions \cite{david_patterson_how_2021}.  However, the actual process of moving towards carbon efficiency is more nuanced than simply improving compute efficiency. In actuality, “operational carbon emissions are a function of both energy and carbon intensity, which is dependent on time and location, and energy is a complex function of several factors which metrics of compute (e.g. FLOPS, number of parameters, and runtime) do not fully capture.” \cite[p.~4]{wright_efficiency_2023}. Therefore compute efficiency does not translate to energy efficiency which does not result in savings in carbon emissions. 

\section{Discussion: Progress beyond Growth} \label{sec:discussion}

Limits to performance, data, and energy supply currently prevent AI from scaling further and are therefore prioritized for innovation and advancements. However, it is the barriers that do \textit{not} represent current limits to growth and are externalized by the AI industry which have the most concerning consequences. Below, we outline these in turn and then discuss how to avoid an overshoot and collapse trajectory by redefining what progress in AI means.

\subsection{Unlimited Capital Expenditure has Consequences} \label{nolimits_capital}

While the lack of revenue generated by the AI industry as compared to the scale of capital expenditure can pose a barrier, there are no apparent limits to growth in investments at the moment. Investments continue to scale despite the lack of profitability of AI companies. We observe some dynamics and feedback loops of the scaling ``hype'' \cite{varoquaux_hype_2024}.

Industry shapes narratives by controlling access to information to build hype and market AI as ``tech-positive'' \cite{whittaker_steep_2021} and transformative. National AI strategies ``talk AI into being'' by framing it as inevitable and revolutionary, and necessary for ``future societal welfare'' \cite{bareis_talking_2022}. The rebranding of LLMs as foundation models is considered a tactic to distance from the negative discourse that surrounds them \cite{whittaker_steep_2021}. The compute divide between academia and industry is also to be noted as industry was responsible for training 81\% of large ML models as of 2022 \cite{besiroglu_compute_2024}. This means these models are developed based on industry motivations and incentives and continue to reify the narrative of AI from industry’s perspective. Prominent actors and groups have superficially advocated for an “AI pause” in order to make AI systems and development safer \cite{future_of_life_institute_pause_2023}. But the display of longtermism \cite{gebru_tescreal_2024} places importance on possible future harms instead of highlighting and recognizing the reality that current AI systems already exhibit and exacerbate inequities and discrimination \cite{gebru_statement_2023}. The main effect of these claims is that their ``extreme risk'' framing reinforces the AI hype by underscoring its  narrative as a powerful, transformative, and inevitable technology. 

%Certain businesses by collecting vast amounts of data can create monopolies and prevent other businesses in the same industry from access to the data. For example, only these companies can alter their terms of conditions to collect data from users using their own software and apps or pay a large fee to access another platform’s content creating a data asymmetry. 

\subsection{Ethics Capture Invisibilizes Social and Ecological Harms} \label{nolimits_ethics}

It is worth noting that the social and ecological costs of continued scaling do not currently feed back into limits to growth, despite increased recognition and organizing. Various forms of \textit{capture} have been identified \cite{whittaker_steep_2021}.

\textbf{Regulatory capture}, “the practice whereby private industry professionals or lobbyists overtake regulatory agencies to serve their own interests” \cite[p.~1]{saltelli_science_2022}, has been discussed since the 1950s. Capture occurs when industry is successfully able to influence policy processes and outcomes. Methods in which capture is performed can be categorized as direct and indirect \cite{wei_how_2024}. AI companies engage in advocacy, agenda-setting, and information management \cite{wei_how_2024} to shape policy content and enforcement. Governments often provide tax breaks to attract data center investments similar to factory investments. Unlike factories, however, data centers require few workers \cite{saijel_kishan_its_2024} and provide limited gains for local communities, which instead face cutbacks to energy capacities. For example, AWS data centers claim to benefit local communities such as adding \$6.4 billion to Oregon’s GDP, creating over 5000 jobs, and providing cooling water from data centers to farmers free of cost, as of 2023 \cite{roger_wehner_5_2023}. However, Oregon’s community members instead face increased electric utilities bills due to data centers’ energy consumption \cite{halper_ais_2024}. An earlier investigation revealed that Apple, Google, Microsoft, and Facebook obtained \$800 million in tax breaks for data centers despite only creating 837 permanent jobs \cite{jeans_data_2021}. Furthermore, several large data center projects initially obscured their identities until the tax incentives were granted \cite{jeans_data_2021} so communities were unaware of the details until it was too late to oppose rezoning and its consequences to property values, property taxes, and school funding \cite{jeans_data_2021}.

\textbf{Academic capture} indirectly influences research agendas \cite{wei_how_2024,young_confronting_2022} and generally takes the form of funding institutions, sponsoring academic conferences or events, inviting certain academics as research experts on panels or boards, or providing compute infrastructure and resources \cite{wei_how_2024}. This is done by these companies to incentivize and influence certain types and topics of research as well as for self-promotion. 

\textbf{Beyond capture}, it is seen that AI companies demonstrate ethicality in a performative manner to appease the larger public. \textit{Ethics washing} (“the practice of feigning ethical consideration to improve how a person or organization is perceived” \cite{noauthor_ethics_nodate}) is one mechanism used to counter and sidestep increased organized resistance while influencing regulatory functions. \textit{Ethics shopping} is “the malpractice of choosing, adapting, or revising ethical principles... from a variety of available offers, in order to retrofit some pre-existing behaviours...” \cite[p.~186]{floridi_translating_2019}. This prevents comparisons between companies and makes it more difficult to parse as the language is mixed and inconsistent. \textit{Ethics dumping} occurs when companies export “...unethical research practices to countries where there are weaker legal and ethical frameworks and enforcing mechanisms.” \cite[p.~189]{floridi_translating_2019}. This is seen particularly with AI data work, such as in the case of business process outsourcing (BPO) companies in Africa who employ data workers that have to parse through disturbing and gruesome content while working under contracts that incentivize productivity targets like speed and accuracy \cite{perrigo_exclusive_2023,satariano_silent_2021}. Many companies offset emissions by purchasing carbon credits that invest in carbon removal and emissions reductions projects, e.g. \cite{esg_today_writing_staff_microsoft_2024,e_s_g_news_meta_2024,george_google_2024,l_google_2024,l_microsoft_2024,melodie_michel_meta_2024,spring_firms_2024}. While these appear to mitigate carbon emissions, ample evidence shows offset credits to be misleading attempts at \textit{greenwashing} \cite{akshat_rathi_how_2024,chapman_are_2023,greenfield_revealed_2023,jennifer_van_evra_carbon_2021,natasha_white_bogus_2023,probst_systematic_2024,rathi_big_2022}. For example, an investigation into Verra revealed 90\% of the rainforest offset credits to be questionable, and 94.9 million carbon credits were claimed for only 5.5 million real emissions reductions \cite{greenfield_revealed_2023}. A more comprehensive study of over 2000 projects shows that only 16\% of carbon credits lead to real emissions reductions \cite{probst_systematic_2024}. 

Despite all these captures, then, the ecological and social costs of AI growth have become hard to ignore as limits to growth become increasingly acute.

\subsection{Limits to Growth, Overshoot, and Collapse} \label{overshoot}

\begin{figure*}
    \centering
    \includegraphics[width=0.85\linewidth]{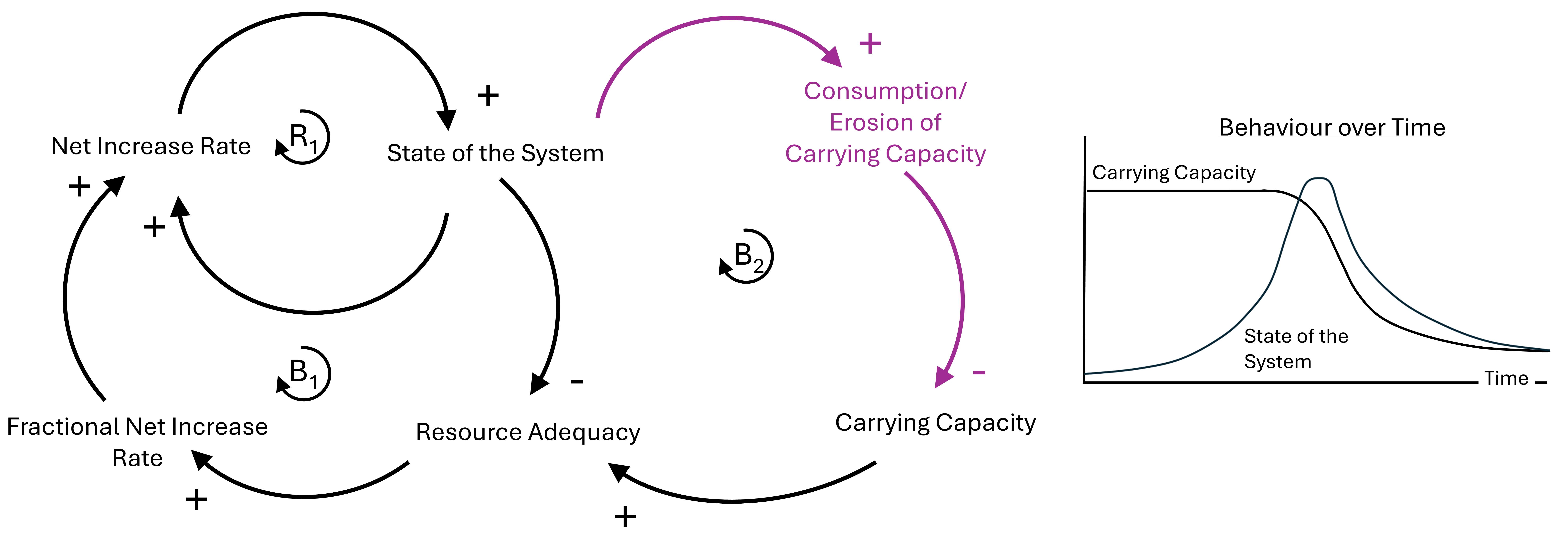}
    \caption{Overshoot and Collapse archetype template, adapted from \cite{sterman_business_2000}. The archetype demonstrates that as the state of the system grows, it encounters limits in resource adequacy. In addition, the high usage of the resource causes erosion of the carrying capacity which results in overshoot and collapse.}
    \label{fig:fig8}
\end{figure*}

Returning to the theme of \textit{Limits to Growth}, consider Fig. \ref{fig:fig8}. The only difference to Fig. \ref{fig:fig1} is an added link between the state of the system and the carrying capacity. This link is based on the recognition that system activity itself can affect the capacity of the containing ecosystem to support continued activities \cite{meadows_limits_1972}, such as ecosystem viability and planetary life support systems \cite{richardson_earth_2023,gupta_just_2024,rockstrom_planetary_2024}. In the anthropocene, computing has reached activity levels with measurable impact on a planetary scale \cite{creutzig_digitalization_2022}.  The crucial difference is that unlike a \textit{limited growth} scenario in which the activity level of a system stagnates, like a saturated market, the erosion of carrying capacity undermines the continued stability of the entire ecosystem and leads either to system collapse or an unstable oscillation trajectory.

Overshoot and collapse patterns can apply in at least two ways. First, model collapse already has been proven with respect to deteriorating data quality, as discussed earlier. In this case, \textit{resource adequacy} would refer to the availability of high-quality training data for LLMs. The proliferation of generated text online erodes that availability as already illustrated by the closure of Wordfreq noticed above \cite{koebler_project_2024}. The collapse of model quality by continued erosion of data quality has been demonstrated empirically \cite{shumailov_ai_2024,hataya_will_2023}. Second, if the generative AI boom continues to drive excessive investments in new hyperscale data centers, their material impact will impact ecosystems on a global scale,  significantly affecting planetary boundaries including atmospheric CO\textsubscript{2}, air quality, freshwater cycles, and pollution \cite{rockstrom_planetary_2009,Steffen__2015}. 

Alternative paths can be opened by \textbf{refocusing}. A \textit{moderate} refocusing could involve moving from resource-intensive progress on model performance to resource-efficiency progress as proposed recently \cite{varoquaux_hype_2024,frugal_ai_2025}. While this is presented as \textit{frugal AI}, it should be noted that a more \textit{radical} refocusing on frugality would shift the focus altogether \textbf{from efficiency to sufficiency} by placing primary emphasis on asking ``what's good enough?''. Nevertheless, frugal AI is a promising avenue for course correction that would much more effectively address limits to growth since it focuses on the primary drivers of accelerating scaling. In other words, it is a high leverage point for system change \cite{meadows_thinking_2011} to open spaces for genuinely innovative development.

\section{Conclusion} \label{conclusion}
Above, we have mapped out the accelerating growth of AI along the lines of technical, economic, ecological, and social dimensions and their interactions. We have illustrated how the mathematical laws of scaling translate into technical development with economic implications of scale subject to capitalist and market dynamics. The resulting economies of scale cause ecological destruction, which is ultimately an ethical concern too.

We explored the dynamics of how these perspectives interact through causal modelling. Drawing on system dynamics archetypes, we argue that the AI industry’s responses to barriers typically attempt to overcome apparent limits in one perspective but fail to account for resulting damages in other perspectives. These damages cause social and ecological harms that are externalized by the AI industry but demonstrate a vital need to realign our priorities around scaling. Lastly, we emphasize the need for refocusing on \textit{sufficient AI} practices to avoid an \textit{overshoot and collapse} trajectory. 

%could talk about the growth metaphor but maybe that's a stretch for here.
Growth always ends -- either by design, or by disaster. Which path we choose has not been decided yet. In Fall 2024, the United Nations Pact for the Future enshrined a commitment to develop measures of progress beyond GDP growth in future agendas and sustainable development policy \cite{un_general_assembly_pact_2024}. As the world comes to finally recognize that pursuing relentless growth is the path to assured disasters, it is time that the AI industry catches on and changes course.

% \begin{acks}
% This research was partially supported by NSERC through RGPIN-2016-06640,  the Canada Foundation for Innovation, and the Ontario Research Foundation.
% \end{acks}

\bibliographystyle{ACM-Reference-Format}
\bibliography{facct25bib_part1,facct25bib_part2,facct25bib_part3}


\begin{thebibliography}{203}

%%% the Web address.

\ifx \showCODEN    \undefined \def \showCODEN     #1{\unskip}     \fi
\ifx \showDOI      \undefined \def \showDOI       #1{#1}\fi
\ifx \showISBNx    \undefined \def \showISBNx     #1{\unskip}     \fi
\ifx \showISBNxiii \undefined \def \showISBNxiii  #1{\unskip}     \fi
\ifx \showISSN     \undefined \def \showISSN      #1{\unskip}     \fi
\ifx \showLCCN     \undefined \def \showLCCN      #1{\unskip}     \fi
\ifx \shownote     \undefined \def \shownote      #1{#1}          \fi
\ifx \showarticletitle \undefined \def \showarticletitle #1{#1}   \fi
\ifx \showURL      \undefined \def \showURL       {\relax}        \fi
% invisible to TeX
\providecommand\bibfield[2]{#2}
\providecommand\bibinfo[2]{#2}
\providecommand\natexlab[1]{#1}
\providecommand\showeprint[2][]{arXiv:#2}

\bibitem[noa({[n.\,d.]})]%
        {noauthor_ethics_nodate}
 \bibinfo{year}{[n.\,d.]}\natexlab{}.
\newblock \bibinfo{title}{Ethics washing}.
\newblock
\newblock
\urldef\tempurl%
\url{https://www.carnegiecouncil.org/explore-engage/key-terms/ethics-washing}
\showURL{%
\tempurl}


\bibitem[cor(2024)]%
        {corps_abandon_esg_2024}
 \bibinfo{year}{2024}\natexlab{}.
\newblock
\newblock
\urldef\tempurl%
\url{https://www.instituteforenergyresearch.org/climate-change/big-corporations-are-abandoning-their-climate-commitments/}
\showURL{%
\tempurl}


\bibitem[noa(2024a)]%
        {noauthor_history-lovers_2024}
 \bibinfo{year}{2024}\natexlab{a}.
\newblock \showarticletitle{A history-lover’s guide to the market panic over {AI}}.
\newblock \bibinfo{journal}{\emph{The Economist}} (\bibinfo{year}{2024}).
\newblock
\showISSN{0013-0613}
\urldef\tempurl%
\url{https://www.economist.com/business/2024/08/06/a-history-lovers-guide-to-the-market-panic-over-ai}
\showURL{%
\tempurl}


\bibitem[noa(2024b)]%
        {noauthor_introducing_2024}
 \bibinfo{year}{2024}\natexlab{b}.
\newblock \bibinfo{title}{Introducing {Llama} 3.1: {Our} most capable models to date}.
\newblock
\newblock
\urldef\tempurl%
\url{https://ai.meta.com/blog/meta-llama-3-1/}
\showURL{%
\tempurl}


\bibitem[noa(2024c)]%
        {noauthor_what_2024}
 \bibinfo{year}{2024}\natexlab{c}.
\newblock \showarticletitle{What happened to the artificial-intelligence revolution?}
\newblock \bibinfo{journal}{\emph{The Economist}} (\bibinfo{year}{2024}).
\newblock
\showISSN{0013-0613}
\urldef\tempurl%
\url{https://www.economist.com/finance-and-economics/2024/07/02/what-happened-to-the-artificial-intelligence-revolution}
\showURL{%
\tempurl}


\bibitem[Abdalla and Abdalla(2021)]%
        {abdalla_grey_2021}
\bibfield{author}{\bibinfo{person}{Mohamed Abdalla} {and} \bibinfo{person}{Moustafa Abdalla}.} \bibinfo{year}{2021}\natexlab{}.
\newblock \showarticletitle{The {Grey} {Hoodie} {Project}: {Big} {Tobacco}, {Big} {Tech}, and the {Threat} on {Academic} {Integrity}}. In \bibinfo{booktitle}{\emph{Proceedings of the 2021 {AAAI}/{ACM} {Conference} on {AI}, {Ethics}, and {Society}}} \emph{(\bibinfo{series}{{AIES} '21})}. \bibinfo{publisher}{Association for Computing Machinery}, \bibinfo{address}{New York, NY, USA}, \bibinfo{pages}{287--297}.
\newblock
\showISBNx{978-1-4503-8473-5}
\urldef\tempurl%
\url{https://doi.org/10.1145/3461702.3462563}
\showDOI{\tempurl}


\bibitem[Acemoglu(2021)]%
        {acemoglu_harms_2021}
\bibfield{author}{\bibinfo{person}{Daron Acemoglu}.} \bibinfo{year}{2021}\natexlab{}.
\newblock \bibinfo{title}{Harms of {AI}}.
\newblock
\newblock
\urldef\tempurl%
\url{https://doi.org/10.3386/w29247}
\showDOI{\tempurl}


\bibitem[Agyeman et~al\mbox{.}(2003)]%
        {agyeman_just_2003}
\bibfield{author}{\bibinfo{person}{Julian Agyeman}, \bibinfo{person}{Robert~Doyle Bullard}, {and} \bibinfo{person}{Bob Evans}.} \bibinfo{year}{2003}\natexlab{}.
\newblock \bibinfo{booktitle}{\emph{Just sustainabilities: {Development} in an unequal world}}.
\newblock \bibinfo{publisher}{MIT press}.
\newblock


\bibitem[{Akshat Rathi}(2024)]%
        {akshat_rathi_google_2024}
\bibfield{author}{\bibinfo{person}{{Akshat Rathi}}.} \bibinfo{year}{2024}\natexlab{}.
\newblock \showarticletitle{Google {Is} {No} {Longer} {Claiming} to {Be} {Carbon} {Neutral}}.
\newblock \bibinfo{journal}{\emph{Bloomberg.com}} (\bibinfo{year}{2024}).
\newblock
\urldef\tempurl%
\url{https://www.bloomberg.com/news/articles/2024-07-08/google-is-no-longer-claiming-to-be-carbon-neutral}
\showURL{%
\tempurl}


\bibitem[{Akshat Rathi} and {Natasha White}(2024)]%
        {akshat_rathi_how_2024}
\bibfield{author}{\bibinfo{person}{{Akshat Rathi}} {and} \bibinfo{person}{{Natasha White}}.} \bibinfo{year}{2024}\natexlab{}.
\newblock \showarticletitle{How {Tech} {Companies} {Are} {Obscuring} {AI}’s {Real} {Carbon} {Footprint}}.
\newblock \bibinfo{journal}{\emph{Bloomberg.com}} (\bibinfo{year}{2024}).
\newblock
\urldef\tempurl%
\url{https://www.bloomberg.com/news/articles/2024-08-21/ai-tech-giants-hide-dirty-energy-with-outdated-carbon-accounting-rules}
\showURL{%
\tempurl}


\bibitem[Al-Sibai(2025)]%
        {Al-Sibai_2025}
\bibfield{author}{\bibinfo{person}{Noor Al-Sibai}.} \bibinfo{year}{2025}\natexlab{}.
\newblock \bibinfo{title}{It Costs So Much to Run ChatGPT That OpenAI Is Losing Money on \$200 ChatGPT Pro Subscriptions}.
\newblock
\newblock
\urldef\tempurl%
\url{https://futurism.com/the-byte/openai-chatgpt-pro-subscription-losing-money}
\showURL{%
\tempurl}


\bibitem[{Alex Irwin-Hunt}(2023)]%
        {alex_irwin-hunt_thirsty_2023}
\bibfield{author}{\bibinfo{person}{{Alex Irwin-Hunt}}.} \bibinfo{year}{2023}\natexlab{}.
\newblock \bibinfo{title}{Thirsty chip facilities under scrutiny in water stressed areas}.
\newblock
\newblock
\urldef\tempurl%
\url{https://www.fdiintelligence.com/content/feature/thirsty-chip-facilities-under-scrutiny-in-water-stressed-areas-82810}
\showURL{%
\tempurl}


\bibitem[Argerich and Patiño-Martínez(2024)]%
        {argerich_measuring_2024}
\bibfield{author}{\bibinfo{person}{Mauricio~Fadel Argerich} {and} \bibinfo{person}{Marta Patiño-Martínez}.} \bibinfo{year}{2024}\natexlab{}.
\newblock \showarticletitle{Measuring and {Improving} the {Energy} {Efficiency} of {Large} {Language} {Models} {Inference}}.
\newblock \bibinfo{journal}{\emph{IEEE Access}}  \bibinfo{volume}{12} (\bibinfo{year}{2024}), \bibinfo{pages}{80194--80207}.
\newblock
\showISSN{2169-3536}
\urldef\tempurl%
\url{https://doi.org/10.1109/ACCESS.2024.3409745}
\showDOI{\tempurl}


\bibitem[Bahri et~al\mbox{.}(2024)]%
        {bahri_explaining_2024}
\bibfield{author}{\bibinfo{person}{Yasaman Bahri}, \bibinfo{person}{Ethan Dyer}, \bibinfo{person}{Jared Kaplan}, \bibinfo{person}{Jaehoon Lee}, {and} \bibinfo{person}{Utkarsh Sharma}.} \bibinfo{year}{2024}\natexlab{}.
\newblock \showarticletitle{Explaining neural scaling laws}.
\newblock \bibinfo{journal}{\emph{Proceedings of the National Academy of Sciences}} \bibinfo{volume}{121}, \bibinfo{number}{27} (\bibinfo{year}{2024}), \bibinfo{pages}{e2311878121}.
\newblock
\showISSN{0027-8424, 1091-6490}
\urldef\tempurl%
\url{https://doi.org/10.1073/pnas.2311878121}
\showDOI{\tempurl}


\bibitem[Bareis and Katzenbach(2022)]%
        {bareis_talking_2022}
\bibfield{author}{\bibinfo{person}{Jascha Bareis} {and} \bibinfo{person}{Christian Katzenbach}.} \bibinfo{year}{2022}\natexlab{}.
\newblock \showarticletitle{Talking {AI} into {Being}: {The} {Narratives} and {Imaginaries} of {National} {AI} {Strategies} and {Their} {Performative} {Politics}}.
\newblock \bibinfo{journal}{\emph{Science, Technology, \& Human Values}} \bibinfo{volume}{47}, \bibinfo{number}{5} (\bibinfo{year}{2022}), \bibinfo{pages}{855--881}.
\newblock
\showISSN{0162-2439}
\urldef\tempurl%
\url{https://doi.org/10.1177/01622439211030007}
\showDOI{\tempurl}


\bibitem[Barendregt et~al\mbox{.}(2021)]%
        {barendregt_defund_2021}
\bibfield{author}{\bibinfo{person}{Wolmet Barendregt}, \bibinfo{person}{Christoph Becker}, \bibinfo{person}{EunJeong Cheon}, \bibinfo{person}{Andrew Clement}, \bibinfo{person}{Pedro Reynolds-Cuéllar}, \bibinfo{person}{Douglas Schuler}, {and} \bibinfo{person}{Lucy Suchman}.} \bibinfo{year}{2021}\natexlab{}.
\newblock \showarticletitle{Defund {Big} {Tech}, {Refund} {Community}}.
\newblock \bibinfo{journal}{\emph{Tech Otherwise}} (\bibinfo{year}{2021}).
\newblock
\urldef\tempurl%
\url{https://doi.org/10.21428/93b2c832.e0100a3f}
\showDOI{\tempurl}


\bibitem[Barr(2023)]%
        {barr_llamas_2023}
\bibfield{author}{\bibinfo{person}{Alistair Barr}.} \bibinfo{year}{2023}\natexlab{}.
\newblock \bibinfo{title}{Llamas don't drink much water. {Meta}'s new {AI} version is damn thirsty.}
\newblock
\newblock
\urldef\tempurl%
\url{https://www.businessinsider.com/meta-llama2-ai-uses-almost-twice-water-as-llama-2023-7}
\showURL{%
\tempurl}


\bibitem[Bashir et~al\mbox{.}(2024)]%
        {bashir_climate_2024}
\bibfield{author}{\bibinfo{person}{Noman Bashir}, \bibinfo{person}{Priya Donti}, \bibinfo{person}{James Cuff}, \bibinfo{person}{Sydney Sroka}, \bibinfo{person}{Marija Ilic}, \bibinfo{person}{Vivienne Sze}, \bibinfo{person}{Christina Delimitrou}, {and} \bibinfo{person}{Elsa Olivetti}.} \bibinfo{year}{2024}\natexlab{}.
\newblock \showarticletitle{The {Climate} and {Sustainability} {Implications} of {Generative} {AI}}.
\newblock \bibinfo{journal}{\emph{An MIT Exploration of Generative AI}} (\bibinfo{year}{2024}).
\newblock
\urldef\tempurl%
\url{https://mit-genai.pubpub.org/pub/8ulgrckc/release/2}
\showURL{%
\tempurl}


\bibitem[Becker(2023)]%
        {becker_insolvent_2023}
\bibfield{author}{\bibinfo{person}{Christoph Becker}.} \bibinfo{year}{2023}\natexlab{}.
\newblock \bibinfo{booktitle}{\emph{Insolvent: {How} to {Reorient} {Computing} for {Just} {Sustainability}}}.
\newblock \bibinfo{publisher}{MIT Press}.
\newblock
\showISBNx{978-0-262-54560-0}
\urldef\tempurl%
\url{https://mitpress.mit.edu/9780262545600/insolvent/}
\showURL{%
\tempurl}


\bibitem[Bender et~al\mbox{.}(2021)]%
        {bender_dangers_2021}
\bibfield{author}{\bibinfo{person}{Emily~M. Bender}, \bibinfo{person}{Timnit Gebru}, \bibinfo{person}{Angelina McMillan-Major}, {and} \bibinfo{person}{Shmargaret Shmitchell}.} \bibinfo{year}{2021}\natexlab{}.
\newblock \showarticletitle{On the Dangers of Stochastic Parrots: Can Language Models Be Too Big? \raisebox{-2pt}{\includegraphics[scale=0.08]{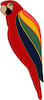}}}. In \bibinfo{booktitle}{\emph{Proceedings of the 2021 {ACM} {Conference} on {Fairness}, {Accountability}, and {Transparency}}} \emph{(\bibinfo{series}{{FAccT} '21})}. \bibinfo{publisher}{Association for Computing Machinery}, \bibinfo{address}{New York, NY, USA}, \bibinfo{pages}{610--623}.
\newblock
\showISBNx{978-1-4503-8309-7}
\urldef\tempurl%
\url{https://doi.org/10.1145/3442188.3445922}
\showDOI{\tempurl}


\bibitem[Besiroglu et~al\mbox{.}(2024)]%
        {besiroglu_compute_2024}
\bibfield{author}{\bibinfo{person}{Tamay Besiroglu}, \bibinfo{person}{Sage~Andrus Bergerson}, \bibinfo{person}{Amelia Michael}, \bibinfo{person}{Lennart Heim}, \bibinfo{person}{Xueyun Luo}, {and} \bibinfo{person}{Neil Thompson}.} \bibinfo{year}{2024}\natexlab{}.
\newblock \bibinfo{title}{The {Compute} {Divide} in {Machine} {Learning}: {A} {Threat} to {Academic} {Contribution} and {Scrutiny}?}
\newblock
\newblock
\urldef\tempurl%
\url{https://doi.org/10.48550/arXiv.2401.02452}
\showDOI{\tempurl}


\bibitem[Bietti(2020)]%
        {bietti_ethics_2020}
\bibfield{author}{\bibinfo{person}{Elettra Bietti}.} \bibinfo{year}{2020}\natexlab{}.
\newblock \showarticletitle{From ethics washing to ethics bashing: a view on tech ethics from within moral philosophy}. In \bibinfo{booktitle}{\emph{Proceedings of the 2020 {Conference} on {Fairness}, {Accountability}, and {Transparency}}} \emph{(\bibinfo{series}{{FAT}* '20})}. \bibinfo{publisher}{Association for Computing Machinery}, \bibinfo{address}{New York, NY, USA}, \bibinfo{pages}{210--219}.
\newblock
\showISBNx{978-1-4503-6936-7}
\urldef\tempurl%
\url{https://doi.org/10.1145/3351095.3372860}
\showDOI{\tempurl}


\bibitem[Birhane(2021)]%
        {birhane_algorithmic_2021}
\bibfield{author}{\bibinfo{person}{Abeba Birhane}.} \bibinfo{year}{2021}\natexlab{}.
\newblock \showarticletitle{Algorithmic injustice: a relational ethics approach}.
\newblock \bibinfo{journal}{\emph{Patterns}} \bibinfo{volume}{2}, \bibinfo{number}{2} (\bibinfo{year}{2021}), \bibinfo{pages}{100205}.
\newblock
\showISSN{2666-3899}
\urldef\tempurl%
\url{https://doi.org/10.1016/j.patter.2021.100205}
\showDOI{\tempurl}


\bibitem[Bommasani et~al\mbox{.}(2022)]%
        {bommasani_opportunities_2022}
\bibfield{author}{\bibinfo{person}{Rishi Bommasani}, \bibinfo{person}{Drew~A. Hudson}, \bibinfo{person}{Ehsan Adeli}, \bibinfo{person}{Russ Altman}, \bibinfo{person}{Simran Arora}, \bibinfo{person}{Sydney~von Arx}, \bibinfo{person}{Michael~S. Bernstein}, \bibinfo{person}{Jeannette Bohg}, \bibinfo{person}{Antoine Bosselut}, \bibinfo{person}{Emma Brunskill}, \bibinfo{person}{Erik Brynjolfsson}, \bibinfo{person}{Shyamal Buch}, \bibinfo{person}{Dallas Card}, \bibinfo{person}{Rodrigo Castellon}, \bibinfo{person}{Niladri Chatterji}, \bibinfo{person}{Annie Chen}, \bibinfo{person}{Kathleen Creel}, \bibinfo{person}{Jared~Quincy Davis}, \bibinfo{person}{Dora Demszky}, \bibinfo{person}{Chris Donahue}, \bibinfo{person}{Moussa Doumbouya}, \bibinfo{person}{Esin Durmus}, \bibinfo{person}{Stefano Ermon}, \bibinfo{person}{John Etchemendy}, \bibinfo{person}{Kawin Ethayarajh}, \bibinfo{person}{Li Fei-Fei}, \bibinfo{person}{Chelsea Finn}, \bibinfo{person}{Trevor Gale}, \bibinfo{person}{Lauren Gillespie}, \bibinfo{person}{Karan
  Goel}, \bibinfo{person}{Noah Goodman}, \bibinfo{person}{Shelby Grossman}, \bibinfo{person}{Neel Guha}, \bibinfo{person}{Tatsunori Hashimoto}, \bibinfo{person}{Peter Henderson}, \bibinfo{person}{John Hewitt}, \bibinfo{person}{Daniel~E. Ho}, \bibinfo{person}{Jenny Hong}, \bibinfo{person}{Kyle Hsu}, \bibinfo{person}{Jing Huang}, \bibinfo{person}{Thomas Icard}, \bibinfo{person}{Saahil Jain}, \bibinfo{person}{Dan Jurafsky}, \bibinfo{person}{Pratyusha Kalluri}, \bibinfo{person}{Siddharth Karamcheti}, \bibinfo{person}{Geoff Keeling}, \bibinfo{person}{Fereshte Khani}, \bibinfo{person}{Omar Khattab}, \bibinfo{person}{Pang~Wei Koh}, \bibinfo{person}{Mark Krass}, \bibinfo{person}{Ranjay Krishna}, \bibinfo{person}{Rohith Kuditipudi}, \bibinfo{person}{Ananya Kumar}, \bibinfo{person}{Faisal Ladhak}, \bibinfo{person}{Mina Lee}, \bibinfo{person}{Tony Lee}, \bibinfo{person}{Jure Leskovec}, \bibinfo{person}{Isabelle Levent}, \bibinfo{person}{Xiang~Lisa Li}, \bibinfo{person}{Xuechen Li}, \bibinfo{person}{Tengyu Ma},
  \bibinfo{person}{Ali Malik}, \bibinfo{person}{Christopher~D. Manning}, \bibinfo{person}{Suvir Mirchandani}, \bibinfo{person}{Eric Mitchell}, \bibinfo{person}{Zanele Munyikwa}, \bibinfo{person}{Suraj Nair}, \bibinfo{person}{Avanika Narayan}, \bibinfo{person}{Deepak Narayanan}, \bibinfo{person}{Ben Newman}, \bibinfo{person}{Allen Nie}, \bibinfo{person}{Juan~Carlos Niebles}, \bibinfo{person}{Hamed Nilforoshan}, \bibinfo{person}{Julian Nyarko}, \bibinfo{person}{Giray Ogut}, \bibinfo{person}{Laurel Orr}, \bibinfo{person}{Isabel Papadimitriou}, \bibinfo{person}{Joon~Sung Park}, \bibinfo{person}{Chris Piech}, \bibinfo{person}{Eva Portelance}, \bibinfo{person}{Christopher Potts}, \bibinfo{person}{Aditi Raghunathan}, \bibinfo{person}{Rob Reich}, \bibinfo{person}{Hongyu Ren}, \bibinfo{person}{Frieda Rong}, \bibinfo{person}{Yusuf Roohani}, \bibinfo{person}{Camilo Ruiz}, \bibinfo{person}{Jack Ryan}, \bibinfo{person}{Christopher Ré}, \bibinfo{person}{Dorsa Sadigh}, \bibinfo{person}{Shiori Sagawa},
  \bibinfo{person}{Keshav Santhanam}, \bibinfo{person}{Andy Shih}, \bibinfo{person}{Krishnan Srinivasan}, \bibinfo{person}{Alex Tamkin}, \bibinfo{person}{Rohan Taori}, \bibinfo{person}{Armin~W. Thomas}, \bibinfo{person}{Florian Tramèr}, \bibinfo{person}{Rose~E. Wang}, \bibinfo{person}{William Wang}, \bibinfo{person}{Bohan Wu}, \bibinfo{person}{Jiajun Wu}, \bibinfo{person}{Yuhuai Wu}, \bibinfo{person}{Sang~Michael Xie}, \bibinfo{person}{Michihiro Yasunaga}, \bibinfo{person}{Jiaxuan You}, \bibinfo{person}{Matei Zaharia}, \bibinfo{person}{Michael Zhang}, \bibinfo{person}{Tianyi Zhang}, \bibinfo{person}{Xikun Zhang}, \bibinfo{person}{Yuhui Zhang}, \bibinfo{person}{Lucia Zheng}, \bibinfo{person}{Kaitlyn Zhou}, {and} \bibinfo{person}{Percy Liang}.} \bibinfo{year}{2022}\natexlab{}.
\newblock \bibinfo{title}{On the {Opportunities} and {Risks} of {Foundation} {Models}}.
\newblock
\newblock
\urldef\tempurl%
\url{https://doi.org/10.48550/arXiv.2108.07258}
\showDOI{\tempurl}


\bibitem[Booth(2024)]%
        {booth_more_2024}
\bibfield{author}{\bibinfo{person}{Robert Booth}.} \bibinfo{year}{2024}\natexlab{}.
\newblock \showarticletitle{More than 140 {Kenya} {Facebook} moderators diagnosed with severe {PTSD}}.
\newblock \bibinfo{journal}{\emph{The Guardian}} (\bibinfo{year}{2024}).
\newblock
\showISSN{0261-3077}
\urldef\tempurl%
\url{https://www.theguardian.com/media/2024/dec/18/kenya-facebook-moderators-sue-after-diagnoses-of-severe-ptsd}
\showURL{%
\tempurl}


\bibitem[Bourzac(2024)]%
        {bourzac_fixing_2024}
\bibfield{author}{\bibinfo{person}{Katherine Bourzac}.} \bibinfo{year}{2024}\natexlab{}.
\newblock \showarticletitle{Fixing {AI}’s energy crisis}.
\newblock \bibinfo{journal}{\emph{Nature}} (\bibinfo{year}{2024}).
\newblock
\urldef\tempurl%
\url{https://doi.org/10.1038/d41586-024-03408-z}
\showDOI{\tempurl}


\bibitem[Burkacky et~al\mbox{.}(2022)]%
        {Ondrej_2022}
\bibfield{author}{\bibinfo{person}{Ondrej Burkacky}, \bibinfo{person}{Julia Dragon}, {and} \bibinfo{person}{Nikolaus Lehmann}.} \bibinfo{year}{2022}\natexlab{}.
\newblock \bibinfo{title}{The semiconductor decade: A trillion-dollar industry}.
\newblock
\newblock
\urldef\tempurl%
\url{https://www.mckinsey.com/industries/semiconductors/our-insights/the-semiconductor-decade-a-trillion-dollar-industry}
\showURL{%
\tempurl}


\bibitem[Börjesson~Rivera et~al\mbox{.}(2014)]%
        {borjesson_rivera_including_2014}
\bibfield{author}{\bibinfo{person}{Miriam Börjesson~Rivera}, \bibinfo{person}{Cecilia Håkansson}, \bibinfo{person}{Åsa Svenfelt}, {and} \bibinfo{person}{Göran Finnveden}.} \bibinfo{year}{2014}\natexlab{}.
\newblock \showarticletitle{Including second order effects in environmental assessments of {ICT}}.
\newblock \bibinfo{journal}{\emph{Environmental Modelling \& Software}}  \bibinfo{volume}{56} (\bibinfo{year}{2014}), \bibinfo{pages}{105--115}.
\newblock
\showISSN{13648152}
\urldef\tempurl%
\url{https://doi.org/10.1016/j.envsoft.2014.02.005}
\showDOI{\tempurl}


\bibitem[Calvert(2024)]%
        {calvert_ai_2024}
\bibfield{author}{\bibinfo{person}{Brian Calvert}.} \bibinfo{year}{2024}\natexlab{}.
\newblock \bibinfo{title}{{AI} already uses as much energy as a small country. {It}’s only the beginning.}
\newblock
\newblock
\urldef\tempurl%
\url{https://www.vox.com/climate/2024/3/28/24111721/climate-ai-tech-energy-demand-rising}
\showURL{%
\tempurl}


\bibitem[Chapman(2023)]%
        {chapman_are_2023}
\bibfield{author}{\bibinfo{person}{Angus Chapman}.} \bibinfo{year}{2023}\natexlab{}.
\newblock \bibinfo{title}{Are carbon offsets all they’re cracked up to be? {We} tracked one from {Kenya} to {England} to find out.}
\newblock
\newblock
\urldef\tempurl%
\url{https://www.vox.com/23817575/carbon-offsets-credits-financialization-ecologi-solutions-scam}
\showURL{%
\tempurl}


\bibitem[Coeckelbergh(2021)]%
        {Coeckelbergh_2021}
\bibfield{author}{\bibinfo{person}{Mark Coeckelbergh}.} \bibinfo{year}{2021}\natexlab{}.
\newblock \showarticletitle{AI for climate: freedom, justice, and other ethical and political challenges}.
\newblock \bibinfo{journal}{\emph{AI and Ethics}} \bibinfo{volume}{1}, \bibinfo{number}{1} (\bibinfo{year}{2021}), \bibinfo{pages}{67–72}.
\newblock
\showISSN{2730-5961}
\urldef\tempurl%
\url{https://doi.org/10.1007/s43681-020-00007-2}
\showDOI{\tempurl}


\bibitem[Couldry and Mejias(2019)]%
        {couldry_data_2019}
\bibfield{author}{\bibinfo{person}{Nick Couldry} {and} \bibinfo{person}{Ulises~A. Mejias}.} \bibinfo{year}{2019}\natexlab{}.
\newblock \showarticletitle{Data {Colonialism}: {Rethinking} {Big} {Data}’s {Relation} to the {Contemporary} {Subject}}.
\newblock \bibinfo{journal}{\emph{Television \& New Media}} \bibinfo{volume}{20}, \bibinfo{number}{4} (\bibinfo{year}{2019}), \bibinfo{pages}{336--349}.
\newblock
\showISSN{1527-4764}
\urldef\tempurl%
\url{https://doi.org/10.1177/1527476418796632}
\showDOI{\tempurl}


\bibitem[Crawford(2024)]%
        {crawford_generative_2024}
\bibfield{author}{\bibinfo{person}{Kate Crawford}.} \bibinfo{year}{2024}\natexlab{}.
\newblock \showarticletitle{Generative {AI}’s environmental costs are soaring — and mostly secret}.
\newblock \bibinfo{journal}{\emph{Nature}} \bibinfo{volume}{626}, \bibinfo{number}{8000} (\bibinfo{year}{2024}), \bibinfo{pages}{693--693}.
\newblock
\urldef\tempurl%
\url{https://doi.org/10.1038/d41586-024-00478-x}
\showDOI{\tempurl}


\bibitem[Creutzig et~al\mbox{.}(2022)]%
        {creutzig_digitalization_2022}
\bibfield{author}{\bibinfo{person}{Felix Creutzig}, \bibinfo{person}{Daron Acemoglu}, \bibinfo{person}{Xuemei Bai}, \bibinfo{person}{Paul~N. Edwards}, \bibinfo{person}{Marie~Josefine Hintz}, \bibinfo{person}{Lynn~H. Kaack}, \bibinfo{person}{Siir Kilkis}, \bibinfo{person}{Stefanie Kunkel}, \bibinfo{person}{Amy Luers}, \bibinfo{person}{Nikola Milojevic-Dupont}, \bibinfo{person}{Dave Rejeski}, \bibinfo{person}{Jürgen Renn}, \bibinfo{person}{David Rolnick}, \bibinfo{person}{Christoph Rosol}, \bibinfo{person}{Daniela Russ}, \bibinfo{person}{Thomas Turnbull}, \bibinfo{person}{Elena Verdolini}, \bibinfo{person}{Felix Wagner}, \bibinfo{person}{Charlie Wilson}, \bibinfo{person}{Aicha Zekar}, {and} \bibinfo{person}{Marius Zumwald}.} \bibinfo{year}{2022}\natexlab{}.
\newblock \showarticletitle{Digitalization and the {Anthropocene}}.
\newblock \bibinfo{journal}{\emph{Annual Review of Environment and Resources}} \bibinfo{volume}{47}, \bibinfo{number}{Volume 47, 2022} (\bibinfo{date}{Oct.} \bibinfo{year}{2022}), \bibinfo{pages}{479--509}.
\newblock
\showISSN{1543-5938, 1545-2050}
\urldef\tempurl%
\url{https://doi.org/10.1146/annurev-environ-120920-100056}
\showDOI{\tempurl}
\newblock
\shownote{Publisher: Annual Reviews}.


\bibitem[{Dan Hendrycks}(2024)]%
        {dan_hendrycks_ai_2024}
\bibfield{author}{\bibinfo{person}{{Dan Hendrycks}}.} \bibinfo{year}{2024}\natexlab{}.
\newblock \bibinfo{booktitle}{\emph{{AI} {Safety}, {Ethics} and {Society}}}.
\newblock


\bibitem[{David Patterson}(2021)]%
        {david_patterson_how_2021}
\bibfield{author}{\bibinfo{person}{{David Patterson}}.} \bibinfo{year}{2021}\natexlab{}.
\newblock \bibinfo{title}{How we’re minimizing {AI}’s carbon footprint}.
\newblock
\newblock
\urldef\tempurl%
\url{https://blog.google/technology/ai/minimizing-carbon-footprint/}
\showURL{%
\tempurl}


\bibitem[Dayarathna et~al\mbox{.}(2015)]%
        {dayarathna_data_2015}
\bibfield{author}{\bibinfo{person}{Miyuru Dayarathna}, \bibinfo{person}{Yonggang Wen}, {and} \bibinfo{person}{Rui Fan}.} \bibinfo{year}{2015}\natexlab{}.
\newblock \showarticletitle{Data center energy consumption modeling: {A} survey}.
\newblock \bibinfo{journal}{\emph{IEEE Communications surveys \& tutorials}} \bibinfo{volume}{18}, \bibinfo{number}{1} (\bibinfo{year}{2015}), \bibinfo{pages}{732--794}.
\newblock
\newblock
\shownote{Publisher: IEEE}.


\bibitem[Deng et~al\mbox{.}(2009)]%
        {deng_imagenet_2009}
\bibfield{author}{\bibinfo{person}{Jia Deng}, \bibinfo{person}{Wei Dong}, \bibinfo{person}{Richard Socher}, \bibinfo{person}{Li-Jia Li}, \bibinfo{person}{Kai Li}, {and} \bibinfo{person}{Li Fei-Fei}.} \bibinfo{year}{2009}\natexlab{}.
\newblock \showarticletitle{{ImageNet}: {A} large-scale hierarchical image database}. In \bibinfo{booktitle}{\emph{2009 {IEEE} {Conference} on {Computer} {Vision} and {Pattern} {Recognition}}}. \bibinfo{pages}{248--255}.
\newblock
\urldef\tempurl%
\url{https://doi.org/10.1109/CVPR.2009.5206848}
\showDOI{\tempurl}


\bibitem[Diamond and Banerjee(2024)]%
        {Diamond_Banerjee_2024}
\bibfield{author}{\bibinfo{person}{N’yoma Diamond} {and} \bibinfo{person}{Soumya Banerjee}.} \bibinfo{year}{2024}\natexlab{}.
\newblock \bibinfo{title}{On the Ethical Considerations of Generative Agents}.
\newblock
\newblock
\urldef\tempurl%
\url{https://doi.org/10.48550/arXiv.2411.19211}
\showDOI{\tempurl}


\bibitem[Dohmatob et~al\mbox{.}(2024)]%
        {dohmatob_tale_2024}
\bibfield{author}{\bibinfo{person}{Elvis Dohmatob}, \bibinfo{person}{Yunzhen Feng}, \bibinfo{person}{Pu Yang}, \bibinfo{person}{Francois Charton}, {and} \bibinfo{person}{Julia Kempe}.} \bibinfo{year}{2024}\natexlab{}.
\newblock \bibinfo{title}{A {Tale} of {Tails}: {Model} {Collapse} as a {Change} of {Scaling} {Laws}}.
\newblock
\newblock
\urldef\tempurl%
\url{https://doi.org/10.48550/arXiv.2402.07043}
\showDOI{\tempurl}
\newblock
\shownote{arXiv:2402.07043 [cs]}.


\bibitem[Domínguez~Hernández et~al\mbox{.}(2024)]%
        {dominguez_hernandez_mapping_2024}
\bibfield{author}{\bibinfo{person}{Andrés Domínguez~Hernández}, \bibinfo{person}{Shyam Krishna}, \bibinfo{person}{Antonella~Maia Perini}, \bibinfo{person}{Michael Katell}, \bibinfo{person}{SJ Bennett}, \bibinfo{person}{Ann Borda}, \bibinfo{person}{Youmna Hashem}, \bibinfo{person}{Semeli Hadjiloizou}, \bibinfo{person}{Sabeehah Mahomed}, \bibinfo{person}{Smera Jayadeva}, \bibinfo{person}{Mhairi Aitken}, {and} \bibinfo{person}{David Leslie}.} \bibinfo{year}{2024}\natexlab{}.
\newblock \showarticletitle{Mapping the individual, social and biospheric impacts of {Foundation} {Models}}. In \bibinfo{booktitle}{\emph{Proceedings of the 2024 {ACM} {Conference} on {Fairness}, {Accountability}, and {Transparency}}} \emph{(\bibinfo{series}{{FAccT} '24})}. \bibinfo{publisher}{Association for Computing Machinery}, \bibinfo{address}{New York, NY, USA}, \bibinfo{pages}{776--796}.
\newblock
\showISBNx{9798400704505}
\urldef\tempurl%
\url{https://doi.org/10.1145/3630106.3658939}
\showDOI{\tempurl}


\bibitem[Dwoskin(2019)]%
        {dwoskin_google_2019}
\bibfield{author}{\bibinfo{person}{Elizabeth Dwoskin}.} \bibinfo{year}{2019}\natexlab{}.
\newblock \showarticletitle{Google reaped millions in tax breaks as it secretly expanded its real estate footprint across the {U}.{S}.}
\newblock \bibinfo{journal}{\emph{Washington Post}} (\bibinfo{year}{2019}).
\newblock
\showISSN{0190-8286}
\urldef\tempurl%
\url{https://www.washingtonpost.com/business/economy/google-reaped-millions-of-tax-breaks-as-it-secretly-expanded-its-real-estate-footprint-across-the-us/2019/02/15/7912e10e-3136-11e9-813a-0ab2f17e305b_story.html}
\showURL{%
\tempurl}


\bibitem[{E. S. G. News}(2024)]%
        {e_s_g_news_meta_2024}
\bibfield{author}{\bibinfo{person}{{E. S. G. News}}.} \bibinfo{year}{2024}\natexlab{}.
\newblock \bibinfo{title}{Meta {Signs} {Long}-{Term} {Deal} with {BTG} {Pactual} for {Up} to 3.9 {Million} {Carbon} {Offset} {Credits} to {Support} {Net} {Zero} {Goals} by 2030}.
\newblock
\newblock
\urldef\tempurl%
\url{https://esgnews.com/meta-signs-long-term-deal-with-btg-pactual-for-up-to-3-9-million-carbon-offset-credits-to-support-net-zero-goals-by-2030/}
\showURL{%
\tempurl}


\bibitem[{Edward Beeching} et~al\mbox{.}(2024)]%
        {edward_beeching_scaling_2024}
\bibfield{author}{\bibinfo{person}{{Edward Beeching}}, \bibinfo{person}{{Lewis Tunstall}}, {and} \bibinfo{person}{{Sasha Rush}}.} \bibinfo{year}{2024}\natexlab{}.
\newblock \bibinfo{title}{Scaling test-time compute}.
\newblock
\newblock
\urldef\tempurl%
\url{https://huggingface.co/spaces/HuggingFaceH4/blogpost-scaling-test-time-compute}
\showURL{%
\tempurl}


\bibitem[{Eleni Kemene} et~al\mbox{.}(2024)]%
        {eleni_kemene_ai_2024}
\bibfield{author}{\bibinfo{person}{{Eleni Kemene}}, \bibinfo{person}{{Bart Valkhof}}, {and} \bibinfo{person}{{Thapelo Tladi}}.} \bibinfo{year}{2024}\natexlab{}.
\newblock \bibinfo{title}{{AI} and energy: {Will} {AI} reduce emissions or increase demand?}
\newblock
\newblock
\urldef\tempurl%
\url{https://www.weforum.org/agenda/2024/07/generative-ai-energy-emissions/}
\showURL{%
\tempurl}


\bibitem[{Epoch AI}(2024)]%
        {epoch_ai_data_2024}
\bibfield{author}{\bibinfo{person}{{Epoch AI}}.} \bibinfo{year}{2024}\natexlab{}.
\newblock \bibinfo{title}{Data on {Notable} {AI} {Models}}.
\newblock
\newblock
\urldef\tempurl%
\url{https://epoch.ai/data/notable-ai-models}
\showURL{%
\tempurl}


\bibitem[{ESG Today Writing Staff}(2024)]%
        {esg_today_writing_staff_microsoft_2024}
\bibfield{author}{\bibinfo{person}{{ESG Today Writing Staff}}.} \bibinfo{year}{2024}\natexlab{}.
\newblock \bibinfo{title}{Microsoft {Buys} 234,000 {Rainforest} {Restoration} {Carbon} {Removal} {Credits} from {Toroto}}.
\newblock
\newblock
\urldef\tempurl%
\url{https://www.esgtoday.com/microsoft-buys-234000-rainforest-restoration-carbon-removal-credits-from-toroto/}
\showURL{%
\tempurl}


\bibitem[Fanger(2024)]%
        {Fanger_2024}
\bibfield{author}{\bibinfo{person}{Ella Fanger}.} \bibinfo{year}{2024}\natexlab{}.
\newblock \showarticletitle{The Dirty Energy Fueling Amazon’s Data Gold Rush}.
\newblock \bibinfo{journal}{\emph{The Nation}} (\bibinfo{year}{2024}).
\newblock
\showISSN{0027-8378}
\urldef\tempurl%
\url{https://www.thenation.com/article/environment/data-centers-virginia-amazon-environment/}
\showURL{%
\tempurl}


\bibitem[Fisher and Streinz(2022)]%
        {fisher_confronting_2022}
\bibfield{author}{\bibinfo{person}{Angelina Fisher} {and} \bibinfo{person}{Thomas Streinz}.} \bibinfo{year}{2022}\natexlab{}.
\newblock \bibinfo{title}{Confronting {Data} {Inequality}}.
\newblock
\newblock
\urldef\tempurl%
\url{https://doi.org/10.2139/ssrn.3825724}
\showDOI{\tempurl}


\bibitem[Floridi(2019)]%
        {floridi_translating_2019}
\bibfield{author}{\bibinfo{person}{Luciano Floridi}.} \bibinfo{year}{2019}\natexlab{}.
\newblock \showarticletitle{Translating {Principles} into {Practices} of {Digital} {Ethics}: {Five} {Risks} of {Being} {Unethical}}.
\newblock \bibinfo{journal}{\emph{Philosophy \& Technology}} \bibinfo{volume}{32}, \bibinfo{number}{2} (\bibinfo{year}{2019}), \bibinfo{pages}{185--193}.
\newblock
\showISSN{2210-5441}
\urldef\tempurl%
\url{https://doi.org/10.1007/s13347-019-00354-x}
\showDOI{\tempurl}


\bibitem[{Frugal AI Challenge Organizing Committee}(2025)]%
        {frugal_ai_2025}
\bibfield{author}{\bibinfo{person}{{Frugal AI Challenge Organizing Committee}}.} \bibinfo{year}{2025}\natexlab{}.
\newblock \bibinfo{title}{Frugal {AI} {Challenge}}.
\newblock
\newblock
\urldef\tempurl%
\url{https://frugal-ai-challenge.github.io/}
\showURL{%
\tempurl}


\bibitem[{Future of Life Institute}(2023)]%
        {future_of_life_institute_pause_2023}
\bibfield{author}{\bibinfo{person}{{Future of Life Institute}}.} \bibinfo{year}{2023}\natexlab{}.
\newblock \bibinfo{title}{Pause {Giant} {AI} {Experiments}: {An} {Open} {Letter}}.
\newblock
\newblock
\urldef\tempurl%
\url{https://futureoflife.org/open-letter/pause-giant-ai-experiments/}
\showURL{%
\tempurl}


\bibitem[Gabbott(2024)]%
        {gabbott_why_2024}
\bibfield{author}{\bibinfo{person}{Miranda Gabbott}.} \bibinfo{year}{2024}\natexlab{}.
\newblock \bibinfo{title}{Why {We} {Don}’t {Know} {AI}'s {True} {Water} {Footprint}}.
\newblock
\newblock
\urldef\tempurl%
\url{https://techpolicy.press/why-we-dont-know-ais-true-water-footprint}
\showURL{%
\tempurl}


\bibitem[Gebru(2020)]%
        {Gebru_2020}
\bibfield{author}{\bibinfo{person}{Timnit Gebru}.} \bibinfo{year}{2020}\natexlab{}.
\newblock \bibinfo{booktitle}{\emph{Race and Gender}}.
\newblock \bibinfo{publisher}{Oxford University Press}, \bibinfo{pages}{0}.
\newblock
\showISBNx{978-0-19-006739-7}
\urldef\tempurl%
\url{https://doi.org/10.1093/oxfordhb/9780190067397.013.16}
\showURL{%
\tempurl}


\bibitem[Gebru et~al\mbox{.}(2023)]%
        {gebru_statement_2023}
\bibfield{author}{\bibinfo{person}{Timnit Gebru}, \bibinfo{person}{Emily~M. Bender}, {and} \bibinfo{person}{Margaret Mitchell}.} \bibinfo{year}{2023}\natexlab{}.
\newblock \bibinfo{title}{Statement from the listed authors of {Stochastic} {Parrots} on the “{AI} pause” letter}.
\newblock
\newblock
\urldef\tempurl%
\url{https://www.dair-institute.org/blog/letter-statement-March2023/}
\showURL{%
\tempurl}


\bibitem[Gebru and Torres(2024)]%
        {gebru_tescreal_2024}
\bibfield{author}{\bibinfo{person}{Timnit Gebru} {and} \bibinfo{person}{Émile~P. Torres}.} \bibinfo{year}{2024}\natexlab{}.
\newblock \showarticletitle{The {TESCREAL} bundle: {Eugenics} and the promise of utopia through artificial general intelligence}.
\newblock \bibinfo{journal}{\emph{First Monday}} (\bibinfo{year}{2024}).
\newblock
\showISSN{1396-0466}
\urldef\tempurl%
\url{https://doi.org/10.5210/fm.v29i4.13636}
\showDOI{\tempurl}


\bibitem[Gelles(2024)]%
        {gelles_is_2024}
\bibfield{author}{\bibinfo{person}{David Gelles}.} \bibinfo{year}{2024}\natexlab{}.
\newblock \showarticletitle{A.{I}.’s {Insatiable} {Appetite} for {Energy}}.
\newblock \bibinfo{journal}{\emph{The New York Times}} (\bibinfo{year}{2024}).
\newblock
\showISSN{0362-4331}
\urldef\tempurl%
\url{https://www.nytimes.com/2024/07/11/climate/artificial-intelligence-energy-usage.html}
\showURL{%
\tempurl}


\bibitem[George(2024)]%
        {george_google_2024}
\bibfield{author}{\bibinfo{person}{Violet George}.} \bibinfo{year}{2024}\natexlab{}.
\newblock \bibinfo{title}{Google {Stops} {Its} {Mass} {Purchase} {Of} {Carbon} {Credits}}.
\newblock
\newblock
\urldef\tempurl%
\url{https://carbonherald.com/google-stops-its-mass-purchase-of-carbon-credits/}
\showURL{%
\tempurl}


\bibitem[Glaser(2019)]%
        {glaser_how_2019}
\bibfield{author}{\bibinfo{person}{April Glaser}.} \bibinfo{year}{2019}\natexlab{}.
\newblock \showarticletitle{How to {Make} {Tech} {Companies} {Actually} {Fight} {Climate} {Change}}.
\newblock \bibinfo{journal}{\emph{Slate}} (\bibinfo{year}{2019}).
\newblock
\showISSN{1091-2339}
\urldef\tempurl%
\url{https://slate.com/technology/2019/09/amazon-climate-walkout-jeff-bezos-employees.html}
\showURL{%
\tempurl}


\bibitem[{Google}(2024)]%
        {google_2024_2024}
\bibfield{author}{\bibinfo{person}{{Google}}.} \bibinfo{year}{2024}\natexlab{}.
\newblock \bibinfo{booktitle}{\emph{2024 {Environmental} {Report}}}.
\newblock \bibinfo{type}{{T}echnical {R}eport}.
\newblock
\urldef\tempurl%
\url{https://www.gstatic.com/gumdrop/sustainability/google-2024-environmental-report.pdf#page=71.15}
\showURL{%
\tempurl}


\bibitem[Govindan(2024)]%
        {govindan_how_2024}
\bibfield{author}{\bibinfo{person}{Kannan Govindan}.} \bibinfo{year}{2024}\natexlab{}.
\newblock \showarticletitle{How {Artificial} {Intelligence} {Drives} {Sustainable} {Frugal} {Innovation}: {A} {Multitheoretical} {Perspective}}.
\newblock \bibinfo{journal}{\emph{IEEE Transactions on Engineering Management}}  \bibinfo{volume}{71} (\bibinfo{year}{2024}), \bibinfo{pages}{638--655}.
\newblock
\showISSN{1558-0040}
\urldef\tempurl%
\url{https://doi.org/10.1109/TEM.2021.3116187}
\showDOI{\tempurl}


\bibitem[Greenfield(2023)]%
        {greenfield_revealed_2023}
\bibfield{author}{\bibinfo{person}{Patrick Greenfield}.} \bibinfo{year}{2023}\natexlab{}.
\newblock \showarticletitle{Revealed: more than 90\% of rainforest carbon offsets by biggest certifier are worthless, analysis shows}.
\newblock \bibinfo{journal}{\emph{The Guardian}} (\bibinfo{year}{2023}).
\newblock
\showISSN{0261-3077}
\urldef\tempurl%
\url{https://www.theguardian.com/environment/2023/jan/18/revealed-forest-carbon-offsets-biggest-provider-worthless-verra-aoe}
\showURL{%
\tempurl}


\bibitem[Grene(2023)]%
        {grene_microsoft_2023}
\bibfield{author}{\bibinfo{person}{Hanna Grene}.} \bibinfo{year}{2023}\natexlab{}.
\newblock \bibinfo{title}{Microsoft highlights innovation in power and utilities}.
\newblock
\newblock
\urldef\tempurl%
\url{https://www.microsoft.com/en-us/industry/blog/energy-and-resources/2023/05/04/microsoft-highlights-innovation-in-power-and-utilities/}
\showURL{%
\tempurl}


\bibitem[Gupta et~al\mbox{.}(2024)]%
        {gupta_just_2024}
\bibfield{author}{\bibinfo{person}{Joyeeta Gupta}, \bibinfo{person}{Xuemei Bai}, \bibinfo{person}{Diana~M. Liverman}, \bibinfo{person}{Johan Rockström}, \bibinfo{person}{Dahe Qin}, \bibinfo{person}{Ben Stewart-Koster}, \bibinfo{person}{Juan~C. Rocha}, \bibinfo{person}{Lisa Jacobson}, \bibinfo{person}{Jesse~F. Abrams}, \bibinfo{person}{Lauren~S. Andersen}, \bibinfo{person}{David I.~Armstrong McKay}, \bibinfo{person}{Govindasamy Bala}, \bibinfo{person}{Stuart~E. Bunn}, \bibinfo{person}{Daniel Ciobanu}, \bibinfo{person}{Fabrice DeClerck}, \bibinfo{person}{Kristie~L. Ebi}, \bibinfo{person}{Lauren Gifford}, \bibinfo{person}{Christopher Gordon}, \bibinfo{person}{Syezlin Hasan}, \bibinfo{person}{Norichika Kanie}, \bibinfo{person}{Timothy~M. Lenton}, \bibinfo{person}{Sina Loriani}, \bibinfo{person}{Awaz Mohamed}, \bibinfo{person}{Nebojsa Nakicenovic}, \bibinfo{person}{David Obura}, \bibinfo{person}{Daniel Ospina}, \bibinfo{person}{Klaudia Prodani}, \bibinfo{person}{Crelis Rammelt}, \bibinfo{person}{Boris
  Sakschewski}, \bibinfo{person}{Joeri Scholtens}, \bibinfo{person}{Thejna Tharammal}, \bibinfo{person}{Detlef~van Vuuren}, \bibinfo{person}{Peter~H. Verburg}, \bibinfo{person}{Ricarda Winkelmann}, \bibinfo{person}{Caroline Zimm}, \bibinfo{person}{Elena Bennett}, \bibinfo{person}{Anders Bjørn}, \bibinfo{person}{Stefan Bringezu}, \bibinfo{person}{Wendy~J. Broadgate}, \bibinfo{person}{Harriet Bulkeley}, \bibinfo{person}{Beatrice Crona}, \bibinfo{person}{Pamela~A. Green}, \bibinfo{person}{Holger Hoff}, \bibinfo{person}{Lei Huang}, \bibinfo{person}{Margot Hurlbert}, \bibinfo{person}{Cristina Y.~A. Inoue}, \bibinfo{person}{Şiir Kılkış}, \bibinfo{person}{Steven~J. Lade}, \bibinfo{person}{Jianguo Liu}, \bibinfo{person}{Imran Nadeem}, \bibinfo{person}{Christopher Ndehedehe}, \bibinfo{person}{Chukwumerije Okereke}, \bibinfo{person}{Ilona~M. Otto}, \bibinfo{person}{Simona Pedde}, \bibinfo{person}{Laura Pereira}, \bibinfo{person}{Lena Schulte-Uebbing}, \bibinfo{person}{J.~David Tàbara}, \bibinfo{person}{Wim~de
  Vries}, \bibinfo{person}{Gail Whiteman}, \bibinfo{person}{Cunde Xiao}, \bibinfo{person}{Xinwu Xu}, \bibinfo{person}{Noelia Zafra-Calvo}, \bibinfo{person}{Xin Zhang}, \bibinfo{person}{Paola Fezzigna}, {and} \bibinfo{person}{Giuliana Gentile}.} \bibinfo{year}{2024}\natexlab{}.
\newblock \showarticletitle{A just world on a safe planet: a {Lancet} {Planetary} {Health}–{Earth} {Commission} report on {Earth}-system boundaries, translations, and transformations}.
\newblock \bibinfo{journal}{\emph{The Lancet Planetary Health}} \bibinfo{volume}{0}, \bibinfo{number}{0} (\bibinfo{date}{Sept.} \bibinfo{year}{2024}).
\newblock
\showISSN{2542-5196}
\urldef\tempurl%
\url{https://doi.org/10.1016/S2542-5196(24)00042-1}
\showDOI{\tempurl}
\newblock
\shownote{Publisher: Elsevier}.


\bibitem[Gupta et~al\mbox{.}(2023)]%
        {gupta_chatgpt_2023}
\bibfield{author}{\bibinfo{person}{Maanak Gupta}, \bibinfo{person}{Charankumar Akiri}, \bibinfo{person}{Kshitiz Aryal}, \bibinfo{person}{Eli Parker}, {and} \bibinfo{person}{Lopamudra Praharaj}.} \bibinfo{year}{2023}\natexlab{}.
\newblock \showarticletitle{From {ChatGPT} to {ThreatGPT}: {Impact} of {Generative} {AI} in {Cybersecurity} and {Privacy}}.
\newblock \bibinfo{journal}{\emph{IEEE Access}}  \bibinfo{volume}{11} (\bibinfo{year}{2023}), \bibinfo{pages}{80218--80245}.
\newblock
\showISSN{2169-3536}
\urldef\tempurl%
\url{https://doi.org/10.1109/ACCESS.2023.3300381}
\showDOI{\tempurl}


\bibitem[Guzzo et~al\mbox{.}(2024)]%
        {Guzzo_2024_rebound}
\bibfield{author}{\bibinfo{person}{D. Guzzo}, \bibinfo{person}{B. Walrave}, \bibinfo{person}{N. Videira}, \bibinfo{person}{I.~C. Oliveira}, {and} \bibinfo{person}{D.~C.~A. Pigosso}.} \bibinfo{year}{2024}\natexlab{}.
\newblock \showarticletitle{Towards a systemic view on rebound effects: Modelling the feedback loops of rebound mechanisms}.
\newblock \bibinfo{journal}{\emph{Ecological Economics}}  \bibinfo{volume}{217} (\bibinfo{year}{2024}), \bibinfo{pages}{108050}.
\newblock
\showISSN{0921-8009}
\urldef\tempurl%
\url{https://doi.org/10.1016/j.ecolecon.2023.108050}
\showDOI{\tempurl}


\bibitem[Hagerty and Rubinov(2019)]%
        {hagerty_global_2019}
\bibfield{author}{\bibinfo{person}{Alexa Hagerty} {and} \bibinfo{person}{Igor Rubinov}.} \bibinfo{year}{2019}\natexlab{}.
\newblock \bibinfo{title}{Global {AI} {Ethics}: {A} {Review} of the {Social} {Impacts} and {Ethical} {Implications} of {Artificial} {Intelligence}}.
\newblock
\newblock
\urldef\tempurl%
\url{https://doi.org/10.48550/arXiv.1907.07892}
\showDOI{\tempurl}


\bibitem[Halper(2024a)]%
        {halper_ais_2024}
\bibfield{author}{\bibinfo{person}{Evan Halper}.} \bibinfo{year}{2024}\natexlab{a}.
\newblock \showarticletitle{{AI}’s hunger for electric power is threatening {U}.{S}. climate goals}.
\newblock \bibinfo{journal}{\emph{Washington Post}} (\bibinfo{year}{2024}).
\newblock
\showISSN{0190-8286}
\urldef\tempurl%
\url{https://www.washingtonpost.com/climate-environment/2024/11/19/ai-cop29-climate-data-centers/}
\showURL{%
\tempurl}


\bibitem[Halper(2024b)]%
        {halper_energy-hungry_2024}
\bibfield{author}{\bibinfo{person}{Evan Halper}.} \bibinfo{year}{2024}\natexlab{b}.
\newblock \showarticletitle{Energy-hungry {AI} firms bet on these moonshot technologies}.
\newblock \bibinfo{journal}{\emph{Washington Post}} (\bibinfo{year}{2024}).
\newblock
\showISSN{0190-8286}
\urldef\tempurl%
\url{https://www.washingtonpost.com/business/2024/12/27/ai-data-centers-energy-nuclear-hydrogen/}
\showURL{%
\tempurl}


\bibitem[Han et~al\mbox{.}(2024)]%
        {han_unpaid_2024}
\bibfield{author}{\bibinfo{person}{Yuelin Han}, \bibinfo{person}{Zhifeng Wu}, \bibinfo{person}{Pengfei Li}, \bibinfo{person}{Adam Wierman}, {and} \bibinfo{person}{Shaolei Ren}.} \bibinfo{year}{2024}\natexlab{}.
\newblock \bibinfo{title}{The {Unpaid} {Toll}: {Quantifying} the {Public} {Health} {Impact} of {AI}}.
\newblock
\newblock
\urldef\tempurl%
\url{https://doi.org/10.48550/arXiv.2412.06288}
\showDOI{\tempurl}


\bibitem[Hao(2024)]%
        {Hao_2024}
\bibfield{author}{\bibinfo{person}{Karen Hao}.} \bibinfo{year}{2024}\natexlab{}.
\newblock \bibinfo{title}{Microsoft’s Hypocrisy on AI}.
\newblock
\newblock
\urldef\tempurl%
\url{https://www.theatlantic.com/technology/archive/2024/09/microsoft-ai-oil-contracts/679804/}
\showURL{%
\tempurl}


\bibitem[Hataya et~al\mbox{.}(2023)]%
        {hataya_will_2023}
\bibfield{author}{\bibinfo{person}{Ryuichiro Hataya}, \bibinfo{person}{Han Bao}, {and} \bibinfo{person}{Hiromi Arai}.} \bibinfo{year}{2023}\natexlab{}.
\newblock \showarticletitle{Will {Large}-scale {Generative} {Models} {Corrupt} {Future} {Datasets}?} \bibinfo{pages}{20555--20565}.
\newblock
\urldef\tempurl%
\url{https://openaccess.thecvf.com/content/ICCV2023/html/Hataya_Will_Large-scale_Generative_Models_Corrupt_Future_Datasets_ICCV_2023_paper.html}
\showURL{%
\tempurl}


\bibitem[Heilweil(2023)]%
        {heilweil_want_2023}
\bibfield{author}{\bibinfo{person}{Rebecca Heilweil}.} \bibinfo{year}{2023}\natexlab{}.
\newblock \showarticletitle{Want to {Win} a {Chip} {War}? {You}’re {Gonna} {Need} a {Lot} of {Water}}.
\newblock \bibinfo{journal}{\emph{Wired}} (\bibinfo{year}{2023}).
\newblock
\showISSN{1059-1028}
\urldef\tempurl%
\url{https://www.wired.com/story/want-to-win-a-chip-war-youre-gonna-need-a-lot-of-water/}
\showURL{%
\tempurl}


\bibitem[Herrman(2024)]%
        {herrman_ai_2024}
\bibfield{author}{\bibinfo{person}{John Herrman}.} \bibinfo{year}{2024}\natexlab{}.
\newblock \bibinfo{title}{{AI} {Investors} {Are} {Starting} to {Wonder}: {Is} {This} {Just} a {Bubble}?}
\newblock
\newblock
\urldef\tempurl%
\url{https://nymag.com/intelligencer/article/ai-investors-are-starting-to-wonder-is-this-just-a-bubble.html}
\showURL{%
\tempurl}


\bibitem[Hilton(2024)]%
        {hilton_taiwan_2024}
\bibfield{author}{\bibinfo{person}{Isabel Hilton}.} \bibinfo{year}{2024}\natexlab{}.
\newblock \showarticletitle{Taiwan {Makes} the {Majority} of the {World}’s {Computer} {Chips}. {Now} {It}’s {Running} {Out} of {Electricity}}.
\newblock \bibinfo{journal}{\emph{Wired}} (\bibinfo{year}{2024}).
\newblock
\showISSN{1059-1028}
\urldef\tempurl%
\url{https://www.wired.com/story/taiwan-makes-the-majority-of-the-worlds-computer-chips-now-its-running-out-of-electricity/}
\showURL{%
\tempurl}


\bibitem[{ILR School and the Aspen Institute}({[n.\,d.]})]%
        {ilr_school_and_the_aspen_institute_how_nodate}
\bibfield{author}{\bibinfo{person}{{ILR School and the Aspen Institute}}.} \bibinfo{year}{[n.\,d.]}\natexlab{}.
\newblock \bibinfo{title}{How many gig workers are there?}
\newblock
\newblock
\urldef\tempurl%
\url{https://www.gigeconomydata.org/basics/how-many-gig-workers-are-there}
\showURL{%
\tempurl}


\bibitem[{International Energy Agency}(2021)]%
        {international_energy_agency_mineral_2021}
\bibfield{author}{\bibinfo{person}{{International Energy Agency}}.} \bibinfo{year}{2021}\natexlab{}.
\newblock \bibinfo{title}{Mineral requirements for clean energy transitions}.
\newblock
\newblock
\urldef\tempurl%
\url{https://www.iea.org/reports/the-role-of-critical-minerals-in-clean-energy-transitions/mineral-requirements-for-clean-energy-transitions}
\showURL{%
\tempurl}


\bibitem[Isaac and Griffith(2024)]%
        {isaac_openai_2024}
\bibfield{author}{\bibinfo{person}{Mike Isaac} {and} \bibinfo{person}{Erin Griffith}.} \bibinfo{year}{2024}\natexlab{}.
\newblock \showarticletitle{{OpenAI} {Is} {Growing} {Fast} and {Burning} {Through} {Piles} of {Money}}.
\newblock \bibinfo{journal}{\emph{The New York Times}} (\bibinfo{year}{2024}).
\newblock
\showISSN{0362-4331}
\urldef\tempurl%
\url{https://www.nytimes.com/2024/09/27/technology/openai-chatgpt-investors-funding.html}
\showURL{%
\tempurl}


\bibitem[Jeans(2021)]%
        {jeans_data_2021}
\bibfield{author}{\bibinfo{person}{David Jeans}.} \bibinfo{year}{2021}\natexlab{}.
\newblock \bibinfo{title}{Data {In} {The} {Dark}: {How} {Big} {Tech} {Secretly} {Secured} \$800 {Million} {In} {Tax} {Breaks} {For} {Data} {Centers}}.
\newblock
\newblock
\urldef\tempurl%
\url{https://www.forbes.com/sites/davidjeans/2021/08/19/data-in-the-dark-how-big-tech-secretly-secured-800-million-in-tax-breaks-for-data-centers/}
\showURL{%
\tempurl}
\newblock
\shownote{Section: Enterprise Tech}.


\bibitem[{Jeff Clabaugh}(2024)]%
        {jeff_clabaugh_northern_2024}
\bibfield{author}{\bibinfo{person}{{Jeff Clabaugh}}.} \bibinfo{year}{2024}\natexlab{}.
\newblock \bibinfo{title}{Northern {Virginia} is again the {No}. 1 data center market, but challenges are mounting}.
\newblock
\newblock
\urldef\tempurl%
\url{https://wtop.com/business-finance/2024/03/northern-virginia-again-ranks-no-1-data-center-market-but-challenges-are-mounting/}
\showURL{%
\tempurl}


\bibitem[Jeng et~al\mbox{.}(2024)]%
        {Jeng_Chang_Duffy_Lam_Maercklein_2024}
\bibfield{author}{\bibinfo{person}{Linda Jeng}, \bibinfo{person}{Wayne Chang}, \bibinfo{person}{Kim Duffy}, \bibinfo{person}{Kristy Lam}, {and} \bibinfo{person}{Elissa Maercklein}.} \bibinfo{year}{2024}\natexlab{}.
\newblock \bibinfo{title}{Chains of Trust: Combatting Synthetic Data Risks of AI}.
\newblock
\newblock
\urldef\tempurl%
\url{https://doi.org/10.2139/ssrn.4854347}
\showDOI{\tempurl}


\bibitem[{Jeni Tennison} and {Tim Davies}(2024)]%
        {jeni_tennison_lets_2024}
\bibfield{author}{\bibinfo{person}{{Jeni Tennison}} {and} \bibinfo{person}{{Tim Davies}}.} \bibinfo{year}{2024}\natexlab{}.
\newblock \bibinfo{title}{Let’s give people power over {AI}}.
\newblock
\newblock
\urldef\tempurl%
\url{https://www.jrf.org.uk/ai-for-public-good/lets-give-people-power-over-ai}
\showURL{%
\tempurl}


\bibitem[{Jennifer Van Evra}(2021)]%
        {jennifer_van_evra_carbon_2021}
\bibfield{author}{\bibinfo{person}{{Jennifer Van Evra}}.} \bibinfo{year}{2021}\natexlab{}.
\newblock \showarticletitle{Carbon offsets might be a dangerous distraction from more effective climate action, experts say}.
\newblock \bibinfo{journal}{\emph{CBC}} (\bibinfo{year}{2021}).
\newblock
\urldef\tempurl%
\url{https://www.cbc.ca/radio/whatonearth/carbon-offsets-might-be-a-dangerous-distraction-from-more-effective-climate-action-experts-say-1.5946764}
\showURL{%
\tempurl}


\bibitem[Kaack et~al\mbox{.}(2022)]%
        {kaack_aligning_2022}
\bibfield{author}{\bibinfo{person}{Lynn~H. Kaack}, \bibinfo{person}{Priya~L. Donti}, \bibinfo{person}{Emma Strubell}, \bibinfo{person}{George Kamiya}, \bibinfo{person}{Felix Creutzig}, {and} \bibinfo{person}{David Rolnick}.} \bibinfo{year}{2022}\natexlab{}.
\newblock \showarticletitle{Aligning artificial intelligence with climate change mitigation}.
\newblock \bibinfo{journal}{\emph{Nature Climate Change}} \bibinfo{volume}{12}, \bibinfo{number}{6} (\bibinfo{year}{2022}), \bibinfo{pages}{518--527}.
\newblock
\showISSN{1758-678X, 1758-6798}
\urldef\tempurl%
\url{https://doi.org/10.1038/s41558-022-01377-7}
\showDOI{\tempurl}


\bibitem[Kaplan et~al\mbox{.}(2020)]%
        {kaplan_scaling_2020}
\bibfield{author}{\bibinfo{person}{Jared Kaplan}, \bibinfo{person}{Sam McCandlish}, \bibinfo{person}{Tom Henighan}, \bibinfo{person}{Tom~B. Brown}, \bibinfo{person}{Benjamin Chess}, \bibinfo{person}{Rewon Child}, \bibinfo{person}{Scott Gray}, \bibinfo{person}{Alec Radford}, \bibinfo{person}{Jeffrey Wu}, {and} \bibinfo{person}{Dario Amodei}.} \bibinfo{year}{2020}\natexlab{}.
\newblock \bibinfo{title}{Scaling {Laws} for {Neural} {Language} {Models}}.
\newblock
\newblock
\urldef\tempurl%
\url{http://arxiv.org/abs/2001.08361}
\showURL{%
\tempurl}
\newblock
\shownote{arXiv:2001.08361 [cs, stat]}.


\bibitem[{Keach Hagey}(2024)]%
        {keach_hagey_sam_2024}
\bibfield{author}{\bibinfo{person}{{Keach Hagey}}.} \bibinfo{year}{2024}\natexlab{}.
\newblock \bibinfo{title}{Sam {Altman} {Seeks} {Trillions} of {Dollars} to {Reshape} {Business} of {Chips} and {AI}}.
\newblock
\newblock
\urldef\tempurl%
\url{https://www.wsj.com/tech/ai/sam-altman-seeks-trillions-of-dollars-to-reshape-business-of-chips-and-ai-89ab3db0}
\showURL{%
\tempurl}


\bibitem[Kim(1992)]%
        {Kim_1992_archetype1}
\bibfield{author}{\bibinfo{person}{Daniel~H. Kim}.} \bibinfo{year}{1992}\natexlab{}.
\newblock \bibinfo{booktitle}{\emph{Systems Archetypes I: Diagnosing Systemic Issues and Designing High-Leverage Interventions}}.
\newblock \bibinfo{publisher}{Pegasus Communications}.
\newblock
\showISBNx{978-1-883823-00-9}


\bibitem[Kim(1994)]%
        {Kim_1994_archetype2}
\bibfield{author}{\bibinfo{person}{Daniel~H. Kim}.} \bibinfo{year}{1994}\natexlab{}.
\newblock \bibinfo{booktitle}{\emph{Systems Archetypes II: Using Systems Archetypes to Take Effective Action}}.
\newblock \bibinfo{publisher}{Pegasus Communications}.
\newblock
\showISBNx{978-1-883823-05-4}


\bibitem[Kim(2000)]%
        {Kim_2000_archetype3}
\bibfield{author}{\bibinfo{person}{Daniel~H. Kim}.} \bibinfo{year}{2000}\natexlab{}.
\newblock \bibinfo{booktitle}{\emph{Systems Archetypes III: Understanding Patterns of Behavior and Delay}}.
\newblock \bibinfo{publisher}{Pegasus Communications}.
\newblock
\showISBNx{978-1-883823-49-8}


\bibitem[Kimball(2024)]%
        {kimball_microsoft_2024}
\bibfield{author}{\bibinfo{person}{Spencer Kimball}.} \bibinfo{year}{2024}\natexlab{}.
\newblock \showarticletitle{Microsoft signs deal to invest more than \$10 billion on renewable energy capacity to power data centers}.
\newblock \bibinfo{journal}{\emph{CNBC}} (\bibinfo{year}{2024}).
\newblock
\urldef\tempurl%
\url{https://www.cnbc.com/2024/05/01/microsoft-brookfield-to-develop-more-than-10point5-gigawatts-of-renewable-energy.html}
\showURL{%
\tempurl}
\newblock
\shownote{Section: Energy}.


\bibitem[Kindig(2024)]%
        {kindig_ai_2024}
\bibfield{author}{\bibinfo{person}{Beth Kindig}.} \bibinfo{year}{2024}\natexlab{}.
\newblock \bibinfo{title}{{AI} {Power} {Consumption}: {Rapidly} {Becoming} {Mission}-{Critical}}.
\newblock
\newblock
\urldef\tempurl%
\url{https://www.forbes.com/sites/bethkindig/2024/06/20/ai-power-consumption-rapidly-becoming-mission-critical/}
\showURL{%
\tempurl}


\bibitem[Kirchenbauer et~al\mbox{.}(2024)]%
        {kirchenbauer_watermark_2024}
\bibfield{author}{\bibinfo{person}{John Kirchenbauer}, \bibinfo{person}{Jonas Geiping}, \bibinfo{person}{Yuxin Wen}, \bibinfo{person}{Jonathan Katz}, \bibinfo{person}{Ian Miers}, {and} \bibinfo{person}{Tom Goldstein}.} \bibinfo{year}{2024}\natexlab{}.
\newblock \bibinfo{title}{A {Watermark} for {Large} {Language} {Models}}.
\newblock
\newblock
\urldef\tempurl%
\url{http://arxiv.org/abs/2301.10226}
\showURL{%
\tempurl}
\newblock
\shownote{arXiv:2301.10226}.


\bibitem[{Kirsten James}(2024)]%
        {kirsten_james_semiconductor_2024}
\bibfield{author}{\bibinfo{person}{{Kirsten James}}.} \bibinfo{year}{2024}\natexlab{}.
\newblock \bibinfo{title}{Semiconductor manufacturing and big tech's water challenge}.
\newblock
\newblock
\urldef\tempurl%
\url{https://www.weforum.org/stories/2024/07/the-water-challenge-for-semiconductor-manufacturing-and-big-tech-what-needs-to-be-done/}
\showURL{%
\tempurl}


\bibitem[Klein and D'Ignazio(2024)]%
        {klein_data_2024}
\bibfield{author}{\bibinfo{person}{Lauren Klein} {and} \bibinfo{person}{Catherine D'Ignazio}.} \bibinfo{year}{2024}\natexlab{}.
\newblock \showarticletitle{Data {Feminism} for {AI}}. In \bibinfo{booktitle}{\emph{Proceedings of the 2024 {ACM} {Conference} on {Fairness}, {Accountability}, and {Transparency}}} \emph{(\bibinfo{series}{{FAccT} '24})}. \bibinfo{publisher}{Association for Computing Machinery}, \bibinfo{address}{New York, NY, USA}, \bibinfo{pages}{100--112}.
\newblock
\showISBNx{9798400704505}
\urldef\tempurl%
\url{https://doi.org/10.1145/3630106.3658543}
\showDOI{\tempurl}


\bibitem[Kleinberg and Raghavan(2021)]%
        {Kleinberg_Raghavan_2021}
\bibfield{author}{\bibinfo{person}{Jon Kleinberg} {and} \bibinfo{person}{Manish Raghavan}.} \bibinfo{year}{2021}\natexlab{}.
\newblock \showarticletitle{Algorithmic monoculture and social welfare}.
\newblock \bibinfo{journal}{\emph{Proceedings of the National Academy of Sciences}} \bibinfo{volume}{118}, \bibinfo{number}{22} (\bibinfo{year}{2021}), \bibinfo{pages}{e2018340118}.
\newblock
\urldef\tempurl%
\url{https://doi.org/10.1073/pnas.2018340118}
\showDOI{\tempurl}


\bibitem[Koebler~·(2024)]%
        {koebler_project_2024}
\bibfield{author}{\bibinfo{person}{Jason Koebler~·}.} \bibinfo{year}{2024}\natexlab{}.
\newblock \bibinfo{title}{Project {Analyzing} {Human} {Language} {Usage} {Shuts} {Down} {Because} ‘{Generative} {AI} {Has} {Polluted} the {Data}’}.
\newblock
\newblock
\urldef\tempurl%
\url{https://www.404media.co/project-analyzing-human-language-usage-shuts-down-because-generative-ai-has-polluted-the-data/}
\showURL{%
\tempurl}


\bibitem[Krizhevsky et~al\mbox{.}(2012)]%
        {krizhevsky_imagenet_2012}
\bibfield{author}{\bibinfo{person}{Alex Krizhevsky}, \bibinfo{person}{Ilya Sutskever}, {and} \bibinfo{person}{Geoffrey~E Hinton}.} \bibinfo{year}{2012}\natexlab{}.
\newblock \showarticletitle{{ImageNet} {Classification} with {Deep} {Convolutional} {Neural} {Networks}}. In \bibinfo{booktitle}{\emph{Advances in {Neural} {Information} {Processing} {Systems}}}, Vol.~\bibinfo{volume}{25}. \bibinfo{publisher}{Curran Associates, Inc.}
\newblock
\urldef\tempurl%
\url{https://proceedings.neurips.cc/paper/2012/hash/c399862d3b9d6b76c8436e924a68c45b-Abstract.html}
\showURL{%
\tempurl}


\bibitem[L(2024a)]%
        {l_google_2024}
\bibfield{author}{\bibinfo{person}{Jennifer L}.} \bibinfo{year}{2024}\natexlab{a}.
\newblock \bibinfo{title}{Google {Ditches} {Carbon} {Offsets}, {Here}'s {Its} {New} {Net} {Zero} {Focus}}.
\newblock
\newblock
\urldef\tempurl%
\url{https://carboncredits.com/google-ditches-carbon-offsets-heres-its-new-net-zero-focus/}
\showURL{%
\tempurl}


\bibitem[L(2024b)]%
        {l_microsoft_2024}
\bibfield{author}{\bibinfo{person}{Jennifer L}.} \bibinfo{year}{2024}\natexlab{b}.
\newblock \bibinfo{title}{Microsoft {Strikes} 2 {Record}-{Breaking} {Carbon} {Credit} {Deals}}.
\newblock
\newblock
\urldef\tempurl%
\url{https://carboncredits.com/microsoft-strikes-2-record-breaking-carbon-credit-deals/}
\showURL{%
\tempurl}


\bibitem[LaCroix and Prince(2023)]%
        {lacroix_deep_2023}
\bibfield{author}{\bibinfo{person}{Travis LaCroix} {and} \bibinfo{person}{Simon J.~D. Prince}.} \bibinfo{year}{2023}\natexlab{}.
\newblock \bibinfo{title}{Deep {Learning} and {Ethics}}.
\newblock
\newblock
\urldef\tempurl%
\url{https://doi.org/10.48550/arXiv.2305.15239}
\showDOI{\tempurl}
\newblock
\shownote{arXiv:2305.15239 [cs]}.


\bibitem[Laurenti et~al\mbox{.}(2016)]%
        {laurenti_unintended_2016}
\bibfield{author}{\bibinfo{person}{Rafael Laurenti}, \bibinfo{person}{Jagdeep Singh}, \bibinfo{person}{Rajib Sinha}, \bibinfo{person}{Josepha Potting}, {and} \bibinfo{person}{Björn Frostell}.} \bibinfo{year}{2016}\natexlab{}.
\newblock \showarticletitle{Unintended {Environmental} {Consequences} of {Improvement} {Actions}: {A} {Qualitative} {Analysis} of {Systems}' {Structure} and {Behavior}}.
\newblock \bibinfo{journal}{\emph{Systems Research and Behavioral Science}} \bibinfo{volume}{33}, \bibinfo{number}{3} (\bibinfo{year}{2016}), \bibinfo{pages}{381--399}.
\newblock
\showISSN{1092-7026}
\urldef\tempurl%
\url{https://doi.org/10.1002/sres.2330}
\showDOI{\tempurl}


\bibitem[Lawson(2024)]%
        {Lawson_2024}
\bibfield{author}{\bibinfo{person}{Alex Lawson}.} \bibinfo{year}{2024}\natexlab{}.
\newblock \showarticletitle{Google to buy nuclear power for AI datacentres in ‘world first’ deal}.
\newblock \bibinfo{journal}{\emph{The Guardian}} (\bibinfo{year}{2024}).
\newblock
\showISSN{0261-3077}
\urldef\tempurl%
\url{https://www.theguardian.com/technology/2024/oct/15/google-buy-nuclear-power-ai-datacentres-kairos-power}
\showURL{%
\tempurl}


\bibitem[Lee(2024)]%
        {lee_synthetic_2024}
\bibfield{author}{\bibinfo{person}{Peter Lee}.} \bibinfo{year}{2024}\natexlab{}.
\newblock \bibinfo{title}{Synthetic {Data} and the {Future} of {AI}}.
\newblock
\newblock
\urldef\tempurl%
\url{https://papers.ssrn.com/abstract=4722162}
\showURL{%
\tempurl}


\bibitem[{Leonardo Nicoletti} et~al\mbox{.}(2024)]%
        {leonardo_nicoletti_ai_2024}
\bibfield{author}{\bibinfo{person}{{Leonardo Nicoletti}}, \bibinfo{person}{{Naureen Malik}}, {and} \bibinfo{person}{{Andre Tartar}}.} \bibinfo{year}{2024}\natexlab{}.
\newblock \bibinfo{title}{{AI} {Power} {Needs} {Threaten} {Billions} in {Damages} for {US} {Households}}.
\newblock
\newblock
\urldef\tempurl%
\url{https://www.bloomberg.com/graphics/2024-ai-power-home-appliances/?utm_source=website&utm_medium=share&utm_campaign=copy}
\showURL{%
\tempurl}


\bibitem[Li et~al\mbox{.}(2023)]%
        {li_making_2023}
\bibfield{author}{\bibinfo{person}{Pengfei Li}, \bibinfo{person}{Jianyi Yang}, \bibinfo{person}{Mohammad~A. Islam}, {and} \bibinfo{person}{Shaolei Ren}.} \bibinfo{year}{2023}\natexlab{}.
\newblock \bibinfo{title}{Making {AI} {Less} "{Thirsty}": {Uncovering} and {Addressing} the {Secret} {Water} {Footprint} of {AI} {Models}}.
\newblock
\newblock
\urldef\tempurl%
\url{http://arxiv.org/abs/2304.03271}
\showURL{%
\tempurl}


\bibitem[Liang et~al\mbox{.}(2021)]%
        {Liang_Wu_Morency_Salakhutdinov_2021}
\bibfield{author}{\bibinfo{person}{Paul~Pu Liang}, \bibinfo{person}{Chiyu Wu}, \bibinfo{person}{Louis-Philippe Morency}, {and} \bibinfo{person}{Ruslan Salakhutdinov}.} \bibinfo{year}{2021}\natexlab{}.
\newblock \showarticletitle{Towards Understanding and Mitigating Social Biases in Language Models}. In \bibinfo{booktitle}{\emph{Proceedings of the 38th International Conference on Machine Learning}}. \bibinfo{publisher}{PMLR}, \bibinfo{pages}{6565–6576}.
\newblock
\urldef\tempurl%
\url{https://proceedings.mlr.press/v139/liang21a.html}
\showURL{%
\tempurl}


\bibitem[Livingstone(2023)]%
        {livingstone_its_2023}
\bibfield{author}{\bibinfo{person}{Grace Livingstone}.} \bibinfo{year}{2023}\natexlab{}.
\newblock \showarticletitle{‘{It}’s pillage’: thirsty {Uruguayans} decry {Google}’s plan to exploit water supply}.
\newblock \bibinfo{journal}{\emph{The Guardian}} (\bibinfo{year}{2023}).
\newblock
\showISSN{0261-3077}
\urldef\tempurl%
\url{https://www.theguardian.com/world/2023/jul/11/uruguay-drought-water-google-data-center}
\showURL{%
\tempurl}


\bibitem[Livingstone(2024)]%
        {livingstone_anger_2024}
\bibfield{author}{\bibinfo{person}{Grace Livingstone}.} \bibinfo{year}{2024}\natexlab{}.
\newblock \showarticletitle{Anger mounts over environmental cost of {Google} datacentre in {Uruguay}}.
\newblock \bibinfo{journal}{\emph{The Guardian}} (\bibinfo{year}{2024}).
\newblock
\showISSN{0261-3077}
\urldef\tempurl%
\url{https://www.theguardian.com/global-development/article/2024/aug/01/uruguay-anger-environmental-cost-google-datacentre-carbon-emissions-toxic-waste-water}
\showURL{%
\tempurl}


\bibitem[Lohn(2023)]%
        {lohn_scaling_2023}
\bibfield{author}{\bibinfo{person}{Andrew Lohn}.} \bibinfo{year}{2023}\natexlab{}.
\newblock \bibinfo{title}{Scaling {AI}}.
\newblock
\newblock
\urldef\tempurl%
\url{https://cset.georgetown.edu/publication/scaling-ai/}
\showURL{%
\tempurl}


\bibitem[Long et~al\mbox{.}(2024)]%
        {Long_Wang_Xiao_Zhao_Ding_Chen_Wang_2024}
\bibfield{author}{\bibinfo{person}{Lin Long}, \bibinfo{person}{Rui Wang}, \bibinfo{person}{Ruixuan Xiao}, \bibinfo{person}{Junbo Zhao}, \bibinfo{person}{Xiao Ding}, \bibinfo{person}{Gang Chen}, {and} \bibinfo{person}{Haobo Wang}.} \bibinfo{year}{2024}\natexlab{}.
\newblock \bibinfo{title}{On LLMs-Driven Synthetic Data Generation, Curation, and Evaluation: A Survey}.
\newblock
\newblock
\urldef\tempurl%
\url{http://arxiv.org/abs/2406.15126}
\showURL{%
\tempurl}


\bibitem[Lorenz et~al\mbox{.}(2024)]%
        {Lorenz__2024}
\bibfield{author}{\bibinfo{person}{Ullrich Lorenz}, \bibinfo{person}{Javad Keypour}, \bibinfo{person}{Nike Sudikatis}, {and} \bibinfo{person}{Jens Konrad}.} \bibinfo{year}{2024}\natexlab{}.
\newblock \showarticletitle{Systemic Properties of Key Production and Consumption Areas - Case studies from the core systems: Food, Energy, Mobility and Housing}.
\newblock \bibinfo{journal}{\emph{ETC ST Report 2024/2, EEA, Copenhagen}} (\bibinfo{year}{2024}).
\newblock
\urldef\tempurl%
\url{https://doi.org/10.5281/ZENODO.12188015}
\showDOI{\tempurl}


\bibitem[Luccioni and Hernandez-Garcia(2023)]%
        {luccioni_counting_2023}
\bibfield{author}{\bibinfo{person}{Alexandra~Sasha Luccioni} {and} \bibinfo{person}{Alex Hernandez-Garcia}.} \bibinfo{year}{2023}\natexlab{}.
\newblock \bibinfo{title}{Counting {Carbon}: {A} {Survey} of {Factors} {Influencing} the {Emissions} of {Machine} {Learning}}.
\newblock
\newblock
\urldef\tempurl%
\url{https://doi.org/10.48550/arXiv.2302.08476}
\showDOI{\tempurl}


\bibitem[Luccioni et~al\mbox{.}(2024a)]%
        {luccioni_power_2024}
\bibfield{author}{\bibinfo{person}{Sasha Luccioni}, \bibinfo{person}{Yacine Jernite}, {and} \bibinfo{person}{Emma Strubell}.} \bibinfo{year}{2024}\natexlab{a}.
\newblock \showarticletitle{Power {Hungry} {Processing}: {Watts} {Driving} the {Cost} of {AI} {Deployment}?}. In \bibinfo{booktitle}{\emph{Proceedings of the 2024 {ACM} {Conference} on {Fairness}, {Accountability}, and {Transparency}}} \emph{(\bibinfo{series}{{FAccT} '24})}. \bibinfo{publisher}{Association for Computing Machinery}, \bibinfo{address}{New York, NY, USA}, \bibinfo{pages}{85--99}.
\newblock
\showISBNx{9798400704505}
\urldef\tempurl%
\url{https://doi.org/10.1145/3630106.3658542}
\showDOI{\tempurl}


\bibitem[Luccioni et~al\mbox{.}(2024b)]%
        {luccioni_environmental_2024}
\bibfield{author}{\bibinfo{person}{Sasha Luccioni}, \bibinfo{person}{Bruna Trevelin}, {and} \bibinfo{person}{Margaret Mitchell}.} \bibinfo{year}{2024}\natexlab{b}.
\newblock \bibinfo{title}{The {Environmental} {Impacts} of {AI} - {Primer}}.
\newblock
\newblock
\urldef\tempurl%
\url{https://huggingface.co/blog/sasha/ai-environment-primer}
\showURL{%
\tempurl}


\bibitem[Malik(2023)]%
        {malik_openais_2023}
\bibfield{author}{\bibinfo{person}{Aisha Malik}.} \bibinfo{year}{2023}\natexlab{}.
\newblock \bibinfo{title}{{OpenAI}'s {ChatGPT} now has 100 million weekly active users}.
\newblock
\newblock
\urldef\tempurl%
\url{https://techcrunch.com/2023/11/06/openais-chatgpt-now-has-100-million-weekly-active-users/}
\showURL{%
\tempurl}


\bibitem[Masanet and Lei(2020)]%
        {Eric_Masanet_Nuoa_Lei_2020}
\bibfield{author}{\bibinfo{person}{Eric Masanet} {and} \bibinfo{person}{Nuoa Lei}.} \bibinfo{year}{2020}\natexlab{}.
\newblock \bibinfo{title}{How Much Energy Do Data Centers Really Use?}
\newblock
\newblock
\urldef\tempurl%
\url{https://energyinnovation.org/2020/03/17/how-much-energy-do-data-centers-really-use/}
\showURL{%
\tempurl}


\bibitem[McQuillan(2022)]%
        {mcquillan_resisting_2022}
\bibfield{author}{\bibinfo{person}{Dan McQuillan}.} \bibinfo{year}{2022}\natexlab{}.
\newblock \bibinfo{booktitle}{\emph{Resisting {AI}: an anti-fascist approach to artificial intelligence}}.
\newblock \bibinfo{publisher}{University Press}, \bibinfo{address}{Bristol}.
\newblock
\showISBNx{978-1-5292-1350-8}


\bibitem[Meadows et~al\mbox{.}(2004)]%
        {meadows_limits_2004}
\bibfield{author}{\bibinfo{person}{Donella Meadows}, \bibinfo{person}{Jorgen Randers}, {and} \bibinfo{person}{Dennis Meadows}.} \bibinfo{year}{2004}\natexlab{}.
\newblock \bibinfo{booktitle}{\emph{Limits to {Growth}: {The} 30-{Year} {Update}}}.
\newblock \bibinfo{publisher}{Chelsea Green Publishing}.
\newblock
\showISBNx{978-1-60358-155-4}


\bibitem[Meadows(2011)]%
        {meadows_thinking_2011}
\bibfield{author}{\bibinfo{person}{Donella~H. Meadows}.} \bibinfo{year}{2011}\natexlab{}.
\newblock \bibinfo{booktitle}{\emph{Thinking in systems: a primer}}.
\newblock \bibinfo{publisher}{Chelsea Green Pub}.
\newblock
\showISBNx{978-1-60358-055-7}


\bibitem[Meadows et~al\mbox{.}(1972)]%
        {meadows_limits_1972}
\bibfield{author}{\bibinfo{person}{Donella~H Meadows}, \bibinfo{person}{Dennis~L Meadows}, {and} \bibinfo{person}{Jørgen Randers}.} \bibinfo{year}{1972}\natexlab{}.
\newblock \showarticletitle{The limits to growth: {A} report for the {Club} of {Rome}’s project on the predicament of mankind}.
\newblock \bibinfo{journal}{\emph{New York: Universe Books}} (\bibinfo{year}{1972}).
\newblock


\bibitem[Mejias and Couldry(2024)]%
        {mejias_data_2024}
\bibfield{author}{\bibinfo{person}{Ulises~A. Mejias} {and} \bibinfo{person}{Nick Couldry}.} \bibinfo{year}{2024}\natexlab{}.
\newblock \bibinfo{booktitle}{\emph{Data {Grab}: {The} {New} {Colonialism} of {Big} {Tech} and {How} to {Fight} {Back}}}.
\newblock \bibinfo{publisher}{University of Chicago Press}.
\newblock
\showISBNx{978-0-226-83231-9}
\urldef\tempurl%
\url{https://www.degruyter.com/document/doi/10.7208/chicago/9780226832319/html}
\showURL{%
\tempurl}


\bibitem[{Melissa Heikkilä} and {Stephanie Arnett}(2024)]%
        {melissa_heikkila_this_2024}
\bibfield{author}{\bibinfo{person}{{Melissa Heikkilä}} {and} \bibinfo{person}{{Stephanie Arnett}}.} \bibinfo{year}{2024}\natexlab{}.
\newblock \bibinfo{title}{This is where the data to build {AI} comes from}.
\newblock
\newblock
\urldef\tempurl%
\url{https://www.technologyreview.com/2024/12/18/1108796/this-is-where-the-data-to-build-ai-comes-from/}
\showURL{%
\tempurl}


\bibitem[{Melodie Michel}(2024)]%
        {melodie_michel_meta_2024}
\bibfield{author}{\bibinfo{person}{{Melodie Michel}}.} \bibinfo{year}{2024}\natexlab{}.
\newblock \bibinfo{title}{Meta to buy millions of carbon credits from {LatAm} forests}.
\newblock
\newblock
\urldef\tempurl%
\url{https://www.csofutures.com/news/meta-to-purchase-millions-of-carbon-credits-from-latin-american-forests/}
\showURL{%
\tempurl}


\bibitem[{Meta Sustainability}(2024)]%
        {meta_sustainability_our_2024}
\bibfield{author}{\bibinfo{person}{{Meta Sustainability}}.} \bibinfo{year}{2024}\natexlab{}.
\newblock \bibinfo{title}{Our approach to clean and renewable energy}.
\newblock
\newblock
\urldef\tempurl%
\url{https://sustainability.atmeta.com/blog/2024/10/14/our-approach-to-clean-and-renewable-energy/}
\showURL{%
\tempurl}


\bibitem[Metz et~al\mbox{.}(2024)]%
        {metz_how_2024}
\bibfield{author}{\bibinfo{person}{Cade Metz}, \bibinfo{person}{Cecilia Kang}, \bibinfo{person}{Sheera Frenkel}, \bibinfo{person}{Stuart~A. Thompson}, {and} \bibinfo{person}{Nico Grant}.} \bibinfo{year}{2024}\natexlab{}.
\newblock \showarticletitle{How {Tech} {Giants} {Cut} {Corners} to {Harvest} {Data} for {A}.{I}.}
\newblock \bibinfo{journal}{\emph{The New York Times}} (\bibinfo{year}{2024}).
\newblock
\showISSN{0362-4331}
\urldef\tempurl%
\url{https://www.nytimes.com/2024/04/06/technology/tech-giants-harvest-data-artificial-intelligence.html}
\showURL{%
\tempurl}


\bibitem[Mickle(2024)]%
        {Mickle_2024}
\bibfield{author}{\bibinfo{person}{Tripp Mickle}.} \bibinfo{year}{2024}\natexlab{}.
\newblock \showarticletitle{Nvidia Doubles Profit as A.I. Chip Sales Soar}.
\newblock \bibinfo{journal}{\emph{The New York Times}} (\bibinfo{year}{2024}).
\newblock
\showISSN{0362-4331}
\urldef\tempurl%
\url{https://www.nytimes.com/2024/11/20/technology/nvidia-earnings-chips-ai.html}
\showURL{%
\tempurl}


\bibitem[Milmo and editor(2024)]%
        {milmo_googles_2024}
\bibfield{author}{\bibinfo{person}{Dan Milmo} {and} \bibinfo{person}{Dan Milmo Global~technology editor}.} \bibinfo{year}{2024}\natexlab{}.
\newblock \showarticletitle{Google’s emissions climb nearly 50\% in five years due to {AI} energy demand}.
\newblock \bibinfo{journal}{\emph{The Guardian}} (\bibinfo{year}{2024}).
\newblock
\showISSN{0261-3077}
\urldef\tempurl%
\url{https://www.theguardian.com/technology/article/2024/jul/02/google-ai-emissions}
\showURL{%
\tempurl}


\bibitem[Mollick(2024)]%
        {mollick_scaling_2024}
\bibfield{author}{\bibinfo{person}{Ethan Mollick}.} \bibinfo{year}{2024}\natexlab{}.
\newblock \bibinfo{title}{Scaling: {The} {State} of {Play} in {AI}}.
\newblock
\newblock
\urldef\tempurl%
\url{https://www.oneusefulthing.org/p/scaling-the-state-of-play-in-ai}
\showURL{%
\tempurl}


\bibitem[Monteith et~al\mbox{.}(2024)]%
        {monteith_artificial_2024}
\bibfield{author}{\bibinfo{person}{Scott Monteith}, \bibinfo{person}{Tasha Glenn}, \bibinfo{person}{John~R. Geddes}, \bibinfo{person}{Peter~C. Whybrow}, \bibinfo{person}{Eric Achtyes}, {and} \bibinfo{person}{Michael Bauer}.} \bibinfo{year}{2024}\natexlab{}.
\newblock \showarticletitle{Artificial intelligence and increasing misinformation}.
\newblock \bibinfo{journal}{\emph{The British Journal of Psychiatry}} \bibinfo{volume}{224}, \bibinfo{number}{2} (\bibinfo{year}{2024}), \bibinfo{pages}{33--35}.
\newblock
\showISSN{0007-1250, 1472-1465}
\urldef\tempurl%
\url{https://doi.org/10.1192/bjp.2023.136}
\showDOI{\tempurl}


\bibitem[Nardi et~al\mbox{.}(2018)]%
        {nardi_computing_2018}
\bibfield{author}{\bibinfo{person}{Bonnie Nardi}, \bibinfo{person}{Bill Tomlinson}, \bibinfo{person}{Donald~J. Patterson}, \bibinfo{person}{Jay Chen}, \bibinfo{person}{Daniel Pargman}, \bibinfo{person}{Barath Raghavan}, {and} \bibinfo{person}{Birgit Penzenstadler}.} \bibinfo{year}{2018}\natexlab{}.
\newblock \showarticletitle{Computing within limits}.
\newblock \bibinfo{journal}{\emph{Commun. ACM}} \bibinfo{volume}{61}, \bibinfo{number}{10} (\bibinfo{year}{2018}), \bibinfo{pages}{86--93}.
\newblock
\showISSN{0001-0782}
\urldef\tempurl%
\url{https://doi.org/10.1145/3183582}
\showDOI{\tempurl}


\bibitem[{Natasha White}(2023)]%
        {natasha_white_bogus_2023}
\bibfield{author}{\bibinfo{person}{{Natasha White}}.} \bibinfo{year}{2023}\natexlab{}.
\newblock \showarticletitle{Bogus {Carbon} {Credits} a '{Pervasive}' {Problem}, {Scientists} {Warn}}.
\newblock \bibinfo{journal}{\emph{TIME}} (\bibinfo{year}{2023}).
\newblock
\urldef\tempurl%
\url{https://time.com/6264772/study-most-carbon-credits-are-bogus/}
\showURL{%
\tempurl}


\bibitem[{Nestor Maslej} et~al\mbox{.}(2024)]%
        {nestor_maslej_ai_2024}
\bibfield{author}{\bibinfo{person}{{Nestor Maslej}}, \bibinfo{person}{{Loredana Fattorini,}}, \bibinfo{person}{{Raymond Perrault}}, \bibinfo{person}{{Vanessa Parli}}, \bibinfo{person}{{Anka Reuel}}, \bibinfo{person}{{Erik Brynjolfsson}}, \bibinfo{person}{{John Etchemendy}}, \bibinfo{person}{{Katrina Ligett}}, \bibinfo{person}{{Terah Lyons}}, \bibinfo{person}{{James Manyika}}, \bibinfo{person}{{Juan Carlos Niebles}}, \bibinfo{person}{{Yoav Shoham}}, \bibinfo{person}{{Russell Wald}}, {and} \bibinfo{person}{{Jack Clark}}.} \bibinfo{year}{2024}\natexlab{}.
\newblock \bibinfo{booktitle}{\emph{The {AI} {Index} 2024 {Annual} {Report}}}.
\newblock \bibinfo{type}{{T}echnical {R}eport}. \bibinfo{institution}{nstitute for Human-Centered AI, Stanford University}, \bibinfo{address}{Stanford, CA}.
\newblock
\urldef\tempurl%
\url{https://aiindex.stanford.edu/report/}
\showURL{%
\tempurl}


\bibitem[Nguyen et~al\mbox{.}(2024)]%
        {nguyen_solar_2024}
\bibfield{author}{\bibinfo{person}{Luan~Thanh Nguyen}, \bibinfo{person}{Shyama Ratnasiri}, \bibinfo{person}{Liam Wagner}, \bibinfo{person}{Dan~The Nguyen}, {and} \bibinfo{person}{Nicholas Rohde}.} \bibinfo{year}{2024}\natexlab{}.
\newblock \showarticletitle{Solar rebound effects: {Short} and long term dynamics}.
\newblock \bibinfo{journal}{\emph{Renewable Energy}}  \bibinfo{volume}{223} (\bibinfo{year}{2024}), \bibinfo{pages}{120051}.
\newblock
\showISSN{0960-1481}
\urldef\tempurl%
\url{https://doi.org/10.1016/j.renene.2024.120051}
\showDOI{\tempurl}


\bibitem[O'Donovan(2024)]%
        {odonovan_fighting_2024}
\bibfield{author}{\bibinfo{person}{Caroline O'Donovan}.} \bibinfo{year}{2024}\natexlab{}.
\newblock \showarticletitle{Fighting back against data centers, one small town at a time}.
\newblock \bibinfo{journal}{\emph{Washington Post}} (\bibinfo{year}{2024}).
\newblock
\showISSN{0190-8286}
\urldef\tempurl%
\url{https://www.washingtonpost.com/technology/2024/10/05/data-center-protest-community-resistance/}
\showURL{%
\tempurl}


\bibitem[Olivo(2024)]%
        {Olivo_2024}
\bibfield{author}{\bibinfo{person}{Antonio Olivo}.} \bibinfo{year}{2024}\natexlab{}.
\newblock \showarticletitle{Internet data centers are fueling drive to old power source: Coal}.
\newblock \bibinfo{journal}{\emph{Washington Post}} (\bibinfo{year}{2024}).
\newblock
\showISSN{0190-8286}
\urldef\tempurl%
\url{https://www.washingtonpost.com/business/interactive/2024/data-centers-internet-power-source-coal/}
\showURL{%
\tempurl}


\bibitem[O'Neil(2017)]%
        {oneil_weapons_2017}
\bibfield{author}{\bibinfo{person}{Cathy O'Neil}.} \bibinfo{year}{2017}\natexlab{}.
\newblock \bibinfo{booktitle}{\emph{Weapons of {Math} {Destruction}: {How} {Big} {Data} {Increases} {Inequality} and {Threatens} {Democracy}}}.
\newblock \bibinfo{publisher}{Crown}.
\newblock
\showISBNx{978-0-553-41883-5}


\bibitem[page(2024)]%
        {Temple_2024}
\bibfield{author}{\bibinfo{person}{James~Templearchive page}.} \bibinfo{year}{2024}\natexlab{}.
\newblock \bibinfo{title}{Google, Amazon and the problem with Big Tech’s climate claims}.
\newblock
\newblock
\urldef\tempurl%
\url{https://www.technologyreview.com/2024/07/17/1095019/google-amazon-and-the-problem-with-big-techs-climate-claims/}
\showURL{%
\tempurl}


\bibitem[Perrigo(2023)]%
        {perrigo_exclusive_2023}
\bibfield{author}{\bibinfo{person}{Billy Perrigo}.} \bibinfo{year}{2023}\natexlab{}.
\newblock \bibinfo{title}{Exclusive: {The} \$2 {Per} {Hour} {Workers} {Who} {Made} {ChatGPT} {Safer}}.
\newblock
\newblock
\urldef\tempurl%
\url{https://time.com/6247678/openai-chatgpt-kenya-workers/}
\showURL{%
\tempurl}


\bibitem[{Pranshu Verma} and {Shelly Tan}(2024)]%
        {pranshu_verma_bottle_2024}
\bibfield{author}{\bibinfo{person}{{Pranshu Verma}} {and} \bibinfo{person}{{Shelly Tan}}.} \bibinfo{year}{2024}\natexlab{}.
\newblock \showarticletitle{A bottle of water per email: the hidden environmental costs of using {AI} chatbots}.
\newblock \bibinfo{journal}{\emph{Washington Post}} (\bibinfo{year}{2024}).
\newblock
\urldef\tempurl%
\url{https://www.washingtonpost.com/technology/2024/09/18/energy-ai-use-electricity-water-data-centers/}
\showURL{%
\tempurl}


\bibitem[Probst et~al\mbox{.}(2024)]%
        {probst_systematic_2024}
\bibfield{author}{\bibinfo{person}{Benedict~S. Probst}, \bibinfo{person}{Malte Toetzke}, \bibinfo{person}{Andreas Kontoleon}, \bibinfo{person}{Laura Díaz~Anadón}, \bibinfo{person}{Jan~C. Minx}, \bibinfo{person}{Barbara~K. Haya}, \bibinfo{person}{Lambert Schneider}, \bibinfo{person}{Philipp~A. Trotter}, \bibinfo{person}{Thales A.~P. West}, \bibinfo{person}{Annelise Gill-Wiehl}, {and} \bibinfo{person}{Volker~H. Hoffmann}.} \bibinfo{year}{2024}\natexlab{}.
\newblock \showarticletitle{Systematic assessment of the achieved emission reductions of carbon crediting projects}.
\newblock \bibinfo{journal}{\emph{Nature Communications}} \bibinfo{volume}{15}, \bibinfo{number}{1} (\bibinfo{year}{2024}).
\newblock
\showISSN{2041-1723}
\urldef\tempurl%
\url{https://doi.org/10.1038/s41467-024-53645-z}
\showDOI{\tempurl}


\bibitem[Rakova and Dobbe(2023)]%
        {rakova_algorithms_2023}
\bibfield{author}{\bibinfo{person}{Bogdana Rakova} {and} \bibinfo{person}{Roel Dobbe}.} \bibinfo{year}{2023}\natexlab{}.
\newblock \showarticletitle{Algorithms as {Social}-{Ecological}-{Technological} {Systems}: an {Environmental} {Justice} {Lens} on {Algorithmic} {Audits}}. In \bibinfo{booktitle}{\emph{Proceedings of the 2023 {ACM} {Conference} on {Fairness}, {Accountability}, and {Transparency}}} \emph{(\bibinfo{series}{{FAccT} '23})}. \bibinfo{publisher}{Association for Computing Machinery}, \bibinfo{address}{New York, NY, USA}, \bibinfo{pages}{491}.
\newblock
\showISBNx{9798400701924}
\urldef\tempurl%
\url{https://doi.org/10.1145/3593013.3594014}
\showDOI{\tempurl}


\bibitem[Rather et~al\mbox{.}(2024)]%
        {rather_breaking_2024}
\bibfield{author}{\bibinfo{person}{Ishfaq~Hussain Rather}, \bibinfo{person}{Sushil Kumar}, {and} \bibinfo{person}{Amir~H. Gandomi}.} \bibinfo{year}{2024}\natexlab{}.
\newblock \showarticletitle{Breaking the data barrier: a review of deep learning techniques for democratizing {AI} with small datasets}.
\newblock \bibinfo{journal}{\emph{Artificial Intelligence Review}} \bibinfo{volume}{57}, \bibinfo{number}{9} (\bibinfo{year}{2024}), \bibinfo{pages}{226}.
\newblock
\showISSN{1573-7462}
\urldef\tempurl%
\url{https://doi.org/10.1007/s10462-024-10859-3}
\showDOI{\tempurl}


\bibitem[Rathi and Bass(2024)]%
        {Akshat_Rathi_Dina_Bass_2024}
\bibfield{author}{\bibinfo{person}{Akshat Rathi} {and} \bibinfo{person}{Dina Bass}.} \bibinfo{year}{2024}\natexlab{}.
\newblock \bibinfo{title}{Microsoft’s AI Investment Imperils Climate Goal As Emissions Jump 30\% - Bloomberg}.
\newblock
\newblock
\urldef\tempurl%
\url{https://www.bloomberg.com/news/articles/2024-05-15/microsoft-s-ai-investment-imperils-climate-goal-as-emissions-jump-30?embedded-checkout=true}
\showURL{%
\tempurl}


\bibitem[Rathi et~al\mbox{.}(2022)]%
        {rathi_big_2022}
\bibfield{author}{\bibinfo{person}{Akshat Rathi}, \bibinfo{person}{Natasha White}, {and} \bibinfo{person}{Demetrios~Pogkas Green}.} \bibinfo{year}{2022}\natexlab{}.
\newblock \showarticletitle{Big {Companies} {Claim} ‘{Carbon} {Neutrality}’ {Using} {Junk} {Carbon} {Offsets}}.
\newblock \bibinfo{journal}{\emph{Bloomberg.com}} (\bibinfo{year}{2022}).
\newblock
\urldef\tempurl%
\url{https://www.bloomberg.com/graphics/2022-carbon-offsets-renewable-energy/}
\showURL{%
\tempurl}


\bibitem[Raworth(2017)]%
        {raworth_doughnut_2017}
\bibfield{author}{\bibinfo{person}{Kate Raworth}.} \bibinfo{year}{2017}\natexlab{}.
\newblock \bibinfo{booktitle}{\emph{Doughnut economics: seven ways to think like a 21st century economist}}.
\newblock \bibinfo{publisher}{Chelsea Green Publishing}, \bibinfo{address}{White River Junction, Vermont}.
\newblock
\showISBNx{978-1-60358-674-0}


\bibitem[{Reed Cassady} and {Sam Peters}(2024)]%
        {reed_cassady_challenge_2024}
\bibfield{author}{\bibinfo{person}{{Reed Cassady}} {and} \bibinfo{person}{{Sam Peters}}.} \bibinfo{year}{2024}\natexlab{}.
\newblock \bibinfo{title}{The challenge of powering {AI}}.
\newblock
\newblock
\urldef\tempurl%
\url{https://www.franklintempleton.ca/en-ca/articles/2024/clearbridge-investments/the-challenge-of-powering-ai}
\showURL{%
\tempurl}


\bibitem[Richardson et~al\mbox{.}(2023)]%
        {richardson_earth_2023}
\bibfield{author}{\bibinfo{person}{Katherine Richardson}, \bibinfo{person}{Will Steffen}, \bibinfo{person}{Wolfgang Lucht}, \bibinfo{person}{Jørgen Bendtsen}, \bibinfo{person}{Sarah~E. Cornell}, \bibinfo{person}{Jonathan~F. Donges}, \bibinfo{person}{Markus Drüke}, \bibinfo{person}{Ingo Fetzer}, \bibinfo{person}{Govindasamy Bala}, \bibinfo{person}{Werner von Bloh}, \bibinfo{person}{Georg Feulner}, \bibinfo{person}{Stephanie Fiedler}, \bibinfo{person}{Dieter Gerten}, \bibinfo{person}{Tom Gleeson}, \bibinfo{person}{Matthias Hofmann}, \bibinfo{person}{Willem Huiskamp}, \bibinfo{person}{Matti Kummu}, \bibinfo{person}{Chinchu Mohan}, \bibinfo{person}{David Nogués-Bravo}, \bibinfo{person}{Stefan Petri}, \bibinfo{person}{Miina Porkka}, \bibinfo{person}{Stefan Rahmstorf}, \bibinfo{person}{Sibyll Schaphoff}, \bibinfo{person}{Kirsten Thonicke}, \bibinfo{person}{Arne Tobian}, \bibinfo{person}{Vili Virkki}, \bibinfo{person}{Lan Wang-Erlandsson}, \bibinfo{person}{Lisa Weber}, {and} \bibinfo{person}{Johan Rockström}.}
  \bibinfo{year}{2023}\natexlab{}.
\newblock \showarticletitle{Earth beyond six of nine planetary boundaries}.
\newblock \bibinfo{journal}{\emph{Science Advances}} \bibinfo{volume}{9}, \bibinfo{number}{37} (\bibinfo{date}{Sept.} \bibinfo{year}{2023}).
\newblock
\urldef\tempurl%
\url{https://doi.org/10.1126/sciadv.adh2458}
\showDOI{\tempurl}


\bibitem[Rillig et~al\mbox{.}(2023)]%
        {Rillig_Ågerstrand_Bi_Gould_Sauerland_2023}
\bibfield{author}{\bibinfo{person}{Matthias~C. Rillig}, \bibinfo{person}{Marlene Ågerstrand}, \bibinfo{person}{Mohan Bi}, \bibinfo{person}{Kenneth~A. Gould}, {and} \bibinfo{person}{Uli Sauerland}.} \bibinfo{year}{2023}\natexlab{}.
\newblock \showarticletitle{Risks and Benefits of Large Language Models for the Environment}.
\newblock \bibinfo{journal}{\emph{Environmental Science \& Technology}} \bibinfo{volume}{57}, \bibinfo{number}{9} (\bibinfo{year}{2023}), \bibinfo{pages}{3464–3466}.
\newblock
\showISSN{0013-936X}
\urldef\tempurl%
\url{https://doi.org/10.1021/acs.est.3c01106}
\showDOI{\tempurl}


\bibitem[Rockström et~al\mbox{.}(2024)]%
        {rockstrom_planetary_2024}
\bibfield{author}{\bibinfo{person}{Johan Rockström}, \bibinfo{person}{Louis Kotzé}, \bibinfo{person}{Svetlana Milutinović}, \bibinfo{person}{Frank Biermann}, \bibinfo{person}{Victor Brovkin}, \bibinfo{person}{Jonathan Donges}, \bibinfo{person}{Jonas Ebbesson}, \bibinfo{person}{Duncan French}, \bibinfo{person}{Joyeeta Gupta}, \bibinfo{person}{Rakhyun Kim}, \bibinfo{person}{Timothy Lenton}, \bibinfo{person}{Dominic Lenzi}, \bibinfo{person}{Nebojsa Nakicenovic}, \bibinfo{person}{Barbara Neumann}, \bibinfo{person}{Fabian Schuppert}, \bibinfo{person}{Ricarda Winkelmann}, \bibinfo{person}{Klaus Bosselmann}, \bibinfo{person}{Carl Folke}, \bibinfo{person}{Wolfgang Lucht}, \bibinfo{person}{David Schlosberg}, \bibinfo{person}{Katherine Richardson}, {and} \bibinfo{person}{Will Steffen}.} \bibinfo{year}{2024}\natexlab{}.
\newblock \showarticletitle{The planetary commons: {A} new paradigm for safeguarding {Earth}-regulating systems in the {Anthropocene}}.
\newblock \bibinfo{journal}{\emph{Proceedings of the National Academy of Sciences}} \bibinfo{volume}{121}, \bibinfo{number}{5} (\bibinfo{date}{Jan.} \bibinfo{year}{2024}), \bibinfo{pages}{e2301531121}.
\newblock
\urldef\tempurl%
\url{https://doi.org/10.1073/pnas.2301531121}
\showDOI{\tempurl}
\newblock
\shownote{Publisher: Proceedings of the National Academy of Sciences}.


\bibitem[Rockström et~al\mbox{.}(2009)]%
        {rockstrom_planetary_2009}
\bibfield{author}{\bibinfo{person}{Johan Rockström}, \bibinfo{person}{Will Steffen}, \bibinfo{person}{Kevin Noone}, \bibinfo{person}{Åsa Persson}, \bibinfo{person}{F.~Stuart Chapin}, \bibinfo{person}{Eric Lambin}, \bibinfo{person}{Timothy~M. Lenton}, \bibinfo{person}{Marten Scheffer}, \bibinfo{person}{Carl Folke}, \bibinfo{person}{Hans~Joachim Schellnhuber}, \bibinfo{person}{Björn Nykvist}, \bibinfo{person}{Cynthia~A. de Wit}, \bibinfo{person}{Terry Hughes}, \bibinfo{person}{Sander van~der Leeuw}, \bibinfo{person}{Henning Rodhe}, \bibinfo{person}{Sverker Sörlin}, \bibinfo{person}{Peter~K. Snyder}, \bibinfo{person}{Robert Costanza}, \bibinfo{person}{Uno Svedin}, \bibinfo{person}{Malin Falkenmark}, \bibinfo{person}{Louise Karlberg}, \bibinfo{person}{Robert~W. Corell}, \bibinfo{person}{Victoria~J. Fabry}, \bibinfo{person}{James Hansen}, \bibinfo{person}{Brian Walker}, \bibinfo{person}{Diana Liverman}, \bibinfo{person}{Katherine Richardson}, \bibinfo{person}{Paul Crutzen}, {and} \bibinfo{person}{Jonathan
  Foley}.} \bibinfo{year}{2009}\natexlab{}.
\newblock \showarticletitle{Planetary {Boundaries}: {Exploring} the {Safe} {Operating} {Space} for {Humanity}}.
\newblock \bibinfo{journal}{\emph{Ecology and Society}} \bibinfo{volume}{14}, \bibinfo{number}{2} (\bibinfo{year}{2009}).
\newblock
\showISSN{1708-3087}
\urldef\tempurl%
\url{https://www.jstor.org/stable/26268316}
\showURL{%
\tempurl}


\bibitem[{Roger Wehner}(2023)]%
        {roger_wehner_5_2023}
\bibfield{author}{\bibinfo{person}{{Roger Wehner}}.} \bibinfo{year}{2023}\natexlab{}.
\newblock \bibinfo{title}{5 ways {AWS} data centers benefit local communities}.
\newblock
\newblock
\urldef\tempurl%
\url{https://www.aboutamazon.com/news/aws/aws-data-center-economic-impact-study}
\showURL{%
\tempurl}


\bibitem[Rogers and Luccioni(2024)]%
        {rogers_position_2024}
\bibfield{author}{\bibinfo{person}{Anna Rogers} {and} \bibinfo{person}{Alexandra~Sasha Luccioni}.} \bibinfo{year}{2024}\natexlab{}.
\newblock \bibinfo{title}{Position: {Key} {Claims} in {LLM} {Research} {Have} a {Long} {Tail} of {Footnotes}}.
\newblock
\newblock
\urldef\tempurl%
\url{https://doi.org/10.48550/arXiv.2308.07120}
\showDOI{\tempurl}


\bibitem[{Saijel Kishan}(2024)]%
        {saijel_kishan_its_2024}
\bibfield{author}{\bibinfo{person}{{Saijel Kishan}}.} \bibinfo{year}{2024}\natexlab{}.
\newblock \showarticletitle{‘{It}’s a {Money} {Loser}’: {Tax} {Breaks} for {Data} {Centers} {Are} {Under} {Fire}}.
\newblock \bibinfo{journal}{\emph{Bloomberg.com}} (\bibinfo{year}{2024}).
\newblock
\urldef\tempurl%
\url{https://www.bloomberg.com/news/articles/2024-05-09/ai-boom-has-some-states-rethinking-subsidies-for-data-centers}
\showURL{%
\tempurl}


\bibitem[Saltelli et~al\mbox{.}(2022)]%
        {saltelli_science_2022}
\bibfield{author}{\bibinfo{person}{Andrea Saltelli}, \bibinfo{person}{Dorothy~J. Dankel}, \bibinfo{person}{Monica Di~Fiore}, \bibinfo{person}{Nina Holland}, {and} \bibinfo{person}{Martin Pigeon}.} \bibinfo{year}{2022}\natexlab{}.
\newblock \showarticletitle{Science, the endless frontier of regulatory capture}.
\newblock \bibinfo{journal}{\emph{Futures}}  \bibinfo{volume}{135} (\bibinfo{year}{2022}), \bibinfo{pages}{102860}.
\newblock
\showISSN{0016-3287}
\urldef\tempurl%
\url{https://doi.org/10.1016/j.futures.2021.102860}
\showDOI{\tempurl}


\bibitem[Samsi et~al\mbox{.}(2023)]%
        {samsi_words_2023}
\bibfield{author}{\bibinfo{person}{Siddharth Samsi}, \bibinfo{person}{Dan Zhao}, \bibinfo{person}{Joseph McDonald}, \bibinfo{person}{Baolin Li}, \bibinfo{person}{Adam Michaleas}, \bibinfo{person}{Michael Jones}, \bibinfo{person}{William Bergeron}, \bibinfo{person}{Jeremy Kepner}, \bibinfo{person}{Devesh Tiwari}, {and} \bibinfo{person}{Vijay Gadepally}.} \bibinfo{year}{2023}\natexlab{}.
\newblock \bibinfo{title}{From {Words} to {Watts}: {Benchmarking} the {Energy} {Costs} of {Large} {Language} {Model} {Inference}}.
\newblock
\newblock
\urldef\tempurl%
\url{https://doi.org/10.48550/arXiv.2310.03003}
\showDOI{\tempurl}


\bibitem[Satariano and Isaac(2021)]%
        {satariano_silent_2021}
\bibfield{author}{\bibinfo{person}{Adam Satariano} {and} \bibinfo{person}{Mike Isaac}.} \bibinfo{year}{2021}\natexlab{}.
\newblock \showarticletitle{The {Silent} {Partner} {Cleaning} {Up} {Facebook} for \$500 {Million} a {Year}}.
\newblock \bibinfo{journal}{\emph{The New York Times}} (\bibinfo{year}{2021}).
\newblock
\showISSN{0362-4331}
\urldef\tempurl%
\url{https://www.nytimes.com/2021/08/31/technology/facebook-accenture-content-moderation.html}
\showURL{%
\tempurl}


\bibitem[Saunders and Tsao(2012)]%
        {saunders_rebound_2012}
\bibfield{author}{\bibinfo{person}{Harry~D. Saunders} {and} \bibinfo{person}{Jeffrey~Y. Tsao}.} \bibinfo{year}{2012}\natexlab{}.
\newblock \showarticletitle{Rebound effects for lighting}.
\newblock \bibinfo{journal}{\emph{Energy Policy}}  \bibinfo{volume}{49} (\bibinfo{date}{Oct.} \bibinfo{year}{2012}), \bibinfo{pages}{477--478}.
\newblock
\showISSN{0301-4215}
\urldef\tempurl%
\url{https://doi.org/10.1016/j.enpol.2012.06.050}
\showDOI{\tempurl}


\bibitem[Sayegh(2024)]%
        {sayegh_billion-dollar_2024}
\bibfield{author}{\bibinfo{person}{Emil Sayegh}.} \bibinfo{year}{2024}\natexlab{}.
\newblock \bibinfo{title}{The {Billion}-{Dollar} {AI} {Gamble}: {Data} {Centers} {As} {The} {New} {High}-{Stakes} {Game}}.
\newblock
\newblock
\urldef\tempurl%
\url{https://www.forbes.com/sites/emilsayegh/2024/09/30/the-billion-dollar-ai-gamble-data-centers-as-the-new-high-stakes-game/}
\showURL{%
\tempurl}


\bibitem[Senge(1997)]%
        {senge_fifth_1997}
\bibfield{author}{\bibinfo{person}{Peter~M Senge}.} \bibinfo{year}{1997}\natexlab{}.
\newblock \showarticletitle{The {Fifth} {Discipline}}.
\newblock \bibinfo{journal}{\emph{Measuring Business Excellence}} \bibinfo{volume}{1}, \bibinfo{number}{3} (\bibinfo{year}{1997}), \bibinfo{pages}{46--51}.
\newblock
\showISSN{1368-3047}
\urldef\tempurl%
\url{https://doi.org/10.1108/eb025496}
\showDOI{\tempurl}


\bibitem[Serokell(2023)]%
        {serokell_what_2023}
\bibfield{author}{\bibinfo{person}{Ivan~Smetannikov Serokell}.} \bibinfo{year}{2023}\natexlab{}.
\newblock \bibinfo{title}{What happens when we run out of data for {AI} models}.
\newblock
\newblock
\urldef\tempurl%
\url{https://venturebeat.com/ai/what-happens-when-we-run-out-of-data-for-ai-models/}
\showURL{%
\tempurl}


\bibitem[Sevilla et~al\mbox{.}(2022)]%
        {sevilla_compute_2022}
\bibfield{author}{\bibinfo{person}{Jaime Sevilla}, \bibinfo{person}{Lennart Heim}, \bibinfo{person}{Anson Ho}, \bibinfo{person}{Tamay Besiroglu}, \bibinfo{person}{Marius Hobbhahn}, {and} \bibinfo{person}{Pablo Villalobos}.} \bibinfo{year}{2022}\natexlab{}.
\newblock \bibinfo{title}{Compute {Trends} {Across} {Three} {Eras} of {Machine} {Learning}}.
\newblock
\newblock
\urldef\tempurl%
\url{https://doi.org/10.48550/arXiv.2202.05924}
\showDOI{\tempurl}


\bibitem[Sevilla et~al\mbox{.}(2024)]%
        {sevilla_can_2024}
\bibfield{author}{\bibinfo{person}{Jaime Sevilla}, \bibinfo{person}{{Tamay Besiroglu}}, \bibinfo{person}{{Ben Cottier}}, \bibinfo{person}{{Josh You}}, \bibinfo{person}{{Edu Roldán}}, \bibinfo{person}{{Pablo Villalobos}}, {and} \bibinfo{person}{{Ege Erdil}}.} \bibinfo{year}{2024}\natexlab{}.
\newblock \bibinfo{booktitle}{\emph{Can {AI} {Scaling} {Continue} {Through} 2030?}}
\newblock \bibinfo{type}{{T}echnical {R}eport}. \bibinfo{institution}{epoch.ai}.
\newblock
\urldef\tempurl%
\url{https://epochai.org/blog/can-ai-scaling-continue-through-2030}
\showURL{%
\tempurl}


\bibitem[Shah and Bender(2022)]%
        {shah_situating_2022}
\bibfield{author}{\bibinfo{person}{Chirag Shah} {and} \bibinfo{person}{Emily~M. Bender}.} \bibinfo{year}{2022}\natexlab{}.
\newblock \showarticletitle{Situating {Search}}. In \bibinfo{booktitle}{\emph{Proceedings of the 2022 {Conference} on {Human} {Information} {Interaction} and {Retrieval}}} \emph{(\bibinfo{series}{{CHIIR} '22})}. \bibinfo{publisher}{Association for Computing Machinery}, \bibinfo{address}{New York, NY, USA}, \bibinfo{pages}{221--232}.
\newblock
\showISBNx{978-1-4503-9186-3}
\urldef\tempurl%
\url{https://doi.org/10.1145/3498366.3505816}
\showDOI{\tempurl}


\bibitem[Shah and Bender(2024)]%
        {shah_envisioning_2024}
\bibfield{author}{\bibinfo{person}{Chirag Shah} {and} \bibinfo{person}{Emily~M. Bender}.} \bibinfo{year}{2024}\natexlab{}.
\newblock \showarticletitle{Envisioning {Information} {Access} {Systems}: {What} {Makes} for {Good} {Tools} and a {Healthy} {Web}?}
\newblock \bibinfo{journal}{\emph{ACM Transactions on the Web}} \bibinfo{volume}{18}, \bibinfo{number}{3} (\bibinfo{year}{2024}), \bibinfo{pages}{1--24}.
\newblock
\showISSN{1559-1131, 1559-114X}
\urldef\tempurl%
\url{https://doi.org/10.1145/3649468}
\showDOI{\tempurl}


\bibitem[{Shaolei Ren}(2023)]%
        {shaolei_ren_how_2023}
\bibfield{author}{\bibinfo{person}{{Shaolei Ren}}.} \bibinfo{year}{2023}\natexlab{}.
\newblock \bibinfo{title}{How much water does {AI} consume? {The} public deserves to know}.
\newblock
\newblock
\urldef\tempurl%
\url{https://oecd.ai/en/wonk/how-much-water-does-ai-consume}
\showURL{%
\tempurl}


\bibitem[Shumailov et~al\mbox{.}(2024)]%
        {shumailov_ai_2024}
\bibfield{author}{\bibinfo{person}{Ilia Shumailov}, \bibinfo{person}{Zakhar Shumaylov}, \bibinfo{person}{Yiren Zhao}, \bibinfo{person}{Nicolas Papernot}, \bibinfo{person}{Ross Anderson}, {and} \bibinfo{person}{Yarin Gal}.} \bibinfo{year}{2024}\natexlab{}.
\newblock \showarticletitle{{AI} models collapse when trained on recursively generated data}.
\newblock \bibinfo{journal}{\emph{Nature}} \bibinfo{volume}{631}, \bibinfo{number}{8022} (\bibinfo{year}{2024}), \bibinfo{pages}{755--759}.
\newblock
\showISSN{0028-0836, 1476-4687}
\urldef\tempurl%
\url{https://doi.org/10.1038/s41586-024-07566-y}
\showDOI{\tempurl}


\bibitem[Similarweb({[n.\,d.]})]%
        {Similarweb}
\bibfield{author}{\bibinfo{person}{Similarweb}.} \bibinfo{year}{[n.\,d.]}\natexlab{}.
\newblock \bibinfo{title}{chatgpt.com Traffic Analytics, Ranking \& Audience [December 2024]}.
\newblock
\newblock
\urldef\tempurl%
\url{https://www.similarweb.com/website/chatgpt.com/}
\showURL{%
\tempurl}


\bibitem[Smuha(2021)]%
        {smuha_beyond_2021}
\bibfield{author}{\bibinfo{person}{Nathalie~A. Smuha}.} \bibinfo{year}{2021}\natexlab{}.
\newblock \bibinfo{title}{Beyond the {Individual}: {Governing} {AI}’s {Societal} {Harm}}.
\newblock
\newblock
\urldef\tempurl%
\url{https://papers.ssrn.com/abstract=3941956}
\showURL{%
\tempurl}


\bibitem[Spring and Spring(2024)]%
        {spring_firms_2024}
\bibfield{author}{\bibinfo{person}{Jake Spring} {and} \bibinfo{person}{Jake Spring}.} \bibinfo{year}{2024}\natexlab{}.
\newblock \showarticletitle{Firms including {Amazon} to buy \$180 million in carbon credits from namesake rainforest}.
\newblock \bibinfo{journal}{\emph{Reuters}} (\bibinfo{year}{2024}).
\newblock
\urldef\tempurl%
\url{https://www.reuters.com/sustainability/firms-including-amazon-buy-180-million-carbon-credits-namesake-rainforest-2024-09-24/}
\showURL{%
\tempurl}


\bibitem[Stapleton et~al\mbox{.}(2016)]%
        {Stapleton_2016}
\bibfield{author}{\bibinfo{person}{Lee Stapleton}, \bibinfo{person}{Steve Sorrell}, {and} \bibinfo{person}{Tim Schwanen}.} \bibinfo{year}{2016}\natexlab{}.
\newblock \showarticletitle{Estimating direct rebound effects for personal automotive travel in Great Britain}.
\newblock \bibinfo{journal}{\emph{Energy Economics}}  \bibinfo{volume}{54} (\bibinfo{year}{2016}), \bibinfo{pages}{313–325}.
\newblock
\showISSN{0140-9883}
\urldef\tempurl%
\url{https://doi.org/10.1016/j.eneco.2015.12.012}
\showDOI{\tempurl}


\bibitem[Steffen et~al\mbox{.}(2015)]%
        {Steffen__2015}
\bibfield{author}{\bibinfo{person}{Will Steffen}, \bibinfo{person}{Katherine Richardson}, \bibinfo{person}{Johan Rockström}, \bibinfo{person}{Sarah~E. Cornell}, \bibinfo{person}{Ingo Fetzer}, \bibinfo{person}{Elena~M. Bennett}, \bibinfo{person}{Reinette Biggs}, \bibinfo{person}{Stephen~R. Carpenter}, \bibinfo{person}{Wim de Vries}, \bibinfo{person}{Cynthia~A. de Wit}, \bibinfo{person}{Carl Folke}, \bibinfo{person}{Dieter Gerten}, \bibinfo{person}{Jens Heinke}, \bibinfo{person}{Georgina~M. Mace}, \bibinfo{person}{Linn~M. Persson}, \bibinfo{person}{Veerabhadran Ramanathan}, \bibinfo{person}{Belinda Reyers}, {and} \bibinfo{person}{Sverker Sörlin}.} \bibinfo{year}{2015}\natexlab{}.
\newblock \showarticletitle{Planetary boundaries: Guiding human development on a changing planet}.
\newblock \bibinfo{journal}{\emph{Science}} \bibinfo{volume}{347}, \bibinfo{number}{6223} (\bibinfo{year}{2015}), \bibinfo{pages}{1259855}.
\newblock
\urldef\tempurl%
\url{https://doi.org/10.1126/science.1259855}
\showDOI{\tempurl}


\bibitem[Sterman(2000)]%
        {sterman_business_2000}
\bibfield{author}{\bibinfo{person}{J Sterman}.} \bibinfo{year}{2000}\natexlab{}.
\newblock \showarticletitle{Business {Dynamics}—{Systems} {Thinking} and {Modeling} for a {Complex} {World}}.
\newblock \bibinfo{journal}{\emph{Journal of the Operational Research Society}} \bibinfo{volume}{53}, \bibinfo{number}{4} (\bibinfo{year}{2000}), \bibinfo{pages}{472--473}.
\newblock
\showISSN{0160-5682, 1476-9360}
\urldef\tempurl%
\url{https://doi.org/10.1057/palgrave.jors.2601336}
\showDOI{\tempurl}


\bibitem[Sutskever(2024)]%
        {Sutskever_2024}
\bibfield{author}{\bibinfo{person}{Ilya Sutskever}.} \bibinfo{year}{2024}\natexlab{}.
\newblock \bibinfo{title}{NeurIPS Test Of Time}.
\newblock
\newblock
\urldef\tempurl%
\url{https://neurips.cc/virtual/2024/test-of-time/105032}
\showURL{%
\tempurl}


\bibitem[Tarnoff and {Kim Tarnoff}(2020)]%
        {tarnoff_tech_2020}
\bibfield{author}{\bibinfo{person}{Ben Tarnoff} {and} \bibinfo{person}{{Kim Tarnoff}}.} \bibinfo{year}{2020}\natexlab{}.
\newblock \bibinfo{title}{Tech {Workers} {Versus} the {Pentagon}}.
\newblock
\newblock
\urldef\tempurl%
\url{https://collectiveaction.tech/2020/tech-workers-versus-the-pentagon/}
\showURL{%
\tempurl}


\bibitem[Tsao et~al\mbox{.}(2010)]%
        {tsao_solid-state_2010}
\bibfield{author}{\bibinfo{person}{J.~Y. Tsao}, \bibinfo{person}{H.~D. Saunders}, \bibinfo{person}{J.~R. Creighton}, \bibinfo{person}{M.~E. Coltrin}, {and} \bibinfo{person}{J.~A. Simmons}.} \bibinfo{year}{2010}\natexlab{}.
\newblock \showarticletitle{Solid-state lighting: an energy-economics perspective}.
\newblock \bibinfo{journal}{\emph{Journal of Physics D: Applied Physics}} \bibinfo{volume}{43}, \bibinfo{number}{35} (\bibinfo{date}{Aug.} \bibinfo{year}{2010}), \bibinfo{pages}{354001}.
\newblock
\showISSN{0022-3727}
\urldef\tempurl%
\url{https://doi.org/10.1088/0022-3727/43/35/354001}
\showDOI{\tempurl}
\newblock
\shownote{Publisher: IOP Publishing}.


\bibitem[Tunstall et~al\mbox{.}(2022)]%
        {tunstall_natural_2022}
\bibfield{author}{\bibinfo{person}{Lewis Tunstall}, \bibinfo{person}{Leandro~von Werra}, \bibinfo{person}{Thomas Wolf}, {and} \bibinfo{person}{Aurélien Géron}.} \bibinfo{year}{2022}\natexlab{}.
\newblock \bibinfo{booktitle}{\emph{Natural language processing with {Transformers}: building language applications with {Hugging} {Face}}}.
\newblock \bibinfo{publisher}{O'Reilly}.
\newblock
\showISBNx{978-1-09-810324-8}


\bibitem[{UN General Assembly}(2024)]%
        {un_general_assembly_pact_2024}
\bibfield{author}{\bibinfo{person}{{UN General Assembly}}.} \bibinfo{year}{2024}\natexlab{}.
\newblock \bibinfo{booktitle}{\emph{The {Pact} for the {Future}}}.
\newblock Number A/RES/79/1 in \bibinfo{series}{Resolutions}. \bibinfo{publisher}{United Nations}, \bibinfo{address}{New York}.
\newblock
\urldef\tempurl%
\url{https://www.un.org/en/summit-of-the-future/pact-for-the-future}
\showURL{%
\tempurl}


\bibitem[Unruh et~al\mbox{.}(2022)]%
        {unruh_human_2022}
\bibfield{author}{\bibinfo{person}{Charlotte~Franziska Unruh}, \bibinfo{person}{Charlotte Haid}, \bibinfo{person}{Fottner Johannes}, {and} \bibinfo{person}{Tim Büthe}.} \bibinfo{year}{2022}\natexlab{}.
\newblock \showarticletitle{Human {Autonomy} in {Algorithmic} {Management}}. In \bibinfo{booktitle}{\emph{Proceedings of the 2022 {AAAI}/{ACM} {Conference} on {AI}, {Ethics}, and {Society}}} \emph{(\bibinfo{series}{{AIES} '22})}. \bibinfo{publisher}{Association for Computing Machinery}, \bibinfo{address}{New York, NY, USA}, \bibinfo{pages}{753--762}.
\newblock
\showISBNx{978-1-4503-9247-1}
\urldef\tempurl%
\url{https://doi.org/10.1145/3514094.3534168}
\showDOI{\tempurl}


\bibitem[{Urs Hölzle}(2022)]%
        {urs_holzle_our_2022}
\bibfield{author}{\bibinfo{person}{{Urs Hölzle}}.} \bibinfo{year}{2022}\natexlab{}.
\newblock \bibinfo{title}{Our commitment to climate-conscious data center cooling}.
\newblock
\newblock
\urldef\tempurl%
\url{https://blog.google/outreach-initiatives/sustainability/our-commitment-to-climate-conscious-data-center-cooling/}
\showURL{%
\tempurl}


\bibitem[van Maanen(2022)]%
        {van_maanen_ai_2022}
\bibfield{author}{\bibinfo{person}{Gijs van Maanen}.} \bibinfo{year}{2022}\natexlab{}.
\newblock \showarticletitle{{AI} {Ethics}, {Ethics} {Washing}, and the {Need} to {Politicize} {Data} {Ethics}}.
\newblock \bibinfo{journal}{\emph{Digital Society}} \bibinfo{volume}{1}, \bibinfo{number}{2} (\bibinfo{year}{2022}), \bibinfo{pages}{9}.
\newblock
\showISSN{2731-4669}
\urldef\tempurl%
\url{https://doi.org/10.1007/s44206-022-00013-3}
\showDOI{\tempurl}


\bibitem[Varantsou(2024)]%
        {varantsou_image_2024}
\bibfield{author}{\bibinfo{person}{Mikhail Varantsou}.} \bibinfo{year}{2024}\natexlab{}.
\newblock \emph{\bibinfo{title}{Image {Rights} in the {Age} of {AI} and {Deep} {Fakes}}}.
\newblock \bibinfo{thesistype}{Ph.\,D. Dissertation}. \bibinfo{school}{Jagiellonian University}.
\newblock
\urldef\tempurl%
\url{https://ruj.uj.edu.pl/entities/publication/b3deb8e2-056f-4e4d-8c30-a847412a20e5}
\showURL{%
\tempurl}


\bibitem[Varoquaux et~al\mbox{.}(2024)]%
        {varoquaux_hype_2024}
\bibfield{author}{\bibinfo{person}{Gaël Varoquaux}, \bibinfo{person}{Alexandra~Sasha Luccioni}, {and} \bibinfo{person}{Meredith Whittaker}.} \bibinfo{year}{2024}\natexlab{}.
\newblock \bibinfo{title}{Hype, {Sustainability}, and the {Price} of the {Bigger}-is-{Better} {Paradigm} in {AI}}.
\newblock
\newblock
\urldef\tempurl%
\url{https://doi.org/10.48550/arXiv.2409.14160}
\showDOI{\tempurl}


\bibitem[Vaswani et~al\mbox{.}(2017)]%
        {vaswani_attention_2017}
\bibfield{author}{\bibinfo{person}{Ashish Vaswani}, \bibinfo{person}{Noam Shazeer}, \bibinfo{person}{Niki Parmar}, \bibinfo{person}{Jakob Uszkoreit}, \bibinfo{person}{Llion Jones}, \bibinfo{person}{Aidan~N Gomez}, \bibinfo{person}{Ł~ukasz Kaiser}, {and} \bibinfo{person}{Illia Polosukhin}.} \bibinfo{year}{2017}\natexlab{}.
\newblock \showarticletitle{Attention is {All} you {Need}}. In \bibinfo{booktitle}{\emph{Advances in {Neural} {Information} {Processing} {Systems}}}, Vol.~\bibinfo{volume}{30}. \bibinfo{publisher}{Curran Associates, Inc.}
\newblock
\urldef\tempurl%
\url{https://papers.neurips.cc/paper/2017/hash/3f5ee243547dee91fbd053c1c4a845aa-Abstract.html}
\showURL{%
\tempurl}


\bibitem[Villalobos et~al\mbox{.}(2024)]%
        {villalobos_will_2024}
\bibfield{author}{\bibinfo{person}{Pablo Villalobos}, \bibinfo{person}{Anson Ho}, \bibinfo{person}{Jaime Sevilla}, \bibinfo{person}{Tamay Besiroglu}, \bibinfo{person}{Lennart Heim}, {and} \bibinfo{person}{Marius Hobbhahn}.} \bibinfo{year}{2024}\natexlab{}.
\newblock \bibinfo{title}{Will we run out of data? {Limits} of {LLM} scaling based on human-generated data}.
\newblock
\newblock
\urldef\tempurl%
\url{https://doi.org/10.48550/arXiv.2211.04325}
\showDOI{\tempurl}


\bibitem[Wang et~al\mbox{.}(2023)]%
        {wang_environmental_2023}
\bibfield{author}{\bibinfo{person}{Qi Wang}, \bibinfo{person}{Nan Huang}, \bibinfo{person}{Zhuo Chen}, \bibinfo{person}{Xiaowen Chen}, \bibinfo{person}{Hanying Cai}, {and} \bibinfo{person}{Yunpeng Wu}.} \bibinfo{year}{2023}\natexlab{}.
\newblock \showarticletitle{Environmental data and facts in the semiconductor manufacturing industry: {An} unexpected high water and energy consumption situation}.
\newblock \bibinfo{journal}{\emph{Water Cycle}}  \bibinfo{volume}{4} (\bibinfo{year}{2023}), \bibinfo{pages}{47--54}.
\newblock
\showISSN{2666-4453}
\urldef\tempurl%
\url{https://doi.org/10.1016/j.watcyc.2023.01.004}
\showDOI{\tempurl}


\bibitem[Wardman(2016)]%
        {Wardman_2016}
\bibfield{author}{\bibinfo{person}{Kellie Wardman}.} \bibinfo{year}{2016}\natexlab{}.
\newblock \bibinfo{title}{Balancing Loops with Delays}.
\newblock
\newblock
\urldef\tempurl%
\url{https://thesystemsthinker.com/balancing-loops-with-delays/}
\showURL{%
\tempurl}


\bibitem[Wei et~al\mbox{.}(2024)]%
        {wei_how_2024}
\bibfield{author}{\bibinfo{person}{Kevin Wei}, \bibinfo{person}{Carson Ezell}, \bibinfo{person}{Nick Gabrieli}, {and} \bibinfo{person}{Chinmay Deshpande}.} \bibinfo{year}{2024}\natexlab{}.
\newblock \showarticletitle{How {Do} {AI} {Companies} “{Fine}-{Tune}” {Policy}? {Examining} {Regulatory} {Capture} in {AI} {Governance}}.
\newblock \bibinfo{journal}{\emph{Proceedings of the AAAI/ACM Conference on AI, Ethics, and Society}}  \bibinfo{volume}{7} (\bibinfo{year}{2024}), \bibinfo{pages}{1539--1555}.
\newblock
\showISSN{3065-8365}
\urldef\tempurl%
\url{https://doi.org/10.1609/aies.v7i1.31745}
\showDOI{\tempurl}


\bibitem[Weidinger et~al\mbox{.}(2022)]%
        {Weidinger_2022}
\bibfield{author}{\bibinfo{person}{Laura Weidinger}, \bibinfo{person}{Jonathan Uesato}, \bibinfo{person}{Maribeth Rauh}, \bibinfo{person}{Conor Griffin}, \bibinfo{person}{Po-Sen Huang}, \bibinfo{person}{John Mellor}, \bibinfo{person}{Amelia Glaese}, \bibinfo{person}{Myra Cheng}, \bibinfo{person}{Borja Balle}, \bibinfo{person}{Atoosa Kasirzadeh}, \bibinfo{person}{Courtney Biles}, \bibinfo{person}{Sasha Brown}, \bibinfo{person}{Zac Kenton}, \bibinfo{person}{Will Hawkins}, \bibinfo{person}{Tom Stepleton}, \bibinfo{person}{Abeba Birhane}, \bibinfo{person}{Lisa~Anne Hendricks}, \bibinfo{person}{Laura Rimell}, \bibinfo{person}{William Isaac}, \bibinfo{person}{Julia Haas}, \bibinfo{person}{Sean Legassick}, \bibinfo{person}{Geoffrey Irving}, {and} \bibinfo{person}{Iason Gabriel}.} \bibinfo{year}{2022}\natexlab{}.
\newblock \showarticletitle{Taxonomy of Risks posed by Language Models}. In \bibinfo{booktitle}{\emph{2022 ACM Conference on Fairness, Accountability, and Transparency}}. \bibinfo{publisher}{ACM}, \bibinfo{address}{Seoul Republic of Korea}, \bibinfo{pages}{214–229}.
\newblock
\showISBNx{978-1-4503-9352-2}
\urldef\tempurl%
\url{https://doi.org/10.1145/3531146.3533088}
\showDOI{\tempurl}


\bibitem[Whitney and Norman(2024)]%
        {whitney_real_2024}
\bibfield{author}{\bibinfo{person}{Cedric~Deslandes Whitney} {and} \bibinfo{person}{Justin Norman}.} \bibinfo{year}{2024}\natexlab{}.
\newblock \showarticletitle{Real {Risks} of {Fake} {Data}: {Synthetic} {Data}, {Diversity}-{Washing} and {Consent} {Circumvention}}. In \bibinfo{booktitle}{\emph{Proceedings of the 2024 {ACM} {Conference} on {Fairness}, {Accountability}, and {Transparency}}} \emph{(\bibinfo{series}{{FAccT} '24})}. \bibinfo{publisher}{Association for Computing Machinery}, \bibinfo{address}{New York, NY, USA}, \bibinfo{pages}{1733--1744}.
\newblock
\showISBNx{9798400704505}
\urldef\tempurl%
\url{https://doi.org/10.1145/3630106.3659002}
\showDOI{\tempurl}


\bibitem[Whittaker(2021)]%
        {whittaker_steep_2021}
\bibfield{author}{\bibinfo{person}{Meredith Whittaker}.} \bibinfo{year}{2021}\natexlab{}.
\newblock \showarticletitle{The steep cost of capture}.
\newblock \bibinfo{journal}{\emph{Interactions}} \bibinfo{volume}{28}, \bibinfo{number}{6} (\bibinfo{year}{2021}), \bibinfo{pages}{50--55}.
\newblock
\showISSN{1072-5520, 1558-3449}
\urldef\tempurl%
\url{https://doi.org/10.1145/3488666}
\showDOI{\tempurl}


\bibitem[Widder and Hicks(2024)]%
        {widder_watching_2024}
\bibfield{author}{\bibinfo{person}{David~Gray Widder} {and} \bibinfo{person}{Mar Hicks}.} \bibinfo{year}{2024}\natexlab{}.
\newblock \bibinfo{title}{Watching the {Generative} {AI} {Hype} {Bubble} {Deflate}}.
\newblock
\newblock
\urldef\tempurl%
\url{https://doi.org/10.48550/arXiv.2408.08778}
\showDOI{\tempurl}


\bibitem[Willenbacher et~al\mbox{.}(2022)]%
        {willenbacher_rebound_2022}
\bibfield{author}{\bibinfo{person}{Martina Willenbacher}, \bibinfo{person}{Torsten Hornauer}, {and} \bibinfo{person}{Volker Wohlgemuth}.} \bibinfo{year}{2022}\natexlab{}.
\newblock \showarticletitle{Rebound {Effects} in {Methods} of {Artificial} {Intelligence}}. In \bibinfo{booktitle}{\emph{Advances and {New} {Trends} in {Environmental} {Informatics}}}, \bibfield{editor}{\bibinfo{person}{Volker Wohlgemuth}, \bibinfo{person}{Stefan Naumann}, \bibinfo{person}{Grit Behrens}, {and} \bibinfo{person}{Hans-Knud Arndt}} (Eds.). \bibinfo{publisher}{Springer International Publishing}, \bibinfo{address}{Cham}, \bibinfo{pages}{73--85}.
\newblock
\showISBNx{978-3-030-88063-7}
\urldef\tempurl%
\url{https://doi.org/10.1007/978-3-030-88063-7_5}
\showDOI{\tempurl}


\bibitem[Williams et~al\mbox{.}(2022)]%
        {williams_exploited_2022}
\bibfield{author}{\bibinfo{person}{Adrienne Williams}, \bibinfo{person}{{Milagros Miceli}}, {and} \bibinfo{person}{{Timnit Gebru}}.} \bibinfo{year}{2022}\natexlab{}.
\newblock \bibinfo{title}{The {Exploited} {Labor} {Behind} {Artificial} {Intelligence}}.
\newblock
\newblock
\urldef\tempurl%
\url{https://www.noemamag.com/the-exploited-labor-behind-artificial-intelligence}
\showURL{%
\tempurl}


\bibitem[Wong(2024)]%
        {wong_silicon_2024}
\bibfield{author}{\bibinfo{person}{Matteo Wong}.} \bibinfo{year}{2024}\natexlab{}.
\newblock \bibinfo{title}{Silicon {Valley}’s {Trillion}-{Dollar} {Leap} of {Faith}}.
\newblock
\newblock
\urldef\tempurl%
\url{https://www.theatlantic.com/technology/archive/2024/07/ai-companies-unprofitable/679278/}
\showURL{%
\tempurl}


\bibitem[Wright et~al\mbox{.}(2023)]%
        {wright_efficiency_2023}
\bibfield{author}{\bibinfo{person}{Dustin Wright}, \bibinfo{person}{Christian Igel}, \bibinfo{person}{Gabrielle Samuel}, {and} \bibinfo{person}{Raghavendra Selvan}.} \bibinfo{year}{2023}\natexlab{}.
\newblock \bibinfo{title}{Efficiency is {Not} {Enough}: {A} {Critical} {Perspective} of {Environmentally} {Sustainable} {AI}}.
\newblock
\newblock
\urldef\tempurl%
\url{https://doi.org/10.48550/arXiv.2309.02065}
\showDOI{\tempurl}


\bibitem[Wyllie et~al\mbox{.}(2024)]%
        {wyllie_fairness_2024}
\bibfield{author}{\bibinfo{person}{Sierra Wyllie}, \bibinfo{person}{Ilia Shumailov}, {and} \bibinfo{person}{Nicolas Papernot}.} \bibinfo{year}{2024}\natexlab{}.
\newblock \showarticletitle{Fairness {Feedback} {Loops}: {Training} on {Synthetic} {Data} {Amplifies} {Bias}}. In \bibinfo{booktitle}{\emph{Proceedings of the 2024 {ACM} {Conference} on {Fairness}, {Accountability}, and {Transparency}}} \emph{(\bibinfo{series}{{FAccT} '24})}. \bibinfo{publisher}{Association for Computing Machinery}, \bibinfo{address}{New York, NY, USA}, \bibinfo{pages}{2113--2147}.
\newblock
\showISBNx{9798400704505}
\urldef\tempurl%
\url{https://doi.org/10.1145/3630106.3659029}
\showDOI{\tempurl}


\bibitem[Xing et~al\mbox{.}(2024)]%
        {xing_when_2024}
\bibfield{author}{\bibinfo{person}{Xiaodan Xing}, \bibinfo{person}{Fadong Shi}, \bibinfo{person}{Jiahao Huang}, \bibinfo{person}{Yinzhe Wu}, \bibinfo{person}{Yang Nan}, \bibinfo{person}{Sheng Zhang}, \bibinfo{person}{Yingying Fang}, \bibinfo{person}{Mike Roberts}, \bibinfo{person}{Carola-Bibiane Schönlieb}, \bibinfo{person}{Javier~Del Ser}, {and} \bibinfo{person}{Guang Yang}.} \bibinfo{year}{2024}\natexlab{}.
\newblock \bibinfo{title}{When {AI} {Eats} {Itself}: {On} the {Caveats} of {AI} {Autophagy}}.
\newblock
\newblock
\urldef\tempurl%
\url{https://doi.org/10.48550/arXiv.2405.09597}
\showDOI{\tempurl}


\bibitem[Xu et~al\mbox{.}(2023)]%
        {xu_combating_2023}
\bibfield{author}{\bibinfo{person}{Danni Xu}, \bibinfo{person}{Shaojing Fan}, {and} \bibinfo{person}{Mohan Kankanhalli}.} \bibinfo{year}{2023}\natexlab{}.
\newblock \showarticletitle{Combating {Misinformation} in the {Era} of {Generative} {AI} {Models}}. In \bibinfo{booktitle}{\emph{Proceedings of the 31st {ACM} {International} {Conference} on {Multimedia}}} \emph{(\bibinfo{series}{{MM} '23})}. \bibinfo{publisher}{Association for Computing Machinery}, \bibinfo{address}{New York, NY, USA}, \bibinfo{pages}{9291--9298}.
\newblock
\showISBNx{9798400701085}
\urldef\tempurl%
\url{https://doi.org/10.1145/3581783.3612704}
\showDOI{\tempurl}


\bibitem[Young et~al\mbox{.}(2022)]%
        {young_confronting_2022}
\bibfield{author}{\bibinfo{person}{Meg Young}, \bibinfo{person}{Michael Katell}, {and} \bibinfo{person}{P.M. Krafft}.} \bibinfo{year}{2022}\natexlab{}.
\newblock \showarticletitle{Confronting {Power} and {Corporate} {Capture} at the {FAccT} {Conference}}. In \bibinfo{booktitle}{\emph{2022 {ACM} {Conference} on {Fairness}, {Accountability}, and {Transparency}}}. \bibinfo{publisher}{ACM}, \bibinfo{address}{Seoul Republic of Korea}, \bibinfo{pages}{1375--1386}.
\newblock
\showISBNx{978-1-4503-9352-2}
\urldef\tempurl%
\url{https://doi.org/10.1145/3531146.3533194}
\showDOI{\tempurl}


\bibitem[Zhao et~al\mbox{.}(2023)]%
        {zhao_invisible_2023}
\bibfield{author}{\bibinfo{person}{Xuandong Zhao}, \bibinfo{person}{Kexun Zhang}, \bibinfo{person}{Zihao Su}, \bibinfo{person}{Saastha Vasan}, \bibinfo{person}{Ilya Grishchenko}, \bibinfo{person}{Christopher Kruegel}, \bibinfo{person}{Giovanni Vigna}, \bibinfo{person}{Yu-Xiang Wang}, {and} \bibinfo{person}{Lei Li}.} \bibinfo{year}{2023}\natexlab{}.
\newblock \bibinfo{title}{Invisible {Image} {Watermarks} {Are} {Provably} {Removable} {Using} {Generative} {AI}}.
\newblock
\newblock
\urldef\tempurl%
\url{https://doi.org/10.48550/arXiv.2306.01953}
\showDOI{\tempurl}
\newblock
\shownote{arXiv:2306.01953 [cs]}.


\bibitem[Zhou et~al\mbox{.}(2023)]%
        {zhou_synthetic_2023}
\bibfield{author}{\bibinfo{person}{Jiawei Zhou}, \bibinfo{person}{Yixuan Zhang}, \bibinfo{person}{Qianni Luo}, \bibinfo{person}{Andrea~G Parker}, {and} \bibinfo{person}{Munmun De~Choudhury}.} \bibinfo{year}{2023}\natexlab{}.
\newblock \showarticletitle{Synthetic {Lies}: {Understanding} {AI}-{Generated} {Misinformation} and {Evaluating} {Algorithmic} and {Human} {Solutions}}. In \bibinfo{booktitle}{\emph{Proceedings of the 2023 {CHI} {Conference} on {Human} {Factors} in {Computing} {Systems}}} \emph{(\bibinfo{series}{{CHI} '23})}. \bibinfo{publisher}{Association for Computing Machinery}, \bibinfo{address}{New York, NY, USA}, \bibinfo{pages}{1--20}.
\newblock
\showISBNx{978-1-4503-9421-5}
\urldef\tempurl%
\url{https://doi.org/10.1145/3544548.3581318}
\showDOI{\tempurl}


\bibitem[Zuboff(2023)]%
        {zuboff_age_2023}
\bibfield{author}{\bibinfo{person}{Shoshana Zuboff}.} \bibinfo{year}{2023}\natexlab{}.
\newblock \showarticletitle{The {Age} of {Surveillance} {Capitalism}}.
\newblock In \bibinfo{booktitle}{\emph{Social {Theory} {Re}-{Wired}} (\bibinfo{edition}{3} ed.)}. \bibinfo{publisher}{Routledge}.
\newblock
\showISBNx{978-1-00-332060-9}


\bibitem[Özsoy(2024)]%
        {ozsoy_energy_2024}
\bibfield{author}{\bibinfo{person}{Tufan Özsoy}.} \bibinfo{year}{2024}\natexlab{}.
\newblock \showarticletitle{The “energy rebound effect” within the framework of environmental sustainability}.
\newblock \bibinfo{journal}{\emph{WIREs Energy and Environment}} \bibinfo{volume}{13}, \bibinfo{number}{2} (\bibinfo{year}{2024}).
\newblock
\showISSN{2041-840X}
\urldef\tempurl%
\url{https://doi.org/10.1002/wene.517}
\showDOI{\tempurl}


\end{thebibliography}

\newpage

%
% If your work has an appendix, this is the place to put it.
\begin{appendix}  \label{appendix}

\section{System Dynamics Archetypes} \label{appendixa}

\begin{center}
\begin{longtable}
{|>{\raggedright\arraybackslash}m{0.35\linewidth}|>{\raggedright\arraybackslash}m{0.25\linewidth}|>{\raggedright\arraybackslash}m{0.4\linewidth}|} \hline 

\caption{Select System Dynamics Archetypes} \\

         Archetype&  Behaviour over Time (BoT)& Description and Example\\ \hline 
\endhead 

\hline \multicolumn{3}{|r|}{{Continued on next page}} \\ \hline
\endfoot

\hline \hline
\endlastfoot

         %% limits to growth archetype
        \includegraphics[width=1\linewidth]{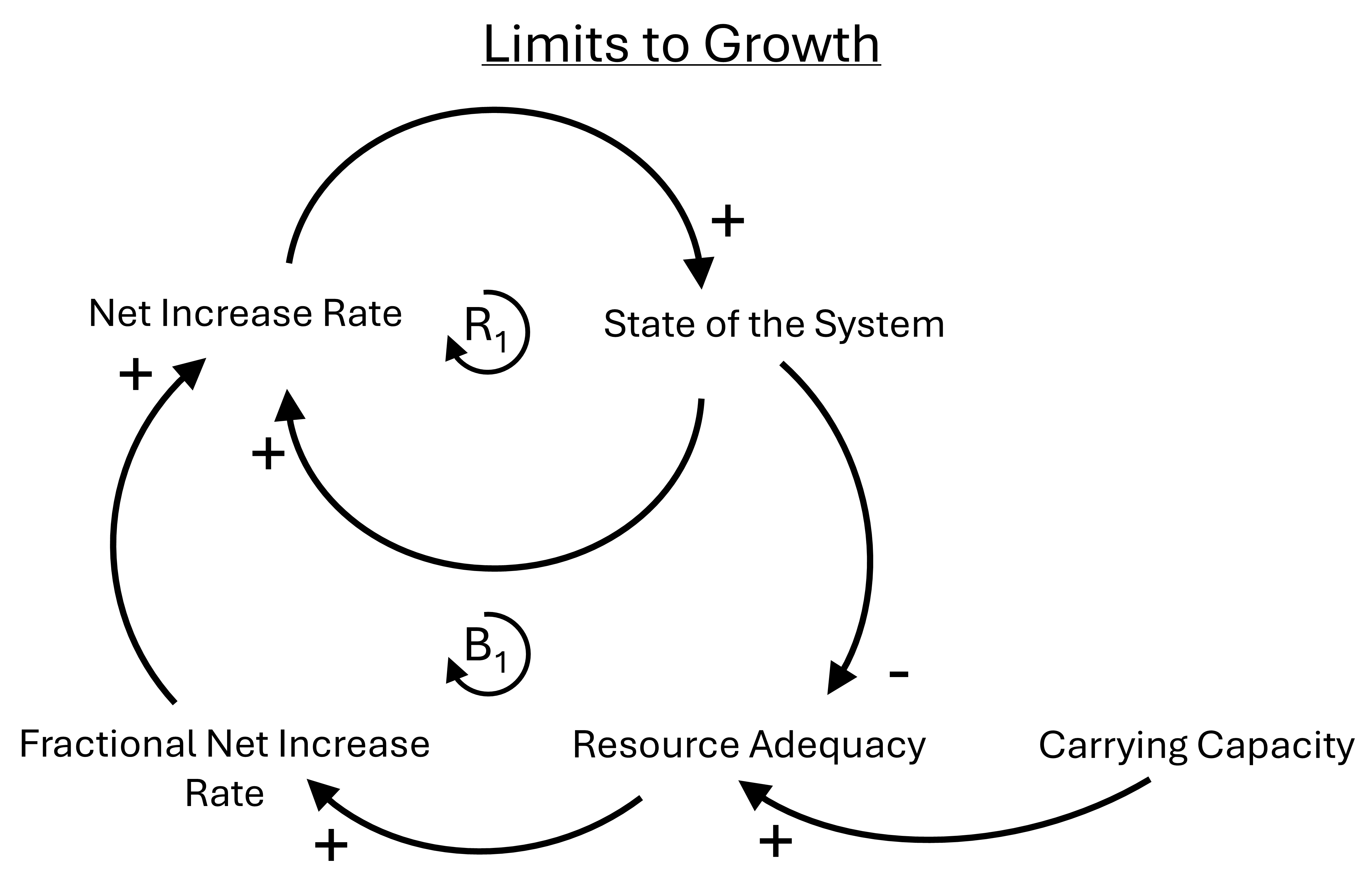}& \includegraphics[width=1\linewidth]{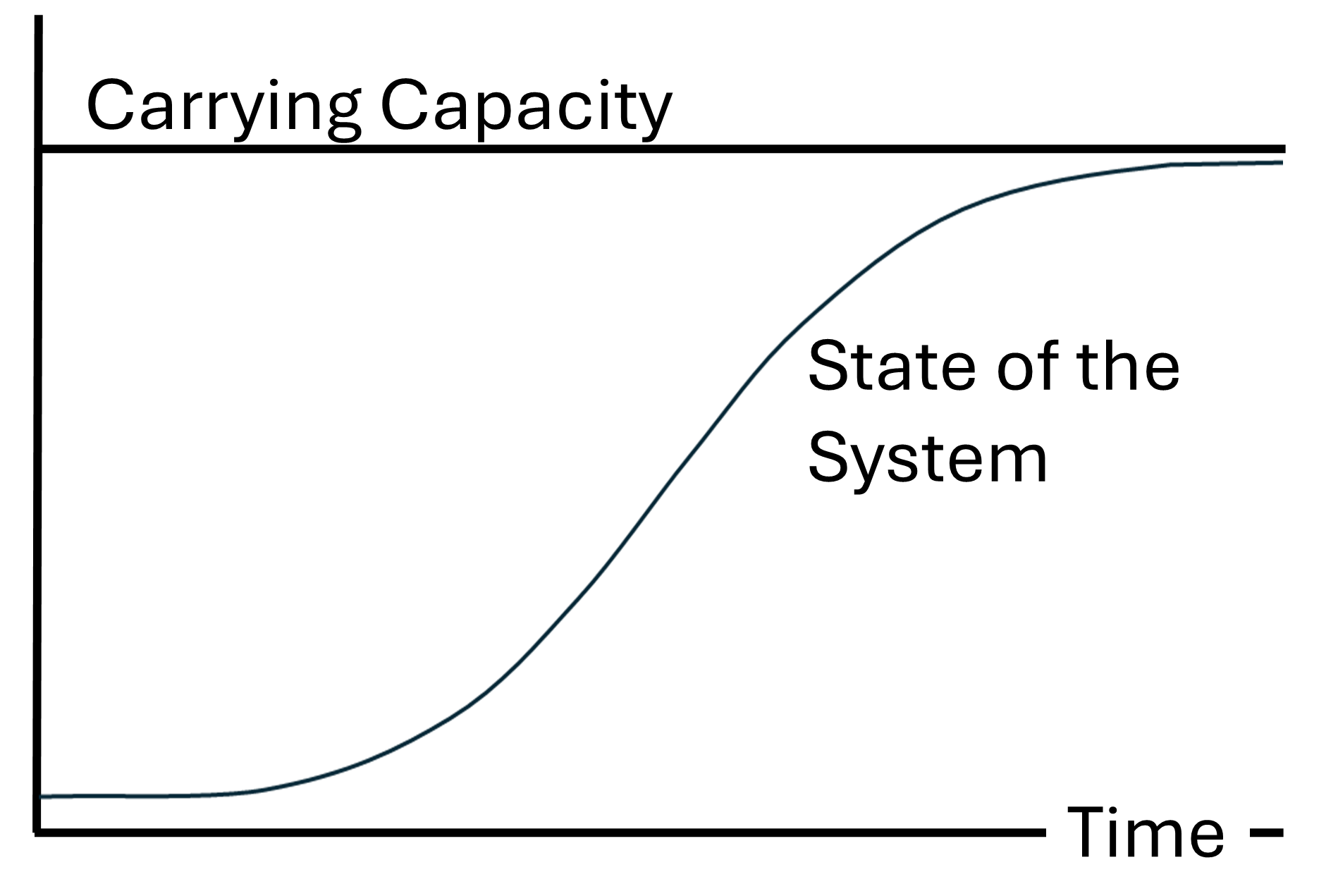}& The \textbf{limits to growth} archetype is used to model ``S-shaped growth'' in which  exponential growth starts fast but then slows as it approaches limits imposed by constraints \cite{sterman_business_2000}. For example, marketing leads to increased sales and thus increased revenue, which increases the marketing budget until the increased sales also increase the need for customer service. The customer service capacity limits the quality of customer service, causing decreased revenue. At this point, increasing the market budget as before will not cause increased revenues \cite{Kim_1994_archetype2}. Exponential growth occurs while the reinforcing loop is dominant. As the resource adequacy declines (customer service capacity), the balancing loop increasingly dominates the dynamics \cite{sterman_business_2000}.  \\ \hline 

         %% rebound effect archetype
        \includegraphics[scale=0.12]{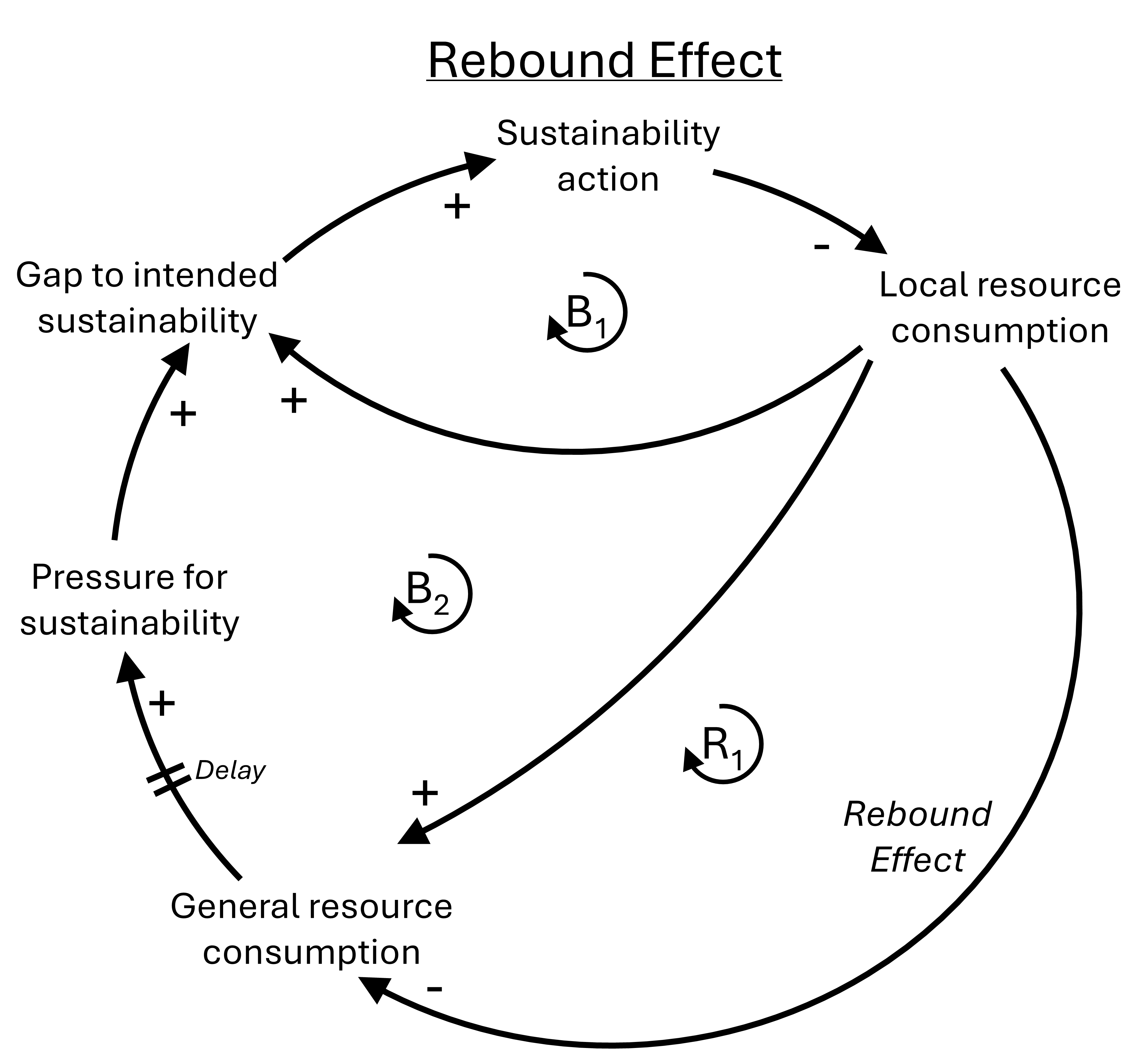}& \includegraphics[width=1\linewidth]{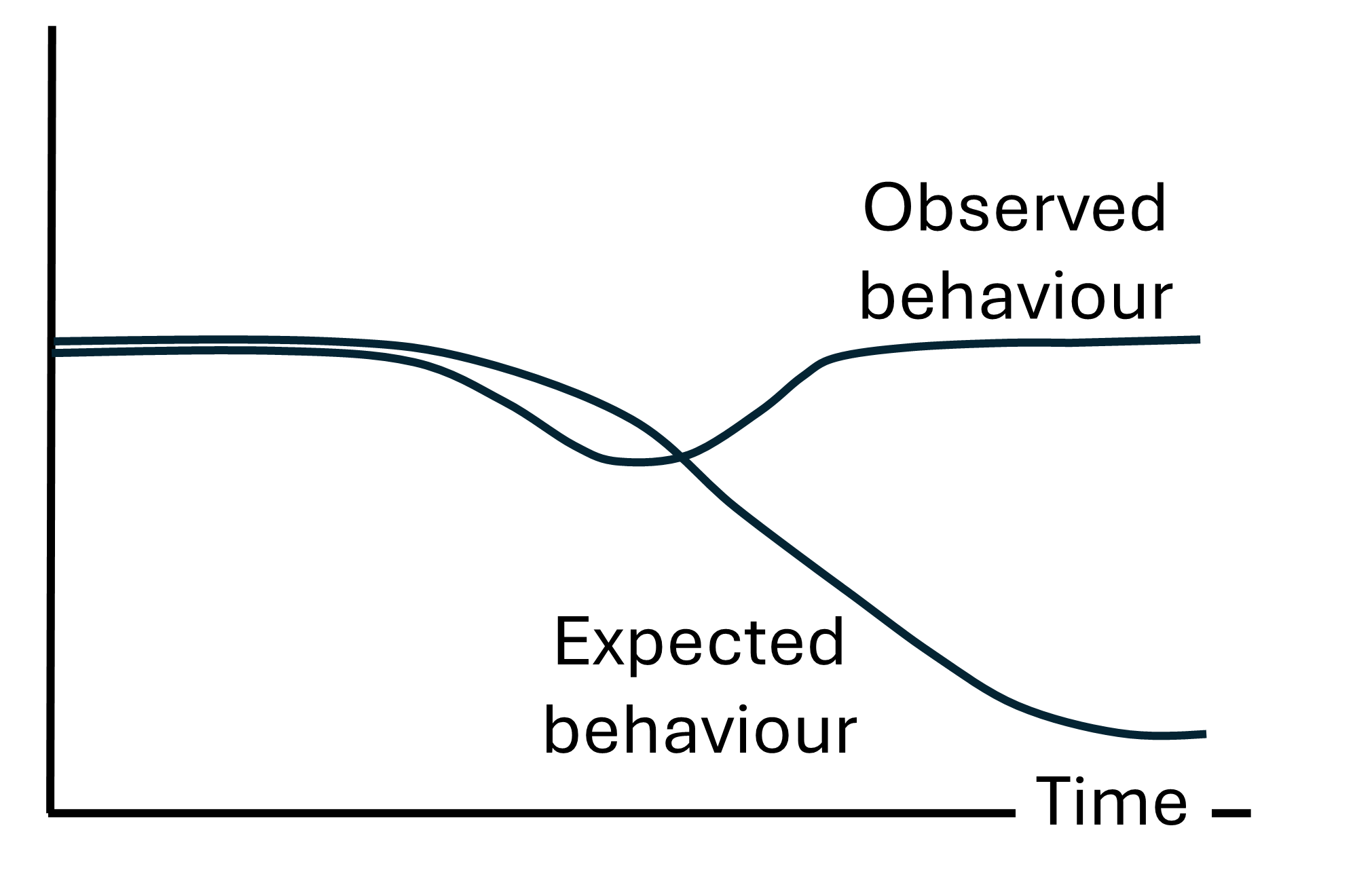} & \textbf{Rebound effects} are not a classic archetype but are a common trajectory for systems. The sample CLD provided is adapted from \cite{Guzzo_2024_rebound} and the BoT from \cite{Lorenz__2024}. For rebound effects, efficiency improvements lead to increased usage of the resource counteracting the efficiency gains. For example, fuel efficiencies can cause more car travel thereby increasing fuel usage \cite{Stapleton_2016}. Rebound effects can occur by several different mechanisms, detailed comprehensively in \cite{Guzzo_2024_rebound}.  \\ \hline 

         %% fixes that fail archetype
         \includegraphics[scale=0.18]{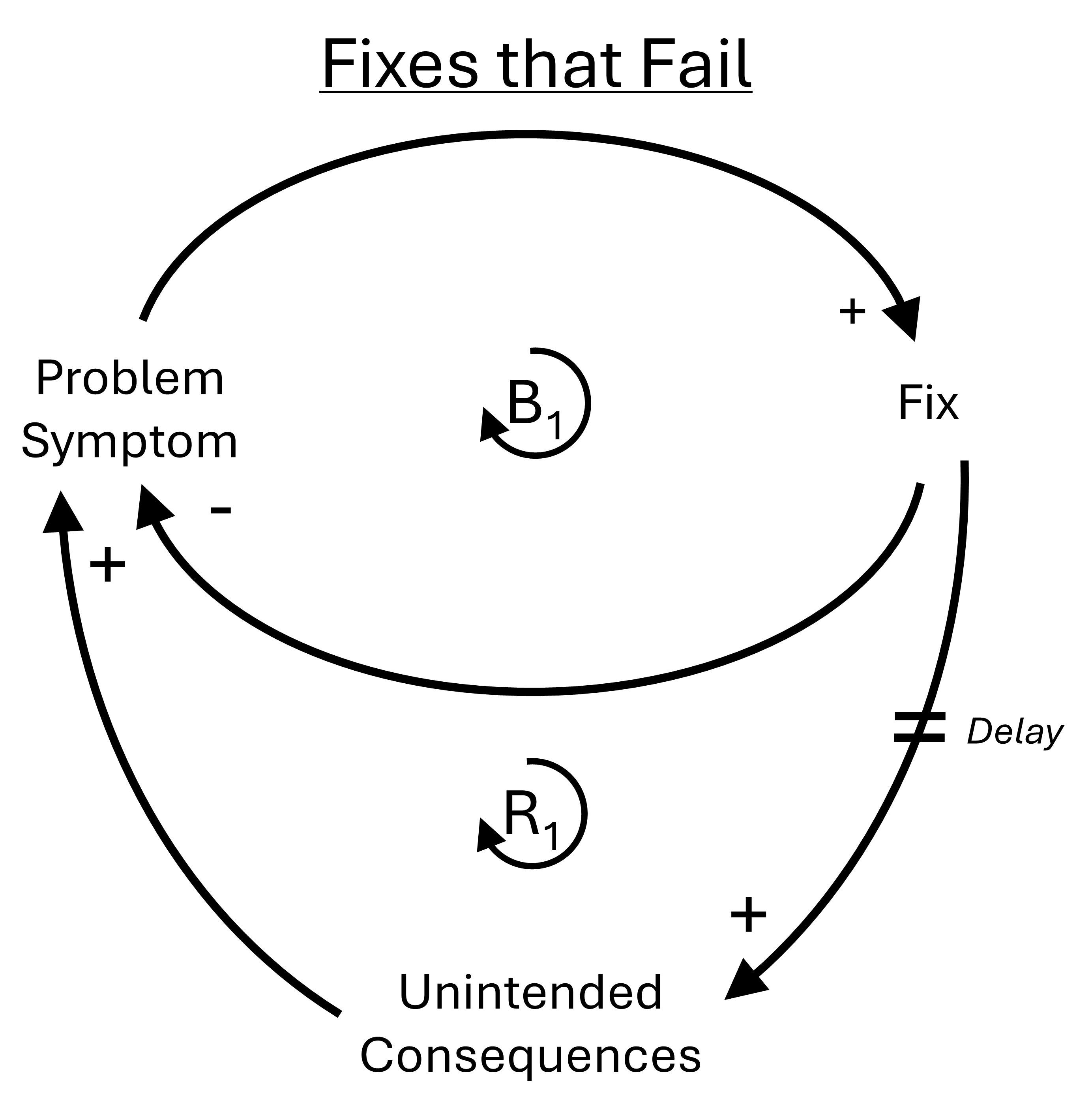}& \includegraphics[width=1\linewidth]{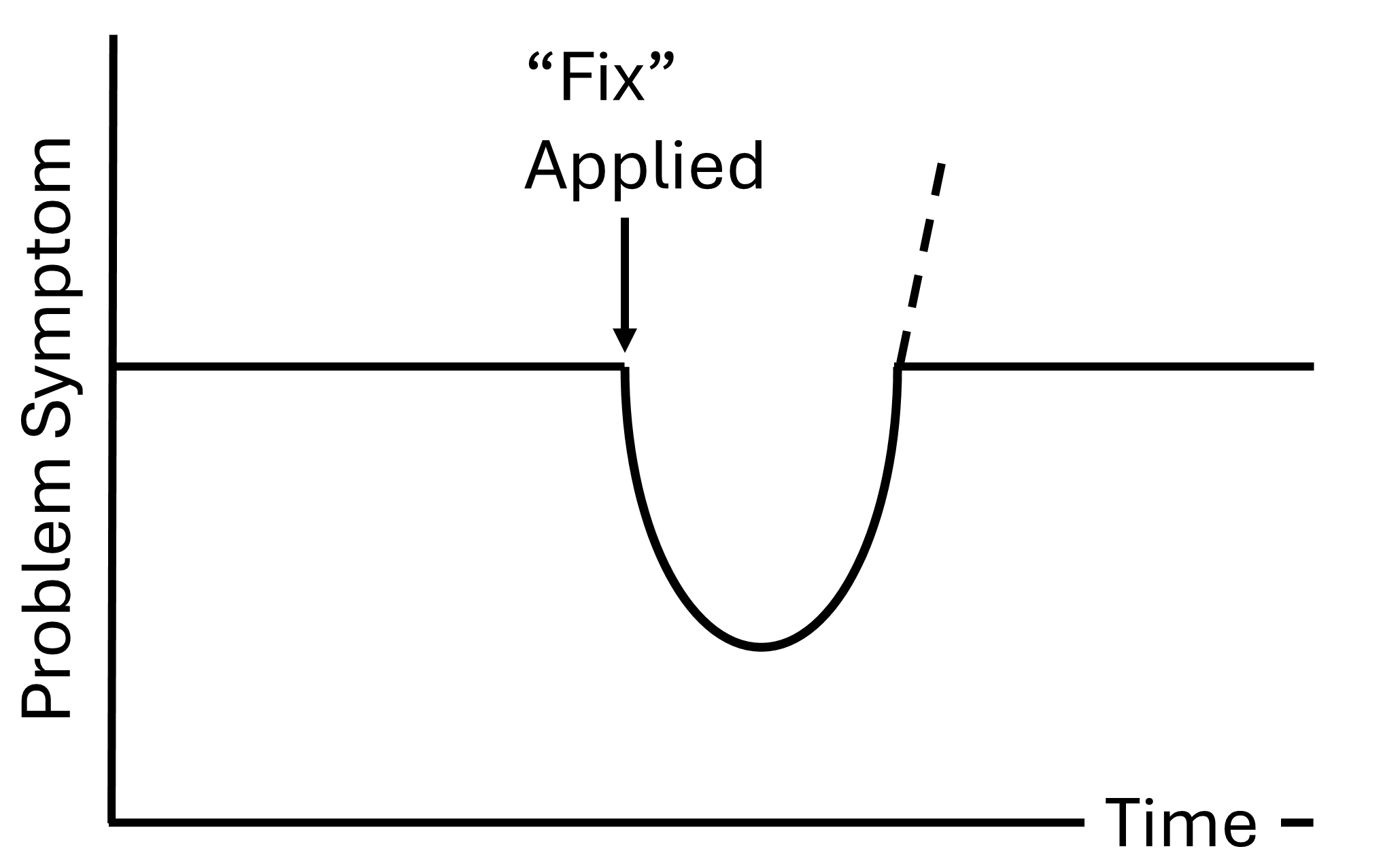}& The \textbf{fixes that fail} archetype warns against the instinct to implement ``quick fixes''. In this archetype, a situation has worsened to a crisis level, leading to the implementation of a quick fix. Rather than solving the issue, the fix causes unintended consequences that over time does not resolve the original problem and can sometimes exacerbate it (shown by the dotted line in the BoT) \cite{Kim_2000_archetype3}. For example, expediting customer orders (fix) to alleviate customer dissatisfaction (problem) causes disruptions to the shipping workflow (unintended consequences) which makes customers unhappy again because of the late orders \cite{Kim_1992_archetype1}.\\ \hline

         %% shifting the burden archetype
         \includegraphics[width=1\linewidth]{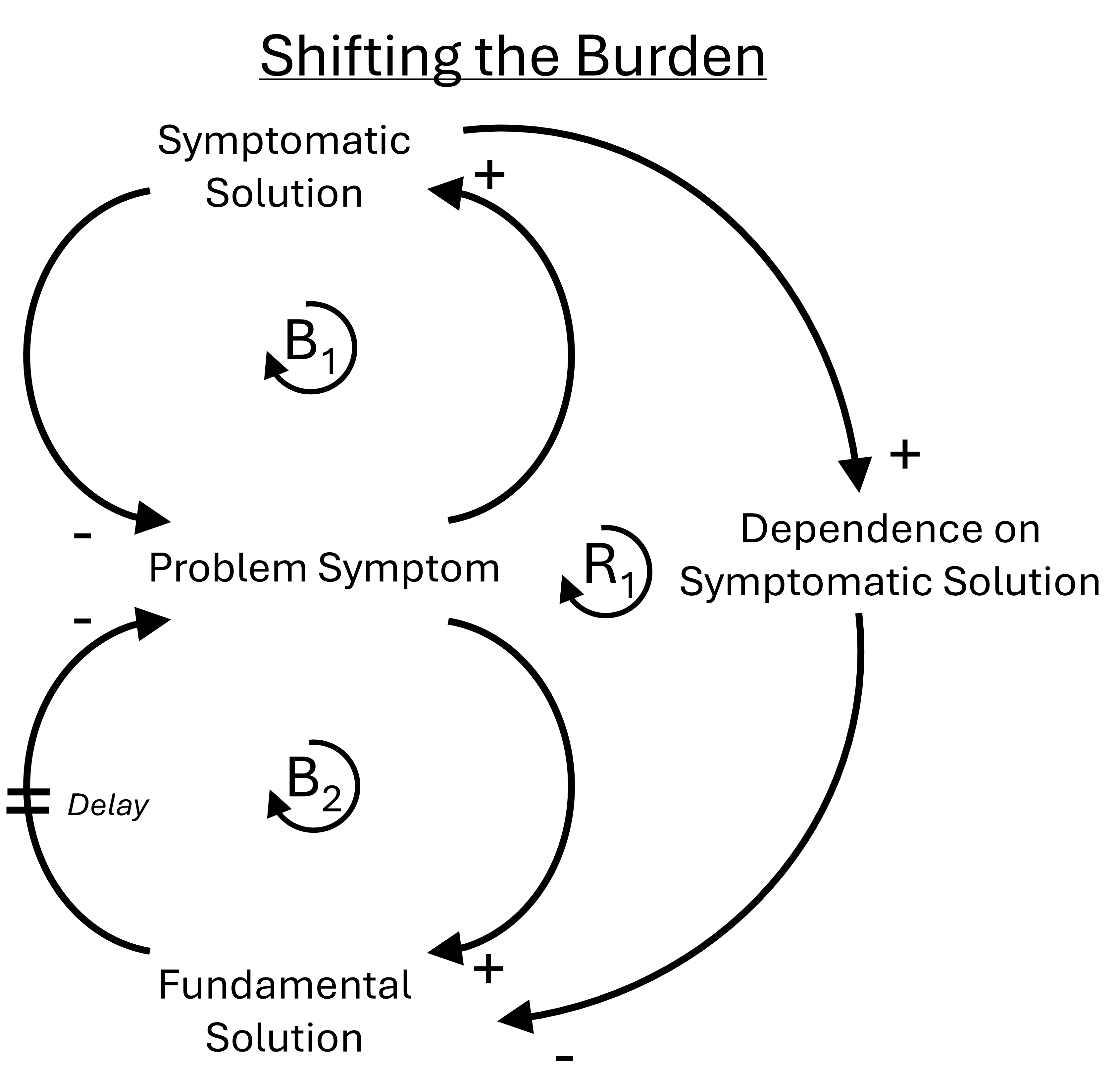}& \includegraphics[width=1\linewidth]{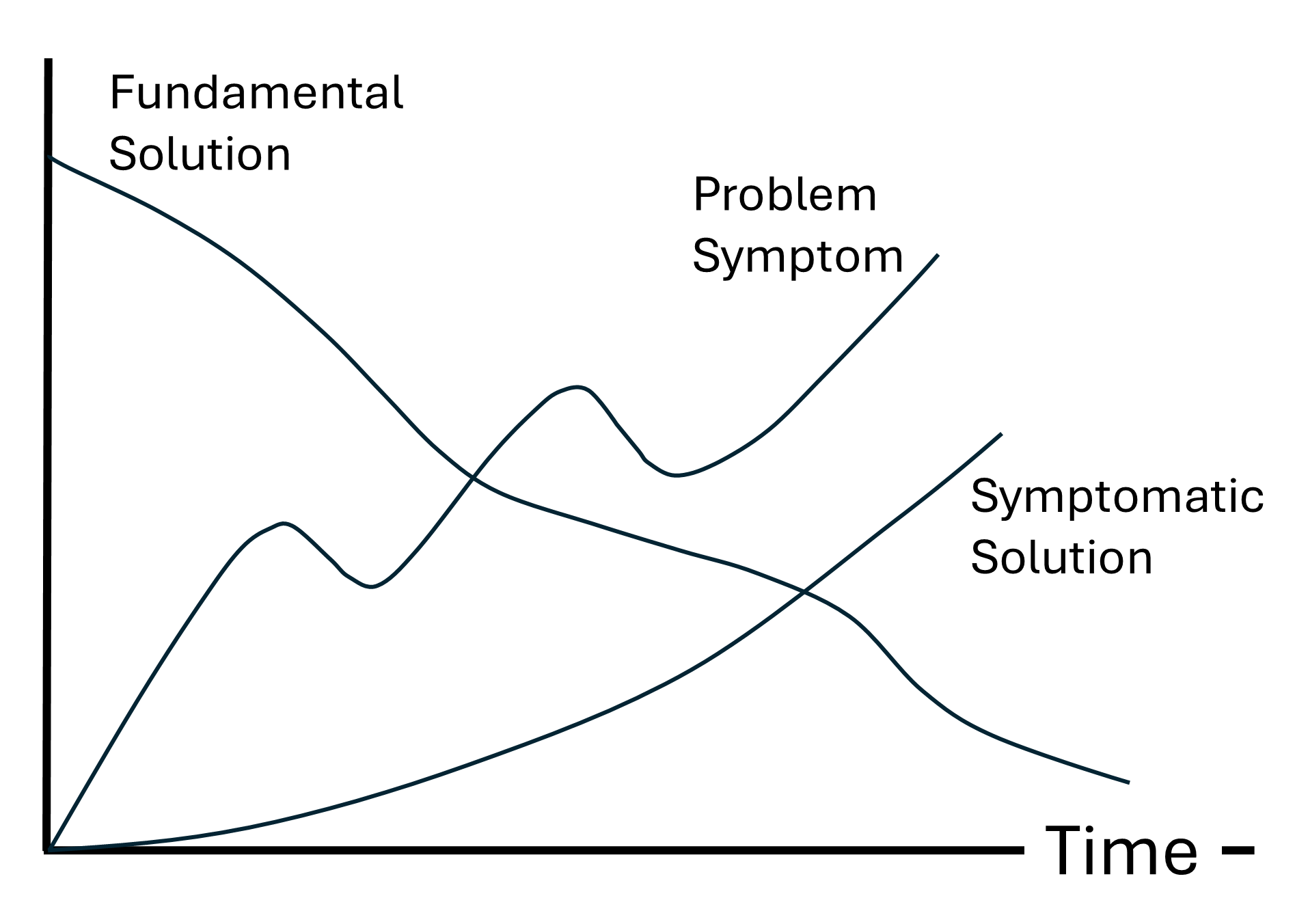}&  Similar to fixes that fail, the \textbf{shifting the burden} archetype models situations where a reaction is needed to an urgent problem resulting in the implementation of a ``quick fix''. A symptomatic solution is applied to resolve a problem which creates a diversion (or sometimes a side effect of dependency on the symptomatic solution) from the fundamental solution \cite{Kim_1992_archetype1}. The difference in the two archetypes is that fixes that fail requires identifying the correct solution whereas shifting the burden requires identifying a ``fundamental capability'' that needs to be established. \cite{Kim_2000_archetype3}. For example, a team depends on one employee who knows how to use the printer. This creates a dependence on that employee who may later quit, leaving the team unable to use the printer.   \\ \hline 

         %% overshoot and oscillate archetype
        \includegraphics[width=1\linewidth]{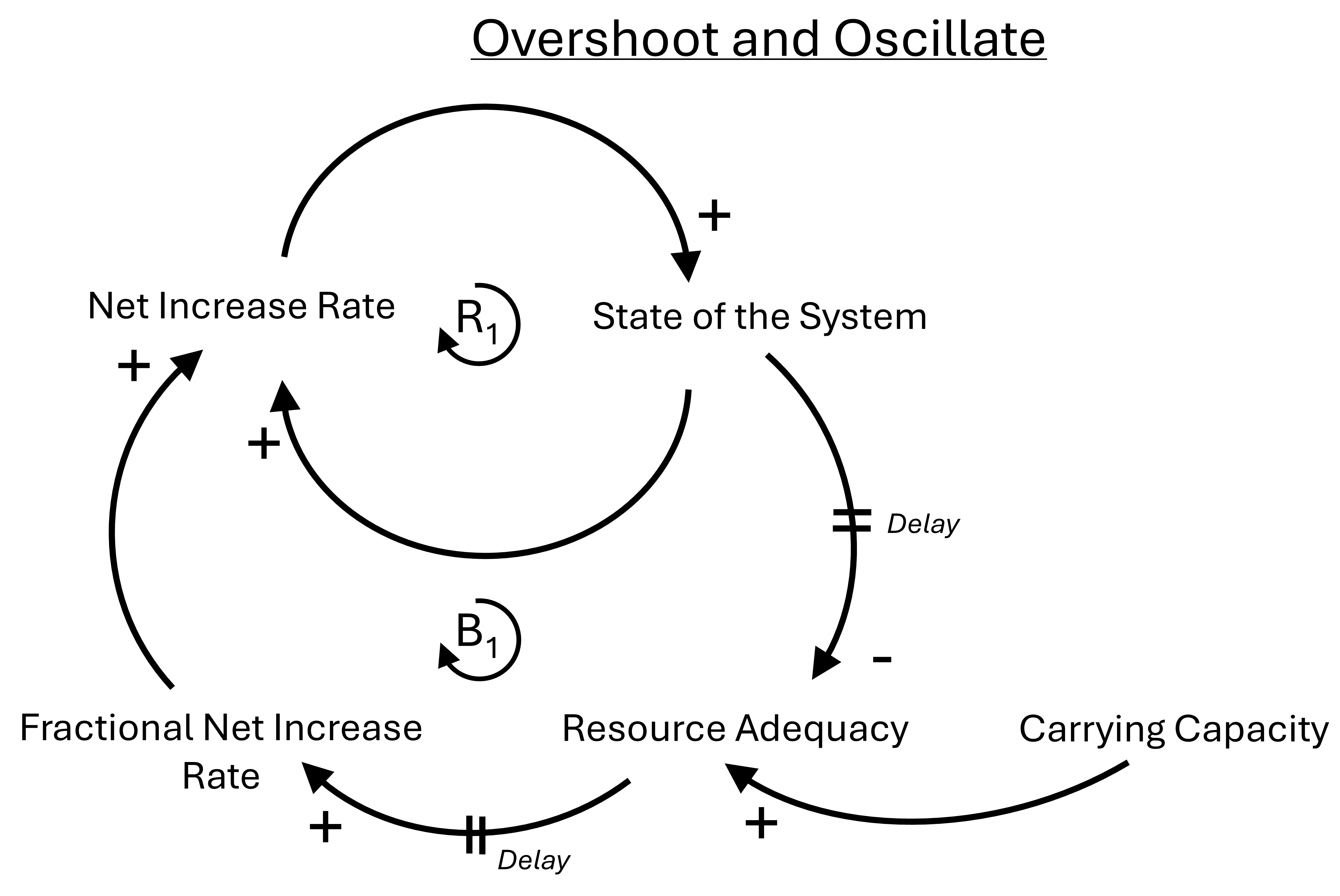}&\includegraphics[width=1\linewidth]{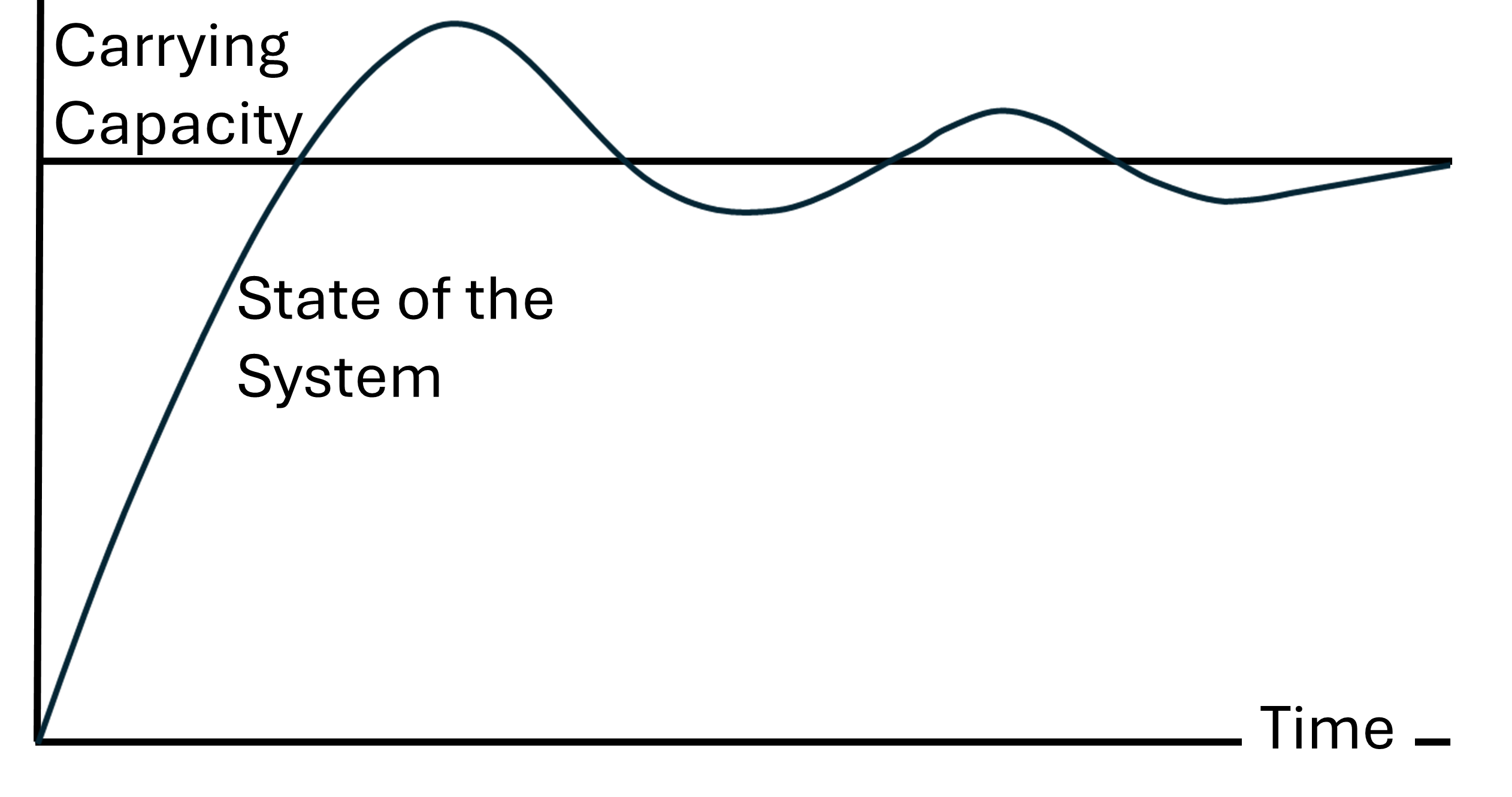}& The \textbf{overshoot and oscillate} archetype occurs when the negative loop in a S-shaped growth has significant time delays. For example, if a company faced increased demand for their product, they would increase the production of that product to meet the growth in demand \cite{Wardman_2016}. However, if production is already at capacity, this creates a delay before production can be scaled. By the time the production is scaled, there is overshoot in the production causing an over-correction in the opposite direction resulting in oscillation \cite{sterman_business_2000}.     \\ \hline

         %% overshoot and collapse archetype
        \includegraphics[width=1\linewidth]{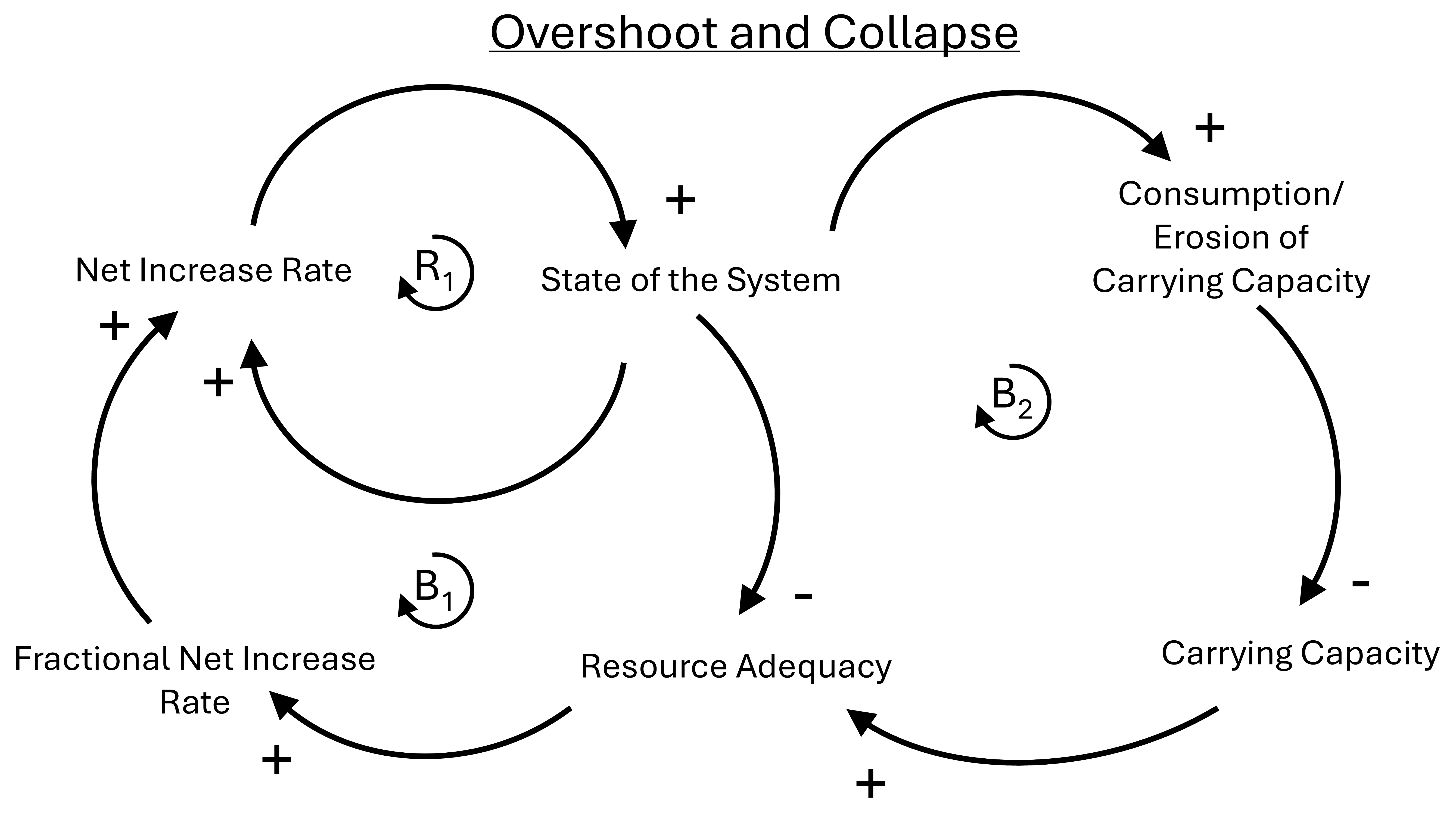}&\includegraphics[width=1\linewidth]{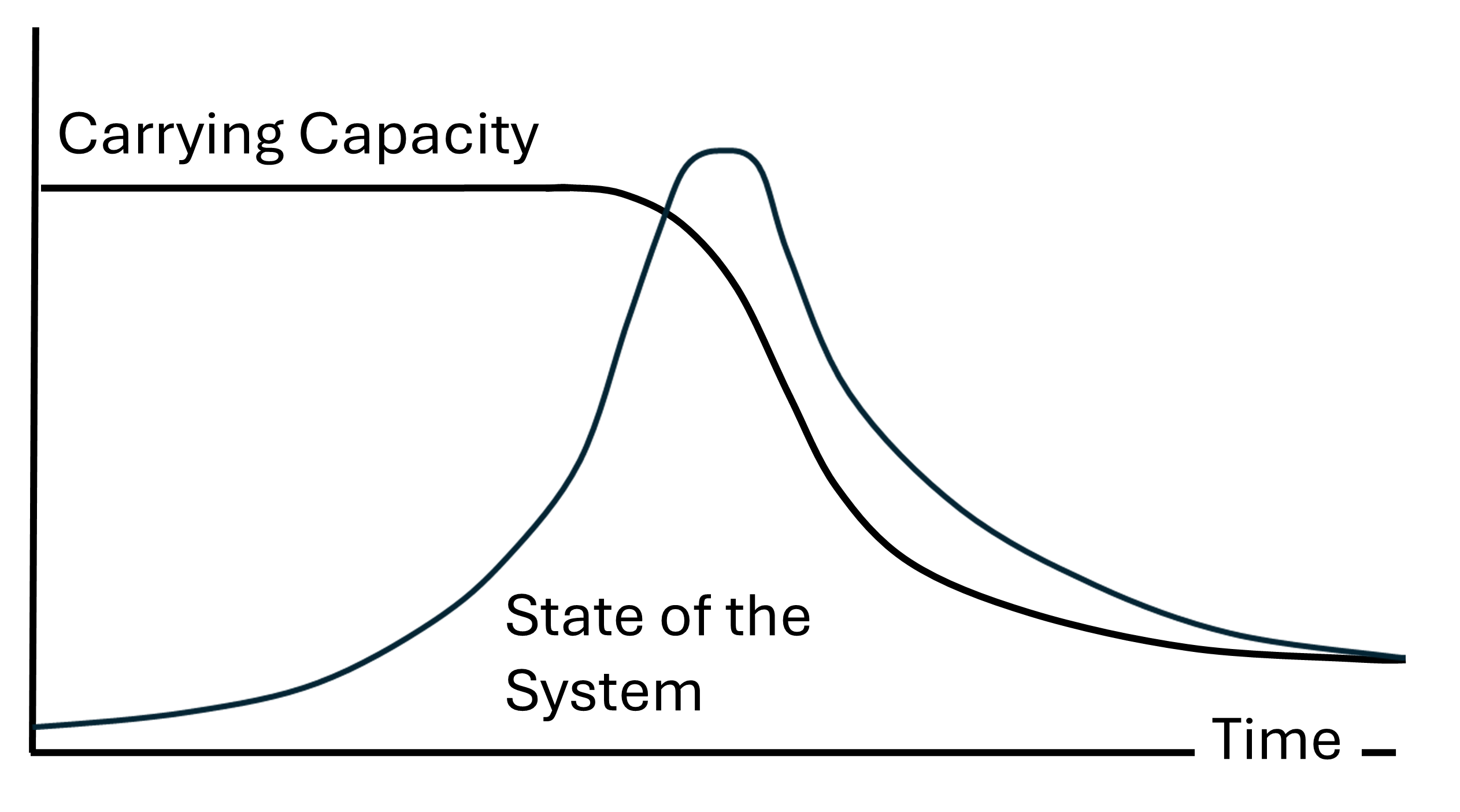}&\textbf{Overshoot and collapse} occurs when the carrying capacity is eroded by the system. An often cited example of this is overfishing causing irreversible erosion of the carrying capacity making it unable to regenerate and thus collapse \cite{sterman_business_2000}. Increasingly this is a prospect for planetary ecosystems. \\ \hline

\end{longtable}
\end{center}

\section{Causal Relationships of AI Scaling} \label{appendixb}

\begin{center}
\begin{longtable}
{|>{\raggedright\arraybackslash}p{0.05\linewidth}|>{\raggedright\arraybackslash}p{0.3\linewidth}|>{\raggedright\arraybackslash}p{0.65\linewidth}|} \hline

\caption{Details on the variables, causal relationships, and rationale for Figs. 4-7}  \\

Loop&  Variables, Causal Link, Polarity& Description and rationale/examples\\ \hline 
          
\endhead 

% \multicolumn{3}{c}%
% {\raggedright\arraybackslash}p{0.05\linewidth}|>{\raggedright\arraybackslash}p{0.3\linewidth}|>{\raggedright\arraybackslash}p{0.65\linewidth}| \hline
%          Loop&  Variables, Causal Link, Polarity& Description and rationale/examples\\ \hline 
          
% \endhead

\hline \multicolumn{3}{|r|}{{Continued on next page}} \\ \hline
\endfoot

\hline \hline
\endlastfoot

 \multicolumn{3}{|c|}{Figure 4}\\\hline
 R\textsubscript{1}& Scale of models →+ Documented model performance&As the scale of models increases, there is an increase in performance gains which results in increased recognized and documented model performance. For example, increased scaling enabled by ImageNet \cite{deng_imagenet_2009} led to the development of AlexNet \cite{krizhevsky_imagenet_2012}. The performance gains achieved by AlexNet lead to a broader documentation and recognition of a new benchmark in model performance \cite{krizhevsky_imagenet_2012}. \\\hline
 R\textsubscript{1}& Documented model performance →+ Capital investment&As documented and modeled performance has increased, capital investment in AI has increased, e.g., \cite{noauthor_what_2024}. \\\hline
 R\textsubscript{1}&Capital investment →+ Scale of models  &As capital investment increases, the scale of models also increases due to investments in R\&D, operations, technical advancements, etc.\\\hline
 R\textsubscript{2}&AI hype →+ Competitive pressure &Increased AI hype leads to more competitive pressure between frontier companies, as they race to achieve the hyped, “transformational” capabilities of AI.\\\hline
 R\textsubscript{2}&Competitive pressure →+ Capital investment  &Increased competitive pressure leads to increased capital investment that is sought by the AI industry in order to scale models. \\\hline
 R\textsubscript{3}&Capital investment →+ AI hype &Increased capital investment leads to increased AI hype because the monetary investment shows that investors believe and support the narrative that AI is transformational and has many benefits and will result in increased profits. \\\hline
 R\textsubscript{3}& AI hype →+ Capital investment &Increased AI hype leads to increased capital investment because increased marketing around AI spurs more investment cycles.\\\hline
 R\textsubscript{4}&Documented model performance →+ AI hype &An increase in documented model performance results in more AI hype because it reinforces the narrative of “bigger-is-better” in AI scaling \cite{varoquaux_hype_2024}. \\\hline
 R\textsubscript{5}& Capital investment →+ Scale of hardware&Higher levels of capital investment lead to increased scale of hardware, such as data centers \cite{wong_silicon_2024}. \\\hline
 R\textsubscript{6}& Scale of models →+ Scale of hardware&As the scale of models increases, the scale of hardware must also increase in order to support it \cite{luccioni_environmental_2024}. \\\hline
 R\textsubscript{6}&Scale of hardware →+ Scale of models &As the scale of hardware increases (e.g., better GPUs), the scale of models also increases (e.g., increased compute) \cite{luccioni_environmental_2024}. \\\hline
 B\textsubscript{1}&Ability to scale models →+ Scale of models&Increased ability to scale models results in more scaling of the models. This means that increasing the ability would enable increased scaling of model parameters, volume of data, and amount of compute \cite{kaplan_scaling_2020}. \\\hline
 B\textsubscript{1}& Scale of models →– Ability to scale models&As the scale of models increases, we have less ability to scale them even further (based on diminishing returns for performance) \cite{lohn_scaling_2023}. \\\hline
 & Capacity of scaling as per scaling laws →+ Ability to scale models&As the capacity of scaling as per scaling laws increases, there is increased ability to scale models. Currently this is a limit due to the scaling laws of AI \cite{dan_hendrycks_ai_2024,kaplan_scaling_2020}.\\\hline
 &Scale of hardware →+ Emissions and natural resource consumption &As the scale of hardware increases, the resulting emissions and natural resource usage also increases. For example, more powerful GPUs enable greater amounts of power consumption \cite{kindig_ai_2024}. \\\hline
 & Scale of models →+ Emissions and natural resource consumption&As the scale of models increases, the resulting emissions and natural resource usage also increases. For example, more parameters lead to more CO\textsubscript{2} emissions \cite{luccioni_environmental_2024}.\\\hline
 & Emissions and natural resource consumption →+ Environmental damages&Increased emissions and natural resource usage results in environmental damages that contribute to breaching planetary boundaries \cite{rockstrom_planetary_2009,creutzig_digitalization_2022}. \\\hline
 & Emissions and natural resource consumption →– Ecological capacity&Increased emissions and natural resource usage results in a decreased ecological capacity, i.e., less energy and water availability.\\\hline 
 & Capital investment →+ Centralization&Increased capital investment leads to increased centralization of power. \\\hline
 & Centralization →+ Social harms&Increased centralization of power results in increased social harms such as but not limited to compute divide \cite{besiroglu_compute_2024}, data grab \cite{couldry_data_2019,mejias_data_2024}, monopolies \cite{couldry_data_2019,mejias_data_2024,melissa_heikkila_this_2024}, capture (e.g., \cite{jeans_data_2021}), and labour exploitation \cite{williams_exploited_2022}.\\\hline
 & Social harms →– Social acceptability&Increased social harms leads to decreased social acceptability as seen with the organized resistance against AI (discussed in Section \ref{social}).\\\hline
 
\multicolumn{3}{|c|}{Figure 5}\\ \hline
R\textsubscript{1}&  Documented model performance →+
Capital Investment& As documented and modeled performance has increased, capital investment in AI has increased, e.g., (\cite{noauthor_what_2024}).\\ \hline 
         R\textsubscript{1}&  Capital investment →+ 
Scale of models&As capital investment increases, the scale of models also increases due to investments in R\&D, operations, technical advancements, etc. \\ \hline
         R\textsubscript{1}&  Scale of models →+
Documented model performance& As the scale of models increases, there is an increase in performance gains which results in increased recognized and documented model performance. For example, increased scaling enabled by ImageNet \cite{deng_imagenet_2009} led to the development of AlexNet \cite{krizhevsky_imagenet_2012}. The performance gains achieved by AlexNet lead to a broader documentation and recognition of a new benchmark in model performance \cite{krizhevsky_imagenet_2012}.  \\ \hline 
         R\textsubscript{2}&  Efforts in improving efficiency →+
Incremental efficiency gains for water and energy&Increased efforts to improve efficiency result in incremental efficiency gains for water and energy, e.g., \cite{argerich_measuring_2024}.  \\ \hline 
         R\textsubscript{2}&  Incremental efficiency gains for water and energy →+
Scale of models
& The increased incremental gains result in further scaling of models (i.e., rebound effect) \cite{willenbacher_rebound_2022}.\\ \hline
         R\textsubscript{2}&  Capital investment →+
Scale of hardware
& Higher levels of capital investment lead to increased scale of hardware, such as data centers \cite{wong_silicon_2024}.\\ \hline
         R\textsubscript{2}&  Scale of hardware →+ 
Emissions and natural resource usage&As the scale of hardware increases, the resulting emissions and natural resource usage also increases \cite{kindig_ai_2024}.\\ \hline
 R\textsubscript{2}& Emissions and natural resource usage →+ 
Efforts in improving efficiency &The increasing emissions and natural resource usage leads to increased efforts in improving efficiency to reduce the cost of energy and water consumption and reduce the environmental impact , e.g., \cite{alex_irwin-hunt_thirsty_2023,david_patterson_how_2021}.\\\hline
 & Emissions and natural resource usage →+ 
Environmental damages&Increased emissions and natural resource usage results in environmental damages that contribute to breaching planetary boundaries \cite{rockstrom_planetary_2009,creutzig_digitalization_2022}. \\\hline
 R\textsubscript{3}&  Scale of models →+
Scale of hardware &As the scale of models increases, the scale of hardware must also increase in order to support it \cite{luccioni_environmental_2024}.\\ \hline
 B\textsubscript{1}& Incremental efficiency gains for water and energy →–
Emissions and natural resource usage&Increased efficiency gains for water and energy lead to decreased usage of water and energy. \\\hline

 \multicolumn{3}{|c|}{Figure 6}\\\hline
 B\textsubscript{1}& Availability of usable training data  →– 
Development of synthetic data&As the availability of usable (high-quality, diverse) training data decreases, development of synthetic data increases \cite{lee_synthetic_2024,Long_Wang_Xiao_Zhao_Ding_Chen_Wang_2024}. \\\hline
 B\textsubscript{1}& Development of synthetic data →+
Availability of usable training data&As more synthetic data is developed, the availability of training data increases.\\\hline
 R\textsubscript{1}& Development of synthetic data →+ Perceived diversity and representativeness&The development of synthetic data leads to an increased \textit{perception} of data diversity and representativeness \cite{whitney_real_2024}. Although in reality, it causes “statistical diversity without representational diversity” \cite{whitney_real_2024}. \\\hline
 R\textsubscript{1}& Perceived diversity and representativeness →+ 
Usage of synthetic data to train models&The increased \textit{perception} of diversity and representativeness leads to its increased usage to train models \cite{whitney_real_2024}. \\\hline
 R\textsubscript{1}& Usage of synthetic data to train models →+ 
Pollution of training data&As more synthetic data is used to train models, it causes increased pollution of training data. This “is a degenerative process affecting generations of learned generative models, in which the data they generate end up polluting the training set of the next generation. Being trained on polluted data, they then misperceive reality.” \cite[p.~755]{shumailov_ai_2024}. \\\hline
 R\textsubscript{1}& Pollution of training data →–
Availability of usable training data&An increase in pollution of training data leads to decreased availability of usable (high-quality, diverse) training data.\\\hline
 R\textsubscript{2}
& Development of synthetic data →+
Usage of synthetic data to train models&As more synthetic data is developed, there is an increase in its usage to train models, e.g., \cite{lee_synthetic_2024,Long_Wang_Xiao_Zhao_Ding_Chen_Wang_2024}.\\\hline
 R\textsubscript{3}& Usage of synthetic data to train models →–
Ability to distinguish between synthetic data and human data&An increase in using synthetic data to train models leads to a decrease in the ability to distinguish between synthetic data and human data due to inadequate ability to watermark \cite{Jeng_Chang_Duffy_Lam_Maercklein_2024,zhao_invisible_2023}.\\\hline
 R\textsubscript{3}& Ability to distinguish between synthetic data and human data →– Pollution of training data&An increased ability to distinguish between synthetic data and human data leads to decreased pollution of training data \cite{xing_when_2024}.\\\hline

 \multicolumn{3}{|c|}{Figure 7}\\\hline
 
 B\textsubscript{1}& Shortfall in energy supply →+ Investment in energy capacity&In recent years, an increase in the gap of energy required to power AI and its supply (in other words, shortfall in energy supply), has led to increased investment in energy capacity development, such as new electricity grids, upgrading electricity grids, and nuclear energy \cite{grene_microsoft_2023,Lawson_2024}.\\\hline
 B\textsubscript{1}&Investment in energy capacity →+ Energy capacity &Increased investment in energy capacity leads to increased energy capacity. \\\hline
 B\textsubscript{1}&Energy capacity →– Shortfall in energy supply &Increased energy capacity leads to decreased shortfall in energy supply, because of the new sources of energy. \\\hline
 R\textsubscript{1} & Energy capacity →+Energy dependency&Increased energy capacity leads to increased dependence on growing amounts and sources of energy, e.g., \cite{wang_environmental_2023}.\\\hline
 R\textsubscript{1} & Energy dependency →+ Energy demand&Increased dependency leads to increased energy demand because of industry priorities to continue scaling AI. \\\hline
 R\textsubscript{1} & Energy demand →+ Shortfall in energy supply&Increased energy demand creates an increased gap or shortfall in energy supply. \\\hline
 B\textsubscript{2}& Energy demand →+ Innovations in frugal AI&A better solution to address increased energy demand would be to increase innovations in frugal AI. \\\hline
 B\textsubscript{2}&Innovations in frugal AI →– Shortfall in energy supply &Increased innovations in frugal AI would more effectively reduce the shortfall in energy supply, as it maximizes energy supply by focusing on making consumption practices more sustainable \cite{govindan_how_2024}. \\\hline
 R\textsubscript{2}&Dependence on scaling energy sources →– Innovations in frugal AI &Energy dependency diverts attention away from innovations in frugal AI. A continued dependence indicates that industry priorities have not shifted from scaling general-use AI to smaller, task-specific AI, thus reducing innovations in frugal AI. \\\hline
\end{longtable}
\end{center}

\end{appendix}

\end{document}